\documentclass[12pt]{article}
\usepackage{amssymb,amsmath,amsthm,bm,graphicx,float}
\usepackage{mathrsfs,dsfont}
\usepackage{hyperref}
\hypersetup{
    colorlinks,
    citecolor=black,
    filecolor=black,
    linkcolor=black,
    urlcolor=black
}
\usepackage{color}

\usepackage{authblk}

\usepackage{natbib}
\usepackage{constants}
\usepackage{enumerate}
\begin{document}
\newcommand{\beq}{\begin{eqnarray}}
\newcommand{\eeq}{\end{eqnarray}}
\newcommand{\beas}{\begin{eqnarray*}}
\newcommand{\enas}{\end{eqnarray*}}
\newcommand{\bea}{\begin{eqnarray}}
\newcommand{\ena}{\end{eqnarray}}
\newcommand{\bms}{\begin{multline*}}
\newcommand{\ems}{\end{multline*}}
\newcommand{\qmq}[1]{\quad \mbox{#1} \quad}
\newcommand{\qm}[1]{\quad \mbox{#1}}
\newcommand{\To}{\rightarrow}
\newcommand{\nn}{\nonumber}
\newcommand{\bbox}{\hfill $\Box$}
\newcommand{\ignore}[1]{}
\newcommand{\bsy}[1]{\boldsymbol{#1}}
\newcommand{\Gg}{\mathcal{G}}
\newcommand{\Bvert}{\left\vert\vphantom{\frac{1}{1}}\right.}
\newcommand{\U}{{\cal U}}
\newcommand{\norm}[1]{\left\lVert#1\right\rVert}
\newcommand{\E}{{\mathbb E}}
\newcommand{\btheta}{{\boldsymbol{\theta}}}
\newcommand{\bTheta}{\Theta}
\newcommand{\bbeta}{\beta}
\newcommand{\logl}{{\cal L}}
\newcommand{\transpose}{\top}
\newcommand{\n}{\eta}
\newtheorem{theorem}{Theorem}[section]
\newtheorem{corollary}{Corollary}[section]
\newtheorem{conjecture}{Conjecture}[section]
\newtheorem{proposition}{Proposition}[section]
\newtheorem{lemma}{Lemma}[section]
\newtheorem{definition}{Definition}[section]
\newtheorem{example}{Example}[section]
\newtheorem{remark}{Remark}[section]
\newtheorem{case}{Case}[section]
\newtheorem{condition}{Condition}[section]
\newtheorem{assumption}{Assumption}[section]


\title{{\bf\Large $M$-estimation and deconvolution in a diffusion model with application to biosensor transdermal blood alcohol monitoring}}

\author{\textsc{Maria Allayioti},$^\dag$ \textsc{Jay Bartroff},$^\circ$ \textsc{Larry Goldstein},$^\dag$ \textsc{Haoxing~Liu},$^\dag$ \textsc{Susan~Luczak},$^\ddag$ and \textsc{Gary Rosen}$^\dag$\\
	\small{Departments of Mathematics$^\dag$ and Psychology,$^\ddag$ \\University of Southern California, Los Angeles, California,\\ Department of Statistics and Data Sciences, University of Texas at Austin$^\circ$\\}
}  
\footnotetext{Key words and phrases: $M$-estimation, regularization, diffusion model, blood alcohol, de-convolution, uniform confidence bounds} 

\date{}
\maketitle

\abstract{We develop $M$-estimation and deconvolution methodology with the goal of making well-founded statistical inference on an individual's blood alcohol level based on noisy  measurements of their skin alcohol content. We first apply our results to a nonlinear least squares estimator of the key parameter that specifies the blood/skin alcohol relation in a diffusion model, and establish its existence, consistency, and asymptotic normality. To make inference on the unknown underlying blood alchohol curve, we develop a basis space deconvolution approach with regulazation, and determine the asymptotic distribution of the error process, thus allowing us to compute uniform confidence bands on the curve.  Simulation studies show agreement between the performance of our curve estimators and their asymptotic distributions at low noise levels, and we apply our methods to a real skin alcohol data set collected via a transdermal biosensor.}
\section{Introduction and background}

Our goal is to model and estimate a human subject's alcohol concentration in the blood~(BAC) or breath\footnote{BAC and BrAC have been recognized as essentially quantitatively indistinguishable up to levels of legal intoxication, see \citet{Swift03}.}
(BrAC) as a function of the alcohol level measured at the skin,
i.e., the transdermal alcohol concentration (TAC), via a biosensor.  Approximately 1\% of the alcohol ingested in the human body is metabolized through the skin \citep[see][]{Swift00}. For decades it has been recognized that the levels of TAC are connected to those of BAC/BrAC, but also that there are challenges in modeling this relationship. Because alcohol has to pass from the blood through the skin to be captured by a TAC sensor placed on the surface of the skin, it is subject to variation across individuals (e.g., skin layer thickness, porosity, tortuosity, etc.) and drinking episodes (e.g., ambient temperature, humidity, subject activity level, skin hydration, vasodilation, etc.). These effects result in a TAC-BAC/BrAC relationship that can be highly variable. Thus TAC devices to date have typically been primarily used only in legal and research settings as abstinence monitors (e.g., in court mandated monitoring of DUI offenders) because of difficulties researchers have found translating raw TAC to the quantity of alcohol in the blood. 

\begin{figure}[H]
    \centering
    \includegraphics[scale=.5]{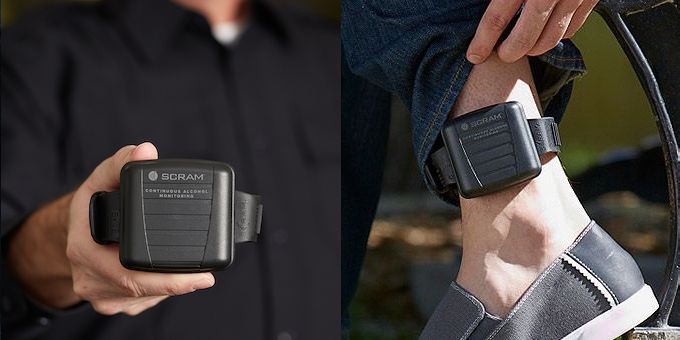}
     \caption{The Alcohol Monitoring System (AMS) Secure Continuous Alcohol Monitoring (SCRAM\textsuperscript{TM}) system. This device was used to gather the data analyzed in Section~\ref{sec:real.data}.}
\label{fig:scram}
\end{figure}

Still, TAC measured by a wearable biosensor device has great potential as a tool to improve personal and public health. It provides a passive, unobtrusive way to collect naturalistic data for extended periods of time. One such device is pictured in Figure~\ref{fig:scram}.  The same is not true about BrAC, which typically must be measured by trained research staff in the laboratory under controlled conditions using a breath analyzer, and thus  is less practical for capturing alcohol levels in the field under real-world conditions. Moreover, the breath analyzer requires a user to be compliant, potentially interferes with naturalistic drinking patterns, and is subject to inaccuracy (e.g., readings too high due to mouth alcohol, or too low due to not properly taking a deep lung breath for a reading). Thus, creating a system that reliably converts  TAC data into estimates of BAC (or BrAC) would greatly benefit the alcohol research and clinical communities who, along with  public health institutes, have been quite interested in such models \citep[see][]{Barnett15,Jung19,NIAAA16,Luczak19}.  Such a tool would dramatically improve the accuracy of field data and the validity of naturalistic studies of alcohol-related health outcomes, disease progression, treatment efficacy, and recovery. A wearable alcohol monitoring device could have consumer appeal as well, helping individuals monitor their own alcohol levels and make better health choices.

Previous work on the TAC-BAC/BrAC relationship began with deterministic models \citep{Banks97, Banks89,Curtain86,Gibson88,Pritchard87,Staffans05,Tucsnak09}  for the ``forward process'' of  the propagation of alcohol from the blood, through the skin, and its measurement by the sensor. Later approaches reversed the forward process to estimate BrAC based on the TAC 
\citep{Dai16,Dumett08,Luczak14,Luczak15,Luczak18,Rosen13,Rosen14,Weiss14}. These studies showed unaccounted for  variation in the TAC-BAC/BrAC relationship and subsequent work began to incorporate uncertainty into the models via a random diffusion equation \citep{Sirlanci17,Sirlanci18,Sirlanci19,Sirlanci19b,Sirlanci19c}. Other statistical modeling approaches include Hill-Kapturczak et al.'s \citeyearpar{Hill-Kapturczak15} regression model for peak BrAC using peak TAC, time of peak TAC, and gender using controlled laboratory data. \citet{Karns-Wright17} examined time delays from peak BrAC to peak TAC. \citet{Webster07,Webster08} used physics-based statistical models for the TAC-BAC/BrAC relationship but ultimatley concluded that, ``due to the highly variable relationship between the BAC and TAC curves, transdermal sensing of real-time BAC using only skin surface measurements may prove to be very challenging'' \citep[][p.\ 463]{Webster08}. 

In this paper, we seek to meet this challenge by using a physics-based statistical model that allows individual, device, and drinking episode level variation by treating the data from each person/device/episode triple as resulting from its own model parameters. We determine the large sample behavior of estimates of  these parameters  and give conditions under which these estimates are consistent and have a limiting normal distribution. We then use those results to give a statistically rigorous asymptotic characterization of the properties of the BAC/BrAC curve estimates obtained from measured TAC, including information on estimation error. As these estimates are made on an individualized basis, they will not be adversely affected when used in a study of a population whose characteristics vary widely. On the other hand, these estimates require individualized calibration over subject, device and environmental conditions.

Further work will generalize our current setting to one where the key model parameters depend on measurable subject and environmental covariates, and, if successful, would help remove much of the burden of calibration. Such an advancement would be an important step forward in the development of reliable and valid quantitative measurement of BAC/BrAC from TAC, of which the current work is the first step. 

The outline of this work is as follows. In Sections \ref{sec:diff.model} and \ref{sec:nonlinear.diffusion} we provide an outline of the partial differential equation diffusion model that drives our inference, and of our least squares approach for the estimation of the unknown vector. 
In Section \ref{sec:Mest} we consider the existence, consistency, and limiting distribution of our least squares estimators in a general $M$-estimation context, and present some examples.
In Section \ref{sec:ApplicationToDiffusion} we apply the results in Section \ref{sec:Mest}  to the diffusion model of Section \ref{sec:diff.model}, and present Theorems \ref{thm:ls.cons.asy.dist.qonly} and \ref{thm:ls.asy.dist.qonly}, which contain our main results on inference for the main parameter ${\bm q}$ of interest, and also for the TAC error variance.
In Section \ref{sec:InferenceBrAC} we apply the results of Section \ref{sec:ApplicationToDiffusion} for making inference on the BrAC curve, and in particular for the construction of uniform error bounds on the resulting curve estimate, and in Section \ref{sec:regularization} we consider regularized versions of our estimator that have improved resilience and stability at larger noise levels. Lastly, in Section \ref{sec:sim.and.data} we evaluate our theoretical results in simulations, and illustrate their behavior using a set of BrAC/TAC observations taken in the lab.

\subsection{Diffusion model}\label{sec:diff.model}
Although our goal is to model a human subject's BAC/BrAC as a function of TAC, the ethanol molecules themselves move in the other direction: from the blood, through the skin, to ultimately be measured by the sensor on the surface of the skin. Thus the relevant physics describe the TAC as a function of BAC/BrAC. We consider a specific model~\eqref{PDE model}  for this transport based on Fick's law of diffusion  \citep[see][]{Smith04} which depends on an unknown, 2-dimensional parameter~$\bm{q}=(q_1,q_2)$.  The result is TAC expressed as a convolution of BAC/BrAC with a kernel or filter, and as a function of the unknown~$\bm{q}$ which we then estimate via nonlinear least squares as described in Section~\ref{sec:nonlinear.diffusion} and whose properties we consider in Section~\ref{sec:ApplicationToDiffusion}. These properties  determine the inferential consequences for BAC/BrAC estimation, and in particular have a large impact on the accuracy of the estimated BraAC curve, as studied in Section \ref{sec:InferenceBrAC}.

Let $x(t,\zeta)$ denote the concentration of ethanol at time~$t \ge 0$ and depth~$\zeta \in [0,1]$ from the skin surface through epidermis, choosing units so that $\mu(t)=x(t,1)$, $t \ge 0$ is the BAC at time~$t$.  A Fick's law-based model \citep[see][Section~2]{Rosen14,Sirlanci19} has  been developed and  used successfully to model data of this type; here we only summarize the main parts. The model specifies $x(t,\zeta)$ as the solution to the  partial differential equation, with boundary condition
\begin{equation}
\label{PDE model}
\frac{\partial x}{\partial t} = q_1 \frac{\partial^2 x}{\partial \zeta^2}, \quad q_1\left. \frac{\partial x}{\partial \zeta}\right|_{\zeta=1}=q_2 \mu(t), \quad q_1 \left.\frac{\partial x}{\partial \zeta}\right|_{\zeta=0}= x \Bvert_{\zeta=0},
\end{equation}
depending on the parameter~$\bm{q}=(q_1,q_2)$.   
The TAC at skin level is then $x(t,0)$. When we want to emphasize dependency on the parameter $\bm{q}$ we will write, for instance, $\mu(t;\bm{q})$. 

The system \eqref{PDE model} with its boundary conditions can be solved in continuous time in terms of unbounded linear operators \citep[see][Section~2]{Sirlanci19}, with solution 
\begin{align}
x(t) &=e^{A(\bm{q})t}x(0) + \int_0^t e^{A(\bm{q})(t-s)}B(\bm{q}) \mu(s)ds. \label{xt.diff.soln.infinite.dim}
\end{align}

In cases we consider, $x(0)$ will be the zero function, that is, observation begins at, or before, the time of first intake of alcohol. By taking a discretization of the distance $\zeta$ from skin level into $k$ steps for some $k$ sufficiently large, the operators in  \eqref{xt.diff.soln.infinite.dim} can be approximated by $k$ dimensional linear operators (i.e., matrices) yielding the approximation to the solution to \eqref{PDE model} given by
\begin{align}
x^{(k)}(t) &=\int_0^t e^{A^{(k)}(\bm{q})(t-s)}B^{(k)}(\bm{q}) \mu(s)ds. \label{xt.diff.soln}
\end{align}
Now fixing, and suppressing in the notation, the level of discretization $k$ , an observation taken at time $t$ can be represented as the linear function of $x(t)$ given by 
\begin{align} 
f_{\mu}(t;\bm{q}) &= \int_0^t Ce^{A(\bm{q})(t-s)}B(\bm{q}) \mu(s)ds, \label{eq:k.is.conv}
\end{align}
plus an additive error term. For observations taken at skin level, the vector $C$ will have a one in its first component, and zeros elsewhere.

The matrices in \eqref{eq:k.is.conv} depend on the unknown parameter $\bm{q}$ as
\begin{equation}
\label{eq:AB.by.q}
A(\bm{q})=q_1D+E \qmq{and} B(\bm{q}) =q_2 F,
\end{equation}
where $C, D, E$, and $F$ are known matrices that result from making the finite-dimensional approximation discussed; methods of computation and consistency for approximating the infinite dimensional solution have been established in \citep[Section 6.2]{Sirlanci19}. 
More precise assumptions and properties of these matrices, and the domain of $\bm{q}$, will be specified in Section~\ref{sec:ApplicationToDiffusion}.

\subsection{Nonlinear least squares estimation}\label{sec:nonlinear.diffusion}

To estimate the parameter ${\bm q}$, we assume that TAC data $\{y_{ij},1 \le i \le n, 1 \le j \le m_i\}$ is collected on a single individual over $n$ different drinking episodes at the $m_i$ times $0 \le t_{i,1}<\cdots<t_{i,m_i} \le T_i$, for given BrAC curves $\mu_i$ on $[0,T_i]$. With ${\bm m}=(m_1,\ldots,m_n)$, the estimator minimizes
\bea \label{eq:Jq.def}
J_{n,{\bm m}}(\bm{q})=\frac{1}{2\sum_{i=1}^n m_i}\sum_{i=1}^n \sum_{j=1}^{m_i} \left( f_{\mu_i}(t_{ij};\bm{q})-y_{ij}\right)^2, 
\ena
where $f_{\mu_i}(t_{ij};\bm{q})$ is given by the right hand side of \eqref{eq:k.is.conv} with $\mu$ replaced by $\mu_i$, the BrAC curve for drinking episode $i$.
The model specified by \eqref{eq:k.is.conv} and \eqref{eq:AB.by.q} is deterministic, but to account for measurement variability, we include additive, homoscedastic errors on the observed values TAC values. The constant variance condition implies that all TAC observations are `equally reliable', and that the error variances, in particular, do not depend on the length of time elapsed since the last observation. For that reason, the least squares objective functions give equal weight to their summands, and when appropriate, weights, inversely proportional to variance, could be included.
We may also allow the length of the time interval $T_i$ of the $i^{th}$ episode, and the location of the sampling times, to be stochastic.

\section{$M$-estimation: Existence, consistency, and limiting distribution}\label{sec:Mest}

In this section we consider $M$-estimation in a general setting that contains what we will require to handle the diffusion model we consider. Our results may be viewed as an extension of existing results on $M$-estimation.  Textbooks that cover $M$-estimation tend to focus on the case  of a univariate parameter  \citep[e.g.,][Chapter~7.2]{Maronna19,Serfling80}, whereas ours covers the  multivariate case. The closest results to ours that we know of are by \citet{Jennrich69}, who obtained similar results but in a setting that is more restrictive in a number of ways.  First, \citet{Jennrich69} considers only least squares estimation whereas our results apply to the more general estimating equation~\eqref{theta-solves}. Second, these previous results only apply to approximate normality and require i.i.d.\ error terms, whereas our Theorem~\ref{mle-normal} can be applied to other limiting distributions and relaxed conditions on the error terms, although our main application is to limiting normality. Finally, these previous results are more restrictive in terms of a number of technical conditions, such as compactness of the parameter space~$\Theta$ which our results do not require, and the existence of ``tail products'' of vectors of observation means and error terms, which our results eschew in favor of more conventional regularity conditions on the score type  function~$\U_n$.

After establishing the notation and setup in Section~\ref{sec:gen.setup}, we state our main results in Section~\ref{sec:gen.main}. In Section~\ref{sec:gen.exs} we provide some general examples of the applications of our results to least squares and maximum likelihood estimation.

\subsection{Set up and summary of results}\label{sec:gen.setup}
 For $n \ge 1$, observed data ${\bm X}^{(n)}$ in a space $\chi^{(n)}$, a parameter space~$\bTheta \subset {\mathbb R}^p$ having non-empty interior, and a function $\U_n: \bTheta \times \chi^{(n)}\rightarrow {\mathbb R}^p$, consider the estimating equation 
\bea \label{theta-solves} \U_n(\btheta) = 0, \quad \btheta \in \bTheta,
\ena 
where the dependence of $\mathcal{U}_n$ on the data is suppressed. 
In our examples $\chi^{(n)}$ will a Euclidean space endowed with a family of densities  $p_n(\bm{x}^{(n)};\btheta), \btheta \in \Theta$ which generate the data from this family with $\btheta=\btheta_0$. 
Two important situations in which the solutions of such equations arise are maximum likelihood and least squares estimation. 

For maximum likelihood, under smoothness conditions on the densities, the maximizer of the log likelihood $\logl_n(\btheta)= \log p_n(\bm{x}^{(n)};\btheta)$ is given as a solution to \eqref{theta-solves} with
\begin{equation}\label{est.eq.MLE}
{\mathcal U}_n(\btheta)= \partial_{\btheta}\logl_n(\btheta;\bm{X}^{(n)}),
\end{equation}
where $\partial_\btheta$ denotes taking derivative with respect to $\btheta$, resulting in a column vector of partial derivatives when $\btheta$ itself is a vector. When the data ${\bm X}^{(n)}$ consists of $n$ independent random vectors ${\bm X}_1,\ldots,{\bm X}_n$ in $\mathbb{R}^d$, each with distribution $p({\bm x}; \btheta_0)$, the space $\chi^{(n)}$ can be identified with $\mathbb{R}^{d\times n}$, and $p_n({\bm x}^{(n)},\btheta)$ is the product of the marginal densities $p({\bm x}_i; \btheta)$ for $i=1,\ldots,n$.

To introduce least squares estimation, suppose that pairs $(\bm{x}_i,y_i) \in \mathbb{R}^d \times \mathbb{R},i=1,\ldots,n$, are observed with distribution depending on $\btheta$  for which 
\begin{align*}
\mathbb{E}_\btheta[y_i|\bm{x}_i]=f_i(\bm{x}_i;\btheta) 
\end{align*}
for $f_i(\bm{x};\btheta)$ in some parametric class of functions. With ${\bm x}^{(n)}=({\bm x}_1,\ldots,{\bm x}_n)$, the least squares estimate of $\btheta$ is given as the minimizer of 
\begin{align*} 
J(\btheta;\bm{x}^{(n)}) = \frac{1}{2n}\sum_{i=1}^n \left(y_i-f_i(\bm{x}_i;\btheta) \right)^2,
\end{align*}
which under smoothness conditions can be obtained via  \eqref{theta-solves} with
\begin{align} \label{eq:least.squares.J}
{\mathcal U}_n(\btheta) = \partial_{\btheta} J(\btheta;\bm{x}^{(n)})=\frac{1}{n}\sum_{i=1}^n \left(f_i(\bm{x}_i;\btheta) -y_i\right) \partial_{\btheta} f_i(\bm{x}_i;\btheta).\end{align}

The aim of the estimating equation $\U_n(\btheta) = 0$
is to provide a value close to the one where the function
$\U_n(\btheta)$ takes the value of 0 in some expected, or asymptotic,
sense. In particular, in Theorem \ref{theorem-consistent-root} we will show, under that when $\U_n(\btheta_0)$ is, under an appropriate scaling, close to zero as
$n \rightarrow \infty$, then the sequence of estimates obtained via the estimating equations will be consistent for the true parameter.

In Theorem \ref{mle-normal}, we will also provide a corresponding limiting distribution for solutions to the estimating equation \eqref{theta-solves}. Let $\U_n(,\btheta)$ have components
$$
\U_n(\btheta)=(\U_{n,j}(\btheta))_{1 \le j \le
p} \quad \mbox{where} \quad \U_{n,j}: {\mathbb R}^n \times \bTheta
\rightarrow {\mathbb R}.
$$

We will write $\U_n'(\btheta)$ as short for  $\partial_{\btheta}\U_n^\top (\btheta) \in \mathbb{R}^{p\times p}$. 
This quantity is the observed information matrix in the case of maximum likelihood estimation, and where we have \eqref{est.eq.MLE} under the
assumption of the existence and continuity of second derivatives of $\logl_n$ for $\btheta \in \bTheta$, its $k,j^{th}$ component is given by
$$
\frac{\partial \U_{n,j}(\btheta)}{\partial
	\theta_k}=
\frac{\partial^2 \logl_n(\btheta)}{\partial
	\theta_k \partial \theta_j} =
\frac{\partial^2 \logl_n (\btheta)}{\partial
	\theta_j \partial \theta_k}=
\frac{\partial \U_{n,k}(\btheta)}{\partial \theta_j}.
$$
In this case, the third condition in \eqref{score-at-true-to-zero}  below is equivalent to the condition that the limiting information matrix $I$ is positive definite.
Tolerating a slight abuse of notation,
we may write $\partial_j$ rather than $\partial_{\theta_j}$ when taking a partial with respect to the $j^{th}$ coordinate variable, and $\partial_j^m$ for the $m^{th}$ order derivative, so for instance denoting the $k,j^{th}$ entry of  $\U_n'(\btheta)$  by $\partial_k \U_{n,j}(\btheta)$.

Over each coordinate $j=1,\ldots,p$, under second order differentiabilty conditions,  we will make use of the second order Taylor expansion of $\U_{n,j}(\btheta)$ around some $\btheta_0 \in \Theta$,
\begin{multline}
\U_{n,j}(\btheta) = \U_{n,j}(\btheta_0) + \sum_{k=1}^p\partial_k \U_{n,j} (\btheta_0) (\theta_k -\theta_{k,0}) \\+ \frac{1}{2} \sum_{1 \le k,l \le p}
(\theta_k -\theta_{k,0})  \partial_{k,l} \U_{n,j}(\btheta_{n,j}^*) (\theta_l -\theta_{l,0}),
\label{U-expand}
\end{multline}
where each $\btheta_{n,j}^*$ lies
on the line segment connecting $\btheta$ and $\btheta_0$.  In the following, we let $\|\cdot\|$ denote the Euclidean norm of a vector, the operator norm of a matrix, and the supremum norm of a function. 

\subsection{Estimating equations, consistency, and asymptotic normality}\label{sec:gen.main}
We now present results that provide conditions for the consistency and existence of a non-trivial limiting distribution for a properly centered and scaled sequence of estimating equation solutions. We also include results on the consistent estimation of parameters on which the asymptotic distribution of our estimate may depend.
\begin{theorem}
\label{theorem-consistent-root} Suppose that $\U_n:\bTheta \times \chi^{(n)} \rightarrow \mathbb{R}^p$ is twice
continuously differentiable in an open set $\bTheta_0 \subset \bTheta$
containing $\btheta_0$, and that there exist a sequence of real numbers $a_n$, a matrix $\Gamma \in
	\mathbb{R}^{p \times p}$ and $\gamma>0$ such that 
\begin{equation}
\label{score-at-true-to-zero} a_n \U_n(\btheta_0)
	\rightarrow_p 0 \qmq{and} 
	a_n \U_n '(\btheta_0) \rightarrow_p \Gamma\qm{as $n\To\infty$} \qmq{with}
	 \inf_{\|\theta \|=1}\theta^\top \Gamma \theta =
	\gamma.
\end{equation}
Suppose further that
for any $\tau \in (0,1)$, that there exists a $K$
	such that for all $n$ sufficiently large, 
	\bea
	\label{U2-uniform-in-prob} P(|a_n \partial_{k,l}\U_{n,j}({\btheta})| \le K,1 \le k,l,j \le p, \btheta
	\in \bTheta_0) &\ge& 1-\tau. 
	\ena Then for any given $\epsilon>0$
	and $\tau \in (0,1)$, for all $n$ sufficiently large, with
	probability at least $1-\tau$ there exists ${\widehat \btheta}_n \in
	\bTheta$ satisfying $\U_n({\widehat \btheta}_n)=0$ and $||{\widehat
		\btheta}_n - \btheta_0|| \le \epsilon$, that is,  a sequence of roots to the estimating equation (\ref{theta-solves})
	consistent for $\btheta_0$.
	
	In addition, for any sequence ${\widehat \btheta}_n \rightarrow_p \btheta_0$, we have
	\bea \label{anUn'-consistent-for-Gamma} a_n\U_n'({\widehat \btheta}_n)
	\rightarrow_p \Gamma, \ena that is, $\Gamma$ can be consistently
	estimated by $a_n\U_n'({\widehat \btheta}_n)$ from any sequence consistent for $\btheta_0$.
\end{theorem}

\noindent {\em Proof:} By replacing $\U_n$ by $a_n \U_n$ and $\btheta$ by
$\btheta-\btheta_0$, we may
assume that the conditions of Theorem \ref{theorem-consistent-root}
hold with $a_n=1$ and $\btheta_0={\bm 0}$. 
For $\delta>0$ let 
\begin{align*}
B_\delta=\{\btheta: \|\btheta\| \le \delta\}.
\end{align*}
For the given $\tau \in (0,1)$, let $K$ and $n_0$ be such that
(\ref{U2-uniform-in-prob}) holds with $\tau$ replaced by $\tau/2$
for $n \ge n_0$. For the given $\epsilon>0$, take $\delta \in (0,
\epsilon)$ such that
$$
B_\delta \subset \bTheta_0 \quad \mbox{and} \quad  C \delta <
\gamma  \quad \mbox{where} \quad C = 2+\frac{Kp^{3/2}}{2}.
$$

By \eqref{score-at-true-to-zero}  there exists $n_1 \ge n_0$ such that
for $n \ge n_1$
\bea
\label{U-control} 
P(\|\U_n({\bm 0})\| < \delta^2) \ge 1-\tau/3 \quad 
P(\|\U_n'({\bm 0})-\Gamma\| < \delta ) \ge 1-\tau/3,
\ena 
and also taken large enough so that \eqref{U2-uniform-in-prob} holds with $\tau$ replaced by $\tau/3$.
By the union bound, all three events hold with probability at least $1-\tau$. 

For $\btheta \in B_\delta$ and $\btheta_{n,j}^*$ given by \eqref{U-expand}, the components of $R_n(\btheta) = (R_{n,1}(\btheta),\ldots,R_{n,p}(\btheta))^\top$  as defined by 
$$
R_{n,j}(\btheta)=\sum_{1 \le k,l \le p}
\theta_k  \partial_{k,l} \U_{n,j}(\btheta_{n,j}^*) \theta_l \qmq{satisfy} |R_{n,j}(\btheta)| \le K \left(\sum_{i=1}^p |\theta_i|\right)^2 \le Kp\|\btheta\|^2.
$$
Then, for $n \ge n_1$, with probability at least $1-\tau$, from
(\ref{U-expand}), (\ref{U-control}) and
(\ref{U2-uniform-in-prob}), 
\beas \|\U_n(\btheta) - \Gamma \btheta\|
&\le& \|\U_n(\btheta) -\U_n'({\bm 0}) \btheta \|+ \|\U_n'({\bm 0}) \btheta -\Gamma \btheta\|\\
&=& \|\U_n({\bm 0})+\frac{1}{2}R_n(\btheta)\|+
\|(\U_n'({\bm 0})-\Gamma) \btheta\|\\
&<& \delta^2 + \frac{Kp^{3/2}}{2} \|\btheta\|^2 + \delta \|\btheta\| \le C
\delta^2, \enas so
$$
\|\btheta^{\top}\U_n(\btheta) - \btheta^\top \Gamma \btheta\| < C
\delta^3.
$$
Hence, if $\|\btheta\|=\delta$, 
\bea \label{eq:inner.product.positive}
 \btheta^{\transpose}\U_n(\btheta)>
\btheta^\transpose \Gamma \btheta - C\delta^3 \ge \gamma
\delta^2-C\delta^3
=\delta^2 (\gamma  - C\delta) > 0.
\ena

Now we argue as in Lemma 2 of \citet{Aitchison58}.
Assume for the sake of contradiction that $\U_n(\btheta)$ does not have a
root in $B_\delta$. Then for $\btheta \in B_\delta$, the function
$f(\btheta)=-\delta \U_n(\btheta)/\|\U_n(\btheta)\|$ continuously
maps $B_\delta$ to itself. By the Brouwer fixed point theorem,
there exists $\bsy{\vartheta} \in B_\delta$, with
$f(\bsy{\vartheta})=\bsy{\vartheta}$. Since $\|f(\btheta)\|=\delta$ for all
$\btheta \in B_\delta$, we have $\|f(\bsy{\vartheta})\|=
\|\bsy{\vartheta}\|=\delta$,  contradicting \eqref{eq:inner.product.positive} via
$\delta^2=\|\bsy{\vartheta}\|^2 = \bsy{\vartheta}^\transpose \bsy{\vartheta} =
\bsy{\vartheta}^\transpose f(\bsy{\vartheta})<0$. Hence $\U_n(\btheta)$ has a root
within $\delta$ of 0, and since $\delta <\epsilon$, therefore
within $\epsilon$, with probability at least $1-\tau$, as required.

To prove (\ref{anUn'-consistent-for-Gamma}), taking $\widehat{\btheta}_n$ to be any consistent sequence for $\btheta_0$, a first order Talyor expansion yields, for all $1 \le j,k \le p$, 
\begin{align*}
\partial_k \U_{n,j}({\widehat \btheta}_n) &= \partial_k \U_{n,j}({\bm 0})+\sum_{l=1}^p \partial_{k,l}\U_{n,j}(\btheta_{n,j}^*){\widehat \btheta_{n,l}}\\
&= \partial_k \U_{n,j}({\bm 0})+Q_{k,n,j}^{\transpose}{\widehat \btheta}_n \qmq{where} Q_{k,n,j}^{\transpose}:=(\partial_{k,1}\U_{n,j}(\btheta_{n,j}^*),\ldots,\partial_{k,p}\U_{n,j}(\btheta_{n,j}^*)),
\end{align*}
where $\btheta_{n,j}^*$ lies along the line segment connecting ${\widehat \btheta}_n$ and ${\bm 0}$. Writing this identity in matrix notation, we have
\begin{align*}
\U_n'({\widehat \btheta}_n)-\U_n'({\bm 0}) = Q_n({\widehat \btheta}_n) \qmq{where} (Q_n(\btheta))_{k,j}= Q_{n,k,j}^{\transpose}\btheta.
\end{align*}

Let $\tau \in (0,1)$
and $\epsilon > 0$ be given, choose $\delta \in (0,\epsilon/K\sqrt{p})$ so
that $B_\delta \subset \bTheta_0$, and let $n_2$ be such
that for all $n \ge n_2$, with probability at least $1-\tau$,
$|\partial_{k,l} \U_n(\btheta)| \le K$ for all $1 \le k,l \le p$ and 
$\|{\widehat \btheta}_n\| \le \delta$.
Then, for $n \ge n_2$ with probability at least $1-\tau$  we have
\begin{align*}
|Q_{n,k,j}^{\transpose} {\widehat \btheta}_n| \le K \sqrt{p}\delta < \epsilon,
\qmq{or equivalently,}
\|\U_n'({\widehat \btheta}_n)-\U_n'({\bm 0})\|_\infty  < \epsilon
\end{align*}
where $\|A\|_\infty=\max_{i,j}|A_{i,j}|$ for $A \in \mathbb{R}^{p\times p}$.
 The claim follows, since
$\epsilon$ and $\tau$ are arbitrary, and $\U_n'({\bm 0}) \rightarrow_p \Gamma$ by assumption.
$\qed$

Our next result provides conditions under which a consistent estimator sequence, properly centered and scaled, converges in distribution. 
\begin{theorem}
	\label{mle-normal} Suppose the sequence of solutions $\widehat
	\btheta_n, n \ge 1$ to (\ref{theta-solves}) is consistent for $\btheta_0$, that \eqref{U2-uniform-in-prob} and the second condition of \eqref{score-at-true-to-zero} hold for some sequence $a_n, n \ge 1$ of real numbers, that 
	the matrix $\Gamma$ in \eqref{score-at-true-to-zero} is non-singular and that $\U_n(\btheta)$ is twice continuously 
	differentiable in an open set $\bTheta_0 \subset \bTheta$
	containing $\btheta_0$. Further, let $b_n$ be a sequence of real
	numbers such that for some random variable $Y$, \bea
	\label{score-converges-in-d} b_n \U_n(\btheta_0) &\rightarrow_d&
	Y. \ena Then
	$$
	\frac{b_n}{a_n}(\widehat \btheta_n - \btheta_0) \rightarrow_d
	-\Gamma^{-1} Y.
	$$
\end{theorem}

\noindent {\em Proof:} As in the proof of Theorem
\ref{theorem-consistent-root}, by replacing $a_n \mathcal{U}_n$ by $\mathcal{U}_n$ we may without loss of generality take
$a_n=1$, and also as done there, take $\btheta_0={\bm 0}$. Since a limit in distribution does not depend on events of vanishingly small probability, by the consistency of ${\widehat \btheta}_n$ and \eqref{U2-uniform-in-prob} we may
assume that for each $n$, sufficiently large, that $\widehat \btheta_n \in
\bTheta_0$, and for some $K$ that $|\partial_{k,j}\U_n(\btheta)| \le K$ for all $1 \le j,k \le p$ and $\btheta \in
\bTheta_0$. For such $n$ the expansion
(\ref{U-expand}) holds, and substituting $\widehat \btheta_n$ for
$\btheta$ and using $\U_n(\widehat \btheta_n)={\bm 0}$ yields \beas
- \U_n({\bm 0}) = (\U_n'({\bm 0}) +
\epsilon_n){\widehat \btheta}_n :=  \Gamma_n {\widehat \btheta}_n \qmq{where} (\epsilon_n)_{j,l} = \frac{1}{2} \sum_{k=1}^p {\widehat \theta}_{n,k} \partial_{k,l} \U_{n,j}(\btheta_{n,j}^*).
\enas
By the Cauchy-Schwarz inequality, 
$$
|(\epsilon_n)_{j,l} | \le \frac{K\sqrt{p}}{2}\|{\widehat \btheta}_n\| \rightarrow_p 0.
$$
Hence $\Gamma_n \rightarrow_p \Gamma$ so that $\Gamma_n^{-1}$
exists with probability tending to 1, and converges in
probability to $\Gamma^{-1}$. Now using
(\ref{score-converges-in-d}), Slutsky's theorem, on an event of probability tending to one as $n$ tends to infinity, 
$$
b_n \widehat \btheta_n = \Gamma_n ^{-1} \left(b_n \Gamma_n  \widehat
\btheta_n \right) = - \Gamma_n ^{-1} \left( b_n \U_n({\bm 0}) \right)
\rightarrow_d -\Gamma^{-1} Y.   
$$ \qed

In the most common case the distributional convergence in \eqref{score-converges-in-d} is to the normal, and shown by applying the Central Limit Theorem to a sum of independent random vectors. This situation is illustrated in the following lemma, in which we  include distributional limits that may have covariance matrices of less than full rank. For a given vector $\bsy{\mu}$ and non-negative definite matrix $\Sigma$,
\begin{align*}
\mbox{we say} \quad \bm{X} \sim {\cal N}(\bsy{\mu},\Sigma) \qmq{when} 
E[e^{\bm{t}^\transpose \bm{X}}] = \exp\left(\frac{1}{2} \bm{t}^\transpose \Sigma \bm{t} + \bm{t}^\transpose \bsy{\mu} \right).
\end{align*}
In particular, in one dimension ${\cal N}(\mu,0)$ is unit mass at $\mu$. 

\begin{lemma} \label{lem:mult.clt}
	Let $\mathcal{A}_\ell, \ell=1,2,\ldots$ be a sequence of arbitrary index sets satisfying $|\mathcal{A}_\ell| \rightarrow \infty$ as $\ell \rightarrow \infty$, and let  $\{\bm{X}_{\ell,a}, a \in \mathcal{A}_\ell\}$ be a collection of $\mathbb{R}^d$ valued independent, mean zero random vectors such that for some matrix $\Sigma$ and some $\delta>0$ 
	\begin{align} \label{eq:MultLind}
		\lim_{\ell \rightarrow \infty}\sum_{a \in \mathcal{A}_\ell} {\rm Var}(\bm{X}_{\ell,a}) = \Sigma \qmq{and} 
		\lim_{\ell \rightarrow \infty}\sum_{a \in \mathcal{A}_\ell, 1 \le k \le d} \E\left[  |X_{\ell,a,k}|^{2+\delta} \right]= 0. 
	\end{align}
	Then 
	\begin{align*}
		{\bm S}_\ell=\sum_{a \in \mathcal{A}_\ell} \bm{X}_{\ell,a} \qmq{satisfies} {\bm S}_\ell \rightarrow {\cal N}(0,\Sigma) \qm{as $\ell \rightarrow \infty$.}
	\end{align*}
\end{lemma}

\noindent \textit{Proof:} We first prove the result in $\mathbb{R}$. By the Lindeberg theorem, (e.g. Theorem 3.4.5,  \citep{durr})if for all $\ell \ge 1$
 the random variables $\{X_{\ell,a}, a \in \mathcal{A}_l\}$ are independent, mean zero, and satisfy 
\begin{align} \label{eq:LindFelR}
	\lim_{\ell \rightarrow \infty}\sum_{a \in \mathcal{A}_\ell} {\rm Var}(X_{\ell,a})=\sigma^2>0,
	\qmq{and for all $\epsilon>0$} \lim_{\ell \rightarrow \infty}\sum_{a \in \mathcal{A}_\ell} E[X_{\ell,a}^2{\bm 1}(|X_{\ell,a}| \ge \epsilon)] =0, 
\end{align}
then $S_\ell \rightarrow_d {\cal N}(0,\sigma^2)$ where $S_\ell = \sum_{a \in \mathcal{A}_\ell} X_{n,i}$. In $\mathbb{R}$, the second condition in \eqref{eq:MultLind} implies the second condition in \eqref{eq:LindFelR}, as for any $\epsilon>0$, with $p=1+\delta/2$ and $q=1+2/\delta$, using H\"older's inequality followed by Markov's, 
\begin{multline*}
	E[X_{\ell,a}^2{\bm 1}(|X_{\ell,a}| \ge \epsilon)] \le E[X_{\ell,a}^{2p}]^{1/p}P(|X_{\ell,a}| \ge \epsilon)^{1/q}\\
	\le E[X_{\ell,a}^{2p}]^{1/p}\left( \frac{E[X_{\ell,a}^{2p}]}{\epsilon^{2p}}\right)^{1/q} = \frac{E[X_{\ell,a}^{2p}]}{\epsilon^{2p/q}} = \frac{E[X_{\ell,a}^{2+\delta}]}{\epsilon^{2p/q}}.
\end{multline*}
Hence, the claim holds in $\mathbb{R}$ when the limiting variance  is positive. When this limit is zero, Chebyshev's inequality yields that $S_\ell \rightarrow_p 0$, and hence $S_\ell$ converges as well to zero in distribution, which is the normal distribution with mean and variance 0.  Hence the conclusion of the lemma holds for $d=1$. 

In general, given a collection of random vectors satisfying the given hypotheses, taking ${\bm v}$ to be of norm 1, the variables $Y_{\ell,a}= {\bm v}^\transpose \bm{X}_{\ell,a}$ for $a \in \mathcal{A}_\ell$ are independent and mean zero for each $\ell$, and satisfy the first condition of \eqref{eq:MultLind} with $\Sigma$ replaced by ${\bm v}^\transpose\Sigma {\bm v}$, and the second condition of \eqref{eq:MultLind} by virtue of this condition holding by assumption for the vector array $\bm{X}_{\ell,a}$, that $\|\bm{v}\|_\infty \le 1$ and hence that 
\begin{align*}
	|Y_{\ell,a}|^{2+\delta} = |{\bm v}^\transpose \bm{X}_{\ell,a}|^{2+\delta} \le  \sum_{a \in \mathcal{A}_\ell,1 \le k  \le d}	|\bm{X}_{\ell,a,k}|^{2+\delta}. 
\end{align*}
As the claim holds in $d=1$ for linear combinations given by any ${\bm v}$ of norm 1, the general result follows by the Cramer-Wold device. \bbox.

\subsection{Examples}\label{sec:gen.exs}
In the section we demonstrate the scope of the results in Section \ref{sec:gen.main} by presenting two applications, one to least squares and the other to maximum likelihood.

The following lemma, a direct application of the dominated convergence theorem, is used to handle the technical matter of interchanges between integration and differentiation
with respect to $\btheta \in \Theta \subset \mathbb{R}^p$. 
\begin{lemma}
\label{partial-int-f=int-partial-f-lemma} Let $f:\mathbb{R}^m \times \Theta \rightarrow {\mathbb R}$ be differentiable with
respect to $\btheta$ in an open set $\Theta_0 \subset \Theta$,
and suppose that there exists $g:{\mathbb R}^m \rightarrow {\mathbb R}$
such that
\begin{align*}
\Bvert \partial_\btheta f(\bm{x};\btheta)\Bvert  \le g({\bm
	x})\qmq{for all $\btheta \in \Theta_0$ and} \int_{{\mathbb R}^m} g(\bm{x})d\bm{x} < \infty.
\end{align*}
Then for all $\btheta \in \Theta_0$,
\bea \label{partial-int-f=int-partial-f} \partial_\btheta \int_{{\mathbb R}^m} f(\bm{x};\btheta) d\bm{x}= \int_{{\mathbb R}^m}
\partial_\btheta f(\bm{x};\btheta) d\bm{x}.
\ena
\end{lemma}

\begin{example}
Least squares estimation. Suppose we observe
\beas
y_i= f(\bm{x}_i, \theta_0) + \epsilon_i\quad i=1,\ldots,n
\enas 
where $f(\bm{x}_i, \theta), \theta \in \Theta \subset \mathbb{R}$ is some specified parametric family of functions; we take a one dimensional parameter here for simplicity. We estimate $\theta_0$ via least squares, minimizing 
\beas
J_n(\theta,{\bm x}^{(n)})=\frac{1}{2n}\sum_{i=1}^n \left( f(\bm{x}_i,\theta)-y_i\right)^2 = \frac{1}{2n}\sum_{i=1}^n \left( f(\bm{x}_i,\theta)-f(\bm{x}_i,\theta_0)-\epsilon_i \right)^2.
\enas 
We assume that $f(\bm{x},\theta)$ has three continuous derivatives with respect to $\theta$ that are uniformly bounded, say by $K$, over some open subset $\Theta_0$ of $\Theta$ that contains $\theta_0$, and that $\epsilon_1,\epsilon_2,\ldots$ are independent random variable distributed as $\epsilon$, a mean zero, variance $\sigma^2$ random variable with $E|\epsilon|^{2+\eta}=\tau^{2+\eta}<\infty$ for some $\eta>0$.

Taking one derivative with respect to $\theta$, we obtain the estimating equation ${\mathcal U}_n(\theta)=0$ where
\begin{multline}
{\mathcal U}_n(\theta)=\frac{1}{n}\sum_{i=1}^n \left( f(\bm{x}_i,\theta)-f(\bm{x}_i,\theta_0)-\epsilon_i \right)\partial_\theta f(\bm{x}_i,\theta) \\
\quad \mbox{so in particular} \quad{\mathcal U}_n(\theta_0)=-\frac{1}{n}\sum_{i=1}^n \epsilon_i \partial_\theta f(\bm{x}_i,\theta_0).\label{eq:score.ls.example}
\end{multline}
The first condition of \eqref{score-at-true-to-zero} of Theorem  \ref{theorem-consistent-root} is satisfied with $a_n=1$, as the errors have zero mean, are uncorrelated and have uniformly bounded variances, implying that $E_{\theta_0}[{\mathcal U}_n(\theta_0)]=0$ and ${\rm Var}_{\theta_0}[{\mathcal U}_n(\theta_0)] \rightarrow 0$. Regarding the second condition of \eqref{score-at-true-to-zero}
taking another derivative, we obtain 
\begin{multline}  \label{Un':sec2example}
{\mathcal U}_n'(\theta_0)=\frac{1}{n}\sum_{i=1}^n \left( \left( \partial_\theta f(\bm{x}_i,\theta)\right)^2 + \left( f(\bm{x}_i,\theta)-f(\bm{x}_i,\theta_0) - \epsilon_i\right)\partial_\theta^2 f(\bm{x}_i,\theta) \right) \Bvert_{\theta_0}\\
=\frac{1}{n}\sum_{i=1}^n  \left( \partial_\theta f(\bm{x}_i,\theta_0)\right)^2 - \frac{1}{n}\sum_{i=1}^n \epsilon_i \partial_\theta^2 f(\bm{x}_i,\theta_0).
\end{multline}

Arguing as for \eqref{eq:score.ls.example}, the second sum tends to zero in probability. If we now take $\bm{x}_i, i=1,2,\ldots$ to be independent random vectors distributed as some $\bm{x}$, then the law of large numbers yields that 
\begin{align}\label{eq:Un'togamma.example}
\frac{1}{n}\sum_{i=1}^n  \left( \partial_\theta f(\bm{x}_i,\theta_0)\right)^2 \rightarrow_p	\gamma = E_{\theta_0}\left( \partial_\theta f(\bm{x}_i,\theta_0)\right)^2, 
\end{align}
showing the second condition of \eqref{score-at-true-to-zero}, and this limit will be positive when $\partial_\theta f(\bm{x},\theta_0)$ is a non-degenerate random variable, thus verifying the final condition in \eqref{score-at-true-to-zero}  in that case. 

It is easy to see that taking another derivative in \eqref{Un':sec2example} yields an average of functions that are bounded over $\Theta_0$, plus a weighted average of the error variables, each one multiplied by some bounded function. As the second weighted average can be seen to be bounded in probability by applying reasoning similar to that used for the score ${\mathcal U}_n(\theta_0)$, condition \eqref{U2-uniform-in-prob} holds.

The only remaining verification needed to invoke Theorem \ref{mle-normal} is to show the properly scaled score at $\theta_0$ has a limiting distribution. Taking $b_n = \sqrt{n}$, we have  
\begin{align*}
	{\rm Var}\left(\frac{1}{\sqrt{n}} {\mathcal U}_n(\theta_0) \right) \rightarrow \sigma^2 \gamma, 	
\end{align*}
by \eqref{eq:Un'togamma.example}, and in addition using the representation of ${\mathcal U}_n(\theta_0)$ from \eqref{eq:score.ls.example},
\begin{align*}
\sum_{i=1}^n \mathbb{E}\left|\frac{\epsilon_i \partial_\theta f(\bm{x}_i,\theta_0)}{\sqrt{n}}\right|^{2+\delta} \le K^{2+\delta} n^{-\delta/2} \tau^{2+\delta} \rightarrow 0.
\end{align*}
Hence, invoking Lemma \ref{lem:mult.clt}, for any consistent sequence of roots,
\begin{align*}
\sqrt{n}\left( \widehat{\theta}_n-\theta_0\right) \rightarrow_d {\cal N}(0,\sigma^2 \gamma^{-1}).
\end{align*}
\end{example}

\begin{example}
Maximum likelihood. Let $p(\bm{x},\btheta), \btheta \in \Theta_0$ be a family of density functions for $\Theta \subset \mathbb{R}^p$, and for some ${\btheta}_0 \in \Theta$, let $\bm{X}_1,\ldots,\bm{X}_n$ be independent random vectors with density $p(\bm{x},{\btheta}_0)$. Let $p(\bm{x},\btheta)$ be three times continuosly differentiable in $\btheta$ with the first two derivatives of $p({\bm x},\btheta)$, and the third derivative of $q({\bm x},\btheta)=\log p({\bm x},\btheta)$,  dominated by an integrable function in some neighborhood $\Theta_0$ of $\btheta_0$. 
Assume further that the Fisher information matrix at $\btheta_0$  is positive definite.

The maximum likelihood estimate of $\btheta_0$ is obtained by maximizing the log likelihood of the data, and hence 
given by a solution to the estimating equation \eqref{theta-solves} with
\begin{align*}
{\mathcal U}_n(\btheta)= \frac{1}{n}\sum_{i=1}^n 
\frac{\partial_{\btheta} p(\bm{X}_i,\btheta)}{p(\bm{X}_i,\btheta)}.
\end{align*}
\end{example}
By Lemma \ref{partial-int-f=int-partial-f-lemma}, for $\btheta \in \Theta_0$ we have 
\begin{align} \label{eq:score.zero.mle}
	\mathbb{E}_{\btheta}[\partial_\btheta \log p(\bm{X},\btheta) ] = \int_{\mathbb{R}}\partial_\btheta p(x,\btheta) dx = \partial_\btheta \int_{\mathbb{R}} p(x,\btheta) dx=0, 
\end{align}
and likewise that the Fisher information $I(\btheta)$ satisfies
\begin{align*}
I(\btheta)=	-\mathbb{E}_{\btheta}[\partial_\btheta^2 \log p(\bm{X},\btheta) ] = {\rm Var}_\btheta(\partial_\btheta \log p(\bm{X},\btheta) ). 
\end{align*}
Hence, by the law of large numbers the first two conditions of \eqref{score-at-true-to-zero} are satisfied with $a_n=1$ and $\Gamma=I(\btheta_0)$, and the last holds by our assumption on the Fisher information. 

Next we show \eqref{U2-uniform-in-prob} is satisfied. Writing $\partial_j$ short for $\partial_{\btheta_j}$,  we may write
\begin{align*}
 {\mathcal U}_n(\btheta)= \frac{1}{n}\sum_{i=1}^n \partial_{\btheta}q({\bm X}_i,\btheta) \qmq{and hence} 
 \partial_{k,l}{\mathcal U}_{n,j}(\btheta) = \frac{1}{n}\sum_{i=1}^n \partial_{k,l,j} q({\bm X}_i,\btheta). 
\end{align*}
Condition \eqref{U2-uniform-in-prob} can be verified by invoking the following uniform strong law of large numbers with $h({\bf x},\btheta)$ applied to the components $\partial_{k,l,j} q({\bm x},\btheta)$.
\begin{theorem}[Le Cam \citeyearpar{LeCam53} \label{thm.unif.strong}Corollary 4.1\ignore{page 26/300}] Let $\Theta$ be a compact metric space and $\chi$ a space on which a probability distribution $F$ is defined. Let $h({\bm x},\btheta)$ be measurable in ${\bm x}$ for each $\btheta \in \Theta$ and continuous in $\btheta$ for almost every ${\bm x}$. Assume there exists $K({\bm x})$ such that $E[K({\bm X})]<\infty$ and $|h({\bm x},\btheta)| \le K({\bm x})$ for all ${\bm x}$ and $\btheta$. Then, with $m(\btheta)=E[h({\bm X},\btheta)]$,
\begin{align*}
	P \left(\lim_{n \rightarrow \infty} \sup_{\theta \in \Theta} \Bvert \frac{1}{n}\sum_{i=1}^n h({\bm X}_i,\btheta) - m(\btheta)\Bvert =0\right)=1,
\end{align*}
where ${\bm X}_1,{\bm X}_2,\ldots$ are independent with distribution $F$. 
\end{theorem}

Lastly, under the given assumptions, the classical central limit theorem yields $$
\sqrt{n}{\mathcal U}_n(\btheta_0)\rightarrow_d {\mathcal N}(0, I(\btheta_0))$$
so that, via Theorem \ref{mle-normal},
$$\sqrt{n}\left( {\widehat \btheta}_n -\btheta_0\right)\rightarrow_d {\mathcal N}(0, I(\btheta_0)^{-1}).$$

For the exponential family
\begin{align*}
	p({\bm x};\btheta) = h({\bm x})\exp \left(
	\eta(\btheta)T({\bm x})-A(\btheta) \right) \qmq{we have}
	q({\bm x};\btheta) = \log h({\bm x}) + \eta(\btheta)T({\bm x})-A(\btheta).
\end{align*}
Hence, the needed conditions are satisfied if $A(\btheta)$ and $\eta(\btheta)$ have three bounded derivatives in some neighborhood of $\btheta_0$, and $E_{\btheta_0} [T({\bm X})]$ exists.

\section{Application to a diffusion equation model}\label{sec:ApplicationToDiffusion}
To more fully specify the output function of the diffusion model arising from PDE \eqref{PDE model} as described in Subsection \ref{sec:diff.model}, consider the parameter space
\begin{align} \label{def:Q0}
{\mathcal Q}= \{(q_1,q_2) \in \mathbb{R}^2: \,\, q_2>0\},
\end{align}
and for given matrices $D,E\in {\mathbb R}^{{k}\times {k}}$,  a vector $F \in {\mathbb R}^{k}$ and  $\bm{q} \in {\mathcal Q}$, recall from \eqref{eq:AB.by.q} that
\begin{equation}
\label{def:A,B}
	A=A(\bm{q})=q_1D+E,\quad B=B(\bm{q}) =q_2 F, 
\end{equation}
and that the TAC at time $t$ is given by
\begin{align} \label{eq:fij.q.only} 
	f_{\mu}(t;\bm{q}) = \int_0^t Ce^{A(t-s)}B \mu(s)ds,
\end{align}
where $C^\transpose \in \mathbb{R}^k$, and $\mu(s)$ is the BAC/BrAC at time $s$. Though our methods work in the given generality, in the physics based model the matrix $A$ will have eigenvalues with negative real parts, and $q_1$ will be strictly positive. The dependence of $f$ on $A,B,C, \mu$ or $\bm{q}$ may be dropped in the following for ease of notation, or included to emphasize some particular feature of interest.

Consider an individual whose data has been collected over $i=1,\ldots,n$ drinking sessions, where the BrAC curve $\mu_i$ for episode $i$ is integrable on $[0,T_i]$, and for some ${\bm q}_0 \in \mathcal{Q}$ and 
 $m_i$ observations of TAC plus a mean zero error
 \begin{align}\label{eq:y=f+e.model}
 	y_{ij}= f_{\mu_i}(t_{i,j};{\bm q}_0)+\epsilon_{i,j},
 \end{align}
 are taken
at the times $0 \le t_{i,1} \le \cdots \le t_{i,m_i} \le T_i \le T$, for some $T>0$. For notational simplicity we may suppress some of the parameters in 
\eqref{eq:y=f+e.model}, for instance, denoting $f_{\mu_i}(t_{i,j};{\bm q})$ by $f_{ij}({\bm q})$, say. 
We encode the observation times of episode $i$ as the probability measure putting mass $1/m_i$ on each observation time, and form the vector of probability measures $\bsy{\nu}_n=(\nu_{1,m_1,},\ldots,\nu_{n,m_n})$. When $n=1$, that is, for the case of a single episode, we drop the index $i$.

For asymptotics, we consider a sequence of experiments indexed by $\ell=1,2,\ldots$, where $n$ and ${\bm m}=(m_1,\ldots,m_n)$ may depend on $\ell$, and hence we may index using $\ell$ in place of $n,{\bm m}$, though this dependence may at times be suppressed in the notation. In the case of a single drinking episode, that is, when $n=1$, we let $\ell=m$. For consistency and asymptotic normality, we require that 
\begin{align}\label{eq:summi->infty}
	\sum_{i=1}^n m_i \rightarrow \infty \qm{as $\ell \rightarrow \infty$.}
\end{align}
In the special case where the number of observations $m_i$ for each $n$ equals a constant~$m$, the requirement \eqref{eq:summi->infty} becomes $nm \rightarrow \infty$, and in the sub-case of a single drinking episode, that $m \rightarrow \infty$.

We apply the methods developed in Section \ref{sec:Mest} to the least squares estimator achieved as a solution to $ {\mathcal U}_\ell(\bm{q})=\bm{0}$ where
\begin{equation}\label{def:J and U}
{\mathcal U}_\ell(\bm{q}) = \partial_{\bm{q}} J_\ell(\bm{q}) \qmq{with}
 J_{\ell}(\bm{q})=\frac{1}{2\sum_{i=1}^n m_i}\sum_{i=1}^n \sum_{j=1}^{m_i} \left( f_{\mu_i}(t_{ij};\bm{q})-y_{ij}\right)^2,
\end{equation}
and $\partial_{\bm q}$ produces the gradient.  For $i \in \{1,2\}$, we continue to let $\partial_i$ denote taking the partial derivative with respect to $q_i$; this notation will extend in the natural way to denote higher order, and mixed partial derivatives. Theorem \ref{thm:ls.cons.asy.dist.qonly} below gives conditions under which the least squares estimate is consistent and has a limiting, asymptotically normal distribution, and as well provides the form of the limiting covariance matrix. Theorem \ref{thm:ls.cons.asy.dist.qonly} is an immediate consequence of Theorems \ref{thm:Gamma} and \ref{thm:ls.asy.dist.qonly}, that verify the conditions of 
Theorems \ref{theorem-consistent-root} and \ref{mle-normal} in the previous section.

We now present our main result regarding the least squares estimator for the diffusion model.

\begin{theorem} \label{thm:ls.cons.asy.dist.qonly}
Suppose the errors $\epsilon_{i,j}$, $1 \le i \le n$, $1 \le j \le m_i$ in model \eqref{eq:y=f+e.model}
are mean zero, uncorrelated and have constant positive variance $\sigma^2$, and let $\mu_i$ and $\nu_{i,n}$ be, respectively, the BrAC curve and empirical measure of the observation times for episode $i=1,\ldots,n$ over $[0,T_i]$, with $\sup_i T_i <\infty$. Assume the existence of the limit
\begin{align} \label{def:Gamma_n.nu}
\Gamma=\lim_{\ell \rightarrow \infty}\Gamma_\ell \qmq{where}	\Gamma_{\ell} = \sum_{i=1}^n  \frac{m_i}{\sum_{k=1}^n m_k} \int_0^{T_i}  \partial_{\bm q}f_{\mu_i}(u;{\bm q}_0)\partial_{\bm q}f_{\mu_i}(u;{\bm q}_0)^\top d\nu_{i,n}. 
\end{align}
and that $\Gamma$ is positive definite. Further, suppose there exists a constant $K$ such that 
\begin{align} \label{eq:L1.brac.unif.bd}
	\|\mu_i\|_1 \le K \qmq{for all $i \ge 1$,}
\end{align}
that is, that the $L^1$ norms of all drinking episode BrAC curves are uniformly bounded. Lastly, suppose that  \eqref{eq:summi->infty} holds. 
Then there exists a consistent sequence of solutions $\widehat{\bm{q}}_\ell$ to the estimating equation ${\mathcal U}_\ell(\bm{q})=0$, given in \eqref{def:J and U}.

If in addition the errors~$\epsilon_{i,j}$ are i.i.d., and for some $\delta>0$ satisfy $E|\epsilon_{i,j}|^{2+\delta} =\tau^{2+\delta}<\infty$, then, along any such consistent sequence $\widehat{\bm{q}}_\ell$, 
\begin{multline} \label{eq:asynom+varestimator}
		\sqrt{m} (\widehat{\bm{q}}_\ell-\bm{q}_0) \rightarrow_d \mathcal{N}\left( {\bm 0}, \sigma^2 \Gamma^{-1} \right) \qmq{where} m= \sum_{i=1}^n m_i\\
		\mbox{and} \quad \widehat{\sigma}_\ell^2 \rightarrow_p \sigma^2 \qmq{where}
		\widehat{\sigma}_\ell^2 = \frac{1}{m}\sum_{i=1}^n \sum_{j=1}^{m_i} \left(y_{ij} - f_{ij}(\widehat{\bm{q}}_\ell) \right)^2.
	\end{multline}
\end{theorem}

When the errors $\epsilon_{i,j}$ in \eqref{eq:y=f+e.model} are Gaussian, then the least squares estimate that minimizes the sum of squares \eqref{eq:Jq.def} is also maximum likelihood. In this case the contribution to the Fisher information from the single observation in \eqref{eq:y=f+e.model} is obtained by taking the covariance matrix of the gradient of the log of the density of the observation,
\begin{multline*}
{\rm Var}\left( \partial_{\bm q}	\log p(y_{ij};f_{ij})\right)
={\rm Var}\left( \partial_{\bm q} \log
\left(
\frac{1}{\sqrt{2 \pi}\sigma}\exp \left(-\frac{1}{2\sigma^2}(y_{ij}-f_{ij})^2
\right)
\right)\right)\\
={\rm Var}\left(\partial_{\bm q} \left( 
-\frac{1}{2\sigma^2}(y_{ij}-f_{ij})^2\right)\right)
= {\rm Var}\left(\frac{\epsilon_{ij}}{\sigma^2}\partial_{\bm q}f_{ij}\right)
= \frac{1}{\sigma^2}\partial_{\bm q}f_{ij}\partial_{\bm q}f_{ij}^\top.
\end{multline*}
 Summing over the observation times yields $m\Gamma_\ell/\sigma^2$ as in \eqref{def:Gamma_n.nu}, hence taking the limit and comparing with the asymptotic variance obtained we see that for normal errors the least squares estimate of ${\bm q}$ achieves the lower bound of the information inequality in an asymptotic sense.

To discuss the satisfaction of condition \eqref{def:Gamma_n.nu} we recall that a sequence of measures $\{\nu_m, m \ge 1\}$ on $\mathbb{R}$ is said to converge weakly to a measure~$\nu$ if for all bounded continuous functions $g:{\mathbb{R}} \rightarrow \mathbb{R}$
\begin{align} \label{eq:nu_m.to.nu}
	\lim_{m\rightarrow \infty}\int_{\mathbb{R}} g(u)d\nu_m = \int_{\mathbb{R}} g(u)d\nu.
\end{align}
By considering components, the same relation holds when $g$ continuously maps $[0,T]$ to the space of matrices of some fixed dimension.  In particular, when $\nu_{i,n}$ has weak limit $\nu_i$ then, as the integrand is bounded and continuous by Lemma \ref{lem:partialsfij.bounded}, the integral in term $i$ of the sum in \eqref{def:Gamma_n.nu} will converge to the same integral with respect to $\nu_i$. 
When $\nu_m$ is the $m^{th}$ element of the sequence of discrete probability measures that gives equal weights to the times $t_{m,1},\ldots,t_{m,m}$ in $[0,T]$ with weak limit $\nu$, then for any continuous function $g: [0,T] \rightarrow \mathbb{R}$,
\begin{align*}  
	\frac{1}{m}\sum_{j=1}^m g(t_{m,j}) = \int_0^T g(u)d\nu_m \rightarrow \int_0^T g(u)d\nu. 
\end{align*}

There are two special cases of note. One is where the distances between consecutive observation times on $[0,T]$ are constant; in this case, the weak limit is the uniform probability  measure on $[0,T]$. A second case is when the observation times are chosen independently according to a probability measure $\nu$ supported on $[0,T]$; in this case, the weak limit in probability is $\nu$, where the  convergence in \eqref{eq:nu_m.to.nu} is in probability, or almost sure if all the observation times are generated successively by a common iid sequence, see \cite{Vara58}.

Consider also the situation where the data from $n$ drinking episodes are independent and identically distributed replicates of the error distribution and  canonical $M,T,\mu,\nu$, where $M$ is the distribution of $m_i, 1 \le i \le n$, making the summands in \eqref{def:Gamma_n.nu} i.i.d. When $T<\infty$ a.s. Lemma \ref{lem:partialsfij.bounded} shows that the integrals in \eqref{def:Gamma_n.nu} are uniformly bounded, and one can show that  as $n \rightarrow \infty$, 

\begin{align*}
	\Gamma_n \rightarrow_p \Gamma = E\left[ \frac{M}{E[M]}
	\int_0^T \partial_{\bm q}f_{\mu}(u;{\bm q}_0)\partial_{\bm q}f_{\mu}(u;{\bm q}_0)^\top d \nu \right],
\end{align*}
where the expectation is taken over  $M,T,\mu$ and $\nu$,
whenever the expectation on the right hand side exists.  See also Assumption \ref{ass:nun.to.nu.C.alpha},  and the discussion following, in Section \ref{sec:InferenceBrAC}.

Before proceeding, we must verify the smoothness of the derivatives of $f_{\mu}(t;\bm{q})$ in 
\eqref{eq:fij.q.only} with respect to $\bm{q}=(q_1,q_2)$. Because of the form of the dependence of the matrix~$A$ on $q_1$ as given in \eqref{def:A,B}, to differentiate $f$ with respect to $q_1$ we will need to consider directional derivatives of matrix exponentials. For square matrices $W$ and $V$ of the same dimension and $u \in \mathbb{R}$, define the first derivative of $e^{uW}$ in direction $V$ by
\begin{align*}
\boldsymbol{{\mathcal D}}_V^1(u,W) = \lim_{h \rightarrow 0}
\frac{\exp(u(W+hV)) - \exp(uW)}{h}.
\end{align*}
We define higher order derivatives $\boldsymbol{{\mathcal D}}_V^k(u,W), k \ge 0$ in the natural way, with $k=0$ returning $e^{uW}$. Now with $A=q_1D+E$ as in \eqref{def:A,B}, we may represent the partial derivative with respect to $q_1$ of $e^{uA}$ as 
\begin{multline*}
\partial_1e^{uA} = \partial_1e^{u(q_1D+E)}= \lim_{h\To 0}\frac{ e^{u((q_1+h)D+E)}-e^{u(q_1D+E)}}{h}= \lim_{h\To 0}\frac{ e^{u(A+hD)}-e^{uA}}{h}\\
=\boldsymbol{{\mathcal D}}_D^1(u,A),
\end{multline*}
and extending to higher order derivatives we obtain
\begin{align}\label{d1neAu=dir.der}
\partial_1^n(e^{(q_1D+E)u}) = \boldsymbol{{\mathcal D}}_D^{n}(u,A).
\end{align}
For any $n \ge 0$, letting $B_n$ be the
$(n+1) \times (n+1)$ block matrix given by 
\begin{align}\label{def:B_n}
 B_n=\left[\begin{array}{ccccc}
W & V  & 0 & \cdots & 0\\
0 & W & V  & \cdots & 0\\
\cdots  & \cdots & \cdots & \cdots & \cdots \\
0 & 0  & \cdots & 0  & W \\
\end{array}
\right],
\end{align}
Theorem 4.13 of \citet{Najfeld95} \cite[see also][for similar applications]{Sirlanci19} provides that
\begin{align}
\label{NajfeldHavel}
 e^{uB_n} = 
\left[
\begin{array}{ccccc}
e^{uW} & \frac{\boldsymbol{{\mathcal D}}_V^1(u,W)}{1!}  & \frac{\boldsymbol{{\mathcal D}}_V^2(u,W)}{2!} & \cdots & \frac{\boldsymbol{{\mathcal D}}_V^n(u,W)}{n!}\\
0 & e^{uW} & \frac{\boldsymbol{{\mathcal D}}_V^1(u,W)}{1!}  & \cdots & \frac{\boldsymbol{{\mathcal D}}_V^{n-1}(u,W)}{(n-1)!}\\
\cdots  & \cdots & \cdots & \cdots & \cdots \\
0 & 0  & \cdots & 0  & e^{uW} \\
\end{array}
\right].
\end{align} 
We now apply \eqref{NajfeldHavel} to obtain bounds on higher order derivatives of the matrix exponential $e^{uA}$ with respect to $q_1$. 
\begin{lemma} \label{lem:NajfeldHave}
Let $W$ and $V$ be square matrices of the same dimension.
Then for all $n \ge 0$ the directional derivative $\boldsymbol{{\mathcal D}}_V^{n}(u,W)$ is analytic in $u$ on $\mathbb{R}$ and satisfies the bound
\begin{align}\label{eq:derv.bound}
\|\boldsymbol{{\mathcal D}}_V^{n}(u,W)\|\le n!\|e^{uB_n}\|
\qm{for all $u \in \mathbb{R}$,}
\end{align}
where $B_n$ is given by (\ref{def:B_n}).

For all $n \ge 0,q_1 \in \mathbb{R}$ and $A=q_1D+E$, the partial derivative $\partial_1^n e^{Au}$ exists, is analytic in $q_1$ and, with $B_n$ given by \eqref{def:B_n} with $W=A$ and $V=D$, 
satisfies
\begin{align*}
	\|\partial_1^n e^{uA}\| \le n!e^{u\|B_n\|} \qmq{with} \sup_{q_1 \in Q}\|B_n\|<\infty
\end{align*}
for $Q$ any bounded subset of $\mathbb{R}$.
\end{lemma}

\noindent {\em Proof:} As the left hand side $e^{uB_n}$  of 
\eqref{NajfeldHavel} is analytic in each component, the matrix on the right hand side must also be analytic, thus yielding the first claim.  Next, for $F$ the submatrix obtained by taking row and column indices ${\bm i},{\bm j}$ of a given matrix $E$, applying an alternate form for the spectral norm in the first equality, we have
$$
\|F\|=\sup_{\|{\bm y}\|=1,\|{\bm x}\|=1}{\bm y}^\top F {\bm x} \le
\sup_{\|{\bm u}\|=1,\|{\bm v}\|=1}{\bm u}^\top E {\bm v} =\|E\|, 
$$
as any value over which the first supremum is taken can be achieved in the second by padding ${\bm x}$ and ${\bm y}$ with zeros in coordinates that are not in ${\bm i}$ and ${\bm j}$, respectively. Hence, inequality \eqref{eq:derv.bound} now follows from \eqref{NajfeldHavel}. The remaining claims now follow in light of \eqref{d1neAu=dir.der} and that $\|B_n\|$ is continuous in $q_1$, and hence bounded on any set with compact closure.
\bbox

%
%
%

We require the following result to handle the derivatives of matrix products.  For $k \ge 0$, ${\cal Q}$ as in \eqref{def:Q0}, we say a matrix $M$ depending on $(\bm{q},u) \in {\cal Q} \times \mathbb{R}$ is $k$-smooth if for any $0 \le a,b \le k$, the mixed partials $\partial_1^a \partial_2^bM$ exist and are continuous for ${\bm q} \in \mathcal{Q}$, and for any bounded subsets $\mathcal{D} \subset \mathcal{Q}$ and $I \subset \mathbb{R}$,
\begin{align*}
\sup_{({\bm q},u) \in \mathcal{D} \times I, 0 \le a,b \le k}
\|\partial_1^a\partial_2^b M\| <\infty.
\end{align*}
We say $M$ is smooth if it is $k$-smooth for all $k \ge 0$.

\begin{lemma}\label{lem:bd.norm.matrix.products}
Let $M_i,i=1,\ldots,d$ be matrices having dimensions such that we may form the product
\begin{align*}
M=\prod_{i=1}^d M_i.
\end{align*}
If $M_1,\ldots,M_d$ are $k$-smooth then so is $M$. 
\end{lemma}

\noindent {\em Proof:} The proof follows directly from the multivariate Leibniz rule that expresses the derivative 	$\partial_1^a\partial_2^b M_i$ for  $0 \le a,b \le k$ as a finite linear combination of products of derivatives of $M_i$, each one with order no greater than $k$,  and recalling that for conformable matrices $\|AB\| \le \|A\| \|B\|$. 
 \bbox

The next lemma provides us with additional smoothness estimates, and the forms of derivatives that later appear. 
\begin{lemma} \label{lem:partialsfij.bounded} 
For $\bm{q} \in {\mathcal Q}$, and 
\begin{align}
	\partial_1( e^{Au}B)
	=\partial_1(e^{Au}) B
	\qmq{and}\label{d1eAuB}
	\partial_2(e^{Au}B)= q_2^{-1}e^{Au}B.
\end{align}	
	
For all $u \in \mathbb{R}$, the matrix function $e^{Au}B$ is smooth in ${\bm q}$. If $\gamma(\cdot)$ is integrable on $[0,T]$, then $f_{\gamma}(t;\bm{q})$ as in  \eqref{eq:fij.q.only} is smooth in ${\bm q}$, continuous for $t \in [0,T]$ and satisfies
\begin{multline*}
\partial_1^a \partial_2^b f_{\gamma}(t;\bm{q})=\left(\int_0^t  C\partial_1^a e^{A(t-s)}F \gamma(s)ds \right) \partial_2^b q_2\\
\qmq{and} \sup_{t \in [0,T]} |\partial_1^a \partial_2^b f_{\gamma}(t;\bm{q})| \le 
 a! e^{T\|B_a\|}
\|\gamma\|_1  \partial_2^b q_2,
\end{multline*}
where $B_n$ is defined in Lemma \ref{lem:NajfeldHave}. 

\end{lemma}
\noindent {\em Proof:} The claims in \eqref{d1eAuB} follow by recalling that $A$ does not depend on $q_2$ and that $B=q_2F$.  That $e^{Au}$ is smooth follows from Lemma \ref{lem:NajfeldHave}, and one easily verifies the smoothness of $B$ directly from \eqref{def:A,B}; hence, the product is smooth by Lemma \ref{lem:bd.norm.matrix.products}. Differentiation under the integral with respect to $q_1$ is then justified by the dominated convergence theorem, from which the smoothness of  $f_{\gamma}(t;\bm{q})$ in ${\bm q}$ and the bound on its partials then follows, in light of Lemma \ref{lem:NajfeldHave}; continuity of $f_{\gamma}(t;\bm{q})$ for $t \in [0,T]$ follows immediately from the integral representation \eqref{eq:fij.q.only}. 
\qed

We present a lemma that is used to verify the conditions of Theorems \ref{theorem-consistent-root} and \ref{mle-normal}.
\begin{lemma} \label{lem:U.V.props}
The score function $\mathcal{U}_\ell$ given by \eqref{def:J and U} may be written as
\begin{align} \label{Uq.decomp}
	{\mathcal U}_\ell(\bm{q})= {\mathcal V}_{\ell,1}(\bm{q})-{\mathcal V}_{\ell,2}(\bm{q})
\end{align}
where
\begin{multline} \label{eq:defV1V2}
	{\mathcal V}_{\ell,1}(\bm{q}) =\frac{1}{\sum_{i=1}^n m_i} \sum_{i=1}^n  \sum_{j=1}^{m_i} \partial_{\bm q}f_{ij}({\bm q})
	\left(f_{ij}(\bm{q})-f_{ij}(\bm{q}_0)\right) \\ 
	\qmq{and}
	{\mathcal V}_{\ell,2}(\bm{q}) =\frac{1}{\sum_{i=1}^n m_i} \sum_{i=1}^n \sum_{j=1}^{m_i} 
	\partial_{\bm q}f_{ij}({\bm q}) \epsilon_{ij}, 
\end{multline}
and 
\begin{multline} \label{eq:U.and.U'.in.V}
		{\mathcal U}_\ell(\bm{q}_0)=-\frac{1}{\sum_{i=1}^n m_i}  \sum_{i=1}^n \sum_{j=1}^{m_i} \partial_{\bm q}f_{ij}({\bm q}_0)  \epsilon_{ij} \\\qmq{and}
	{\mathcal U}_\ell'(\bm{q}_0)=  \Gamma_\ell - \frac{1}{\sum_{i=1}^n m_i} \sum_{i=1}^n\sum_{j=1}^{m_i}
	\partial_{\bm q}^2f_{ij}({\bm q}_0).
\end{multline}
When the errors $\epsilon_{ij}, 1 \le i \le n, 1 \le j \le m_i$ have mean zero with common variance $\sigma^2$ and are uncorrelated, then 
\begin{align} \label{eq:Gamma_n.1}
{\rm Var}({\mathcal U}_\ell(\bm{q}_0))
	= \frac{\sigma^2}{\sum_{i=1}^n m_i} \Gamma_\ell.
\end{align}

\end{lemma}

\noindent {\em Proof:} The decomposition \eqref{Uq.decomp} and the expressions in \eqref{eq:defV1V2} that result follow directly from \eqref{eq:y=f+e.model}, and then the first expression in \eqref{eq:U.and.U'.in.V} may be obtained by evaluation at $\bm{q}_0$. Identity \eqref{eq:Gamma_n.1} now follows from the first expression in \eqref{eq:U.and.U'.in.V}, which yields that 
$$
{\rm Var}({\mathcal U}_\ell(\bm{q}_0))=  \frac{\sigma^2}{(\sum_{i=1}^n m_i)^2} \sum_{i=1}^n  \sum_{j=1}^{m_i} \partial_{\bm q}f_{ij}({\bm q}_0)\partial_{\bm q}f_{ij}({\bm q}_0)^\top,
$$
followed by \eqref{def:Gamma_n.nu}. 
Differentiation now yields 
\begin{multline*} 
	{\mathcal V}_{\ell,1}'(\bm{q}_0)=\frac{1}{\sum_{i=1}^n m_i} \sum_{i=1}^n\sum_{j=1}^{m_i}
	\partial_{\bm q}f_{ij}({\bm q}_0)\partial_{\bm q}f_{ij}({\bm q}_0)^\top 
	\\ \quad \mbox{and}
	\quad {\mathcal V}_{\ell,2}'(\bm{q}_0)=\frac{1}{\sum_{i=1}^n m_i} \sum_{i=1}^n\sum_{j=1}^{m_i}
	\partial_{\bm q}^2f_{ij}({\bm q}_0)
	\epsilon_{ij},
\end{multline*}
thus yielding the second expression in \eqref{eq:U.and.U'.in.V} in light of \eqref{def:Gamma_n.nu}. \qed

 \begin{theorem} \label{thm:Gamma} 
Suppose the errors $\epsilon_{i,j}, 1 \le i \le n, 1 \le j \le m_i$ in model \eqref{eq:y=f+e.model} are mean zero, uncorrelated and have constant positive variance $\sigma^2$. Assume in addition that the limit $\Gamma$ in \eqref{def:Gamma_n.nu} exists and is positive definite, and that \eqref{eq:summi->infty} holds. Then with $\mathcal{U}_\ell$ given by \eqref{def:J and U}, the hypotheses \eqref{score-at-true-to-zero}  of Theorem 
\ref{theorem-consistent-root} are satisfied with $\Gamma$ as in \eqref{def:Gamma_n.nu}, $a_n=1$ and any bounded neighborhood $\Theta_0 \subset \mathcal{Q}$ of ${\bm q}_0$.
If in addition there exists a constant $K$ such that \eqref{eq:L1.brac.unif.bd} holds, then 
 \eqref{U2-uniform-in-prob} also holds. 
\end{theorem}

\noindent {\em Proof:}  Let  $\Theta_0$ be any bounded neighborhood of ${\bm q}_0$. By Lemma \ref{lem:partialsfij.bounded}, the partial derivatives of $f_{ij}({\bm q}):=f_{\mu_i}(t_{i,j};\bm{q})$ of \eqref{eq:fij.q.only} of all orders exist, and are continuous and uniformly bounded over $\Theta_0$. Hence $\mathcal{U}_\ell$ is twice continuously differentiable, with uniformly bounded derivatives, over $\Theta_0$. 

The first condition in \eqref{score-at-true-to-zero} holds as  ${\mathbb E} [{\mathcal U}_\ell(\bm{q}_0)]=0$ via 
the first identity in \eqref{eq:U.and.U'.in.V} of Lemma \ref{lem:U.V.props}, and by \eqref{eq:Gamma_n.1} in that same lemma, which yields that ${\rm Var}({\mathcal U}_\ell(\bm{q}_0)) \rightarrow 0$ by virtue of  \eqref{def:Gamma_n.nu} and \eqref{eq:summi->infty}.

Focusing now on the the second identity in \eqref{eq:U.and.U'.in.V}, to show the second condition in \eqref{score-at-true-to-zero}, as the limit $\Gamma$ in \eqref{def:Gamma_n.nu} exists,  it suffices that the second term in that identity tends to zero in probability. In fact, the variance of this mean zero term tends to zero, since the components of the matrix $\partial_{\bm q}^2 f_{ij}({\bm q}_0)$ are uniformly bounded on $\Theta_0$ by Lemma \ref{lem:partialsfij.bounded} and multiply mean zero, uncorrelated random variables with uniformly bounded variances.
The matrix $\Gamma$ is positive definite by assumption, so the last condition in \eqref{score-at-true-to-zero}  holds. 

Lastly, we show that 
inequality \eqref{U2-uniform-in-prob} is satisfied. From the decomposition \eqref{Uq.decomp}
we see that we may write $\partial_{k,l}\U_{\ell,r}({\bm q})$ as a difference of the form 
\begin{align*}
R_\ell-S_\ell:=	\frac{1}{\sum_{i=1}^n m_i}\sum_{i=1}^n  \sum_{j=1}^{m_i}g_{1,ij}(\bm{q}) - \frac{1}{\sum_{i=1}^n m_i}\sum_{i=1}^n
\sum_{j=1}^{m_i}g_{2,ij}(\bm{q})\epsilon_{i,j}
\end{align*}
where, by Lemmas \ref{lem:bd.norm.matrix.products},  \ref{lem:partialsfij.bounded} and \eqref{eq:L1.brac.unif.bd}, there exists $K_1$ such that
\begin{align*}
	\sup_{\bm{q} \in \Theta_0, p \in \{1,2\}}\Bvert
	g_{p,ij}(\bm{q})
	\Bvert \le K_1.
\end{align*}
Hence, for the first component, 
\begin{align*}
|R_\ell|=	\left| \frac{1}{\sum_{i=1}^n m_i}\sum_{i=1}^n \sum_{j=1}^{m_i}g_{1,ij}(\bm{q}) \right|  \le \frac{1}{\sum_{i=1}^n m_i}\sum_{i=1}^n \sum_{j=1}^{m_i}K_1 = K_1,
\end{align*}
while for the second component, 
\begin{align*}
{\rm Var}(S_\ell) \le \frac{\sigma^2}{(\sum_{i=1}^n m_i)^2}\sum_{i=1}^n 
	 \sum_{j=1}^{m_i} K_1^2 \le \frac{\sigma^2  K_1^2}{\sum_{i=1}^n m_i} \rightarrow 0. 
\end{align*}
Hence, for any $\tau \in (0,1)$, by Chebyshev's inequality, we may pick $K_2$ such that $P(|S_\ell| \ge K_2) \le \tau/8$ for all $n \ge 1$. 
Thus, setting $K=K_1+K_2$, we obtain, for all $n \ge 1$,
\begin{multline*}
	P\left( |R_n-S_n|>K, \bm{q} \in \Theta_0\right) \le P\left( |R_n|+|S_n|>K, \bm{q} \in \Theta_0\right)\\ 
	\le P\left( K_1+|S_n|>K_1+K_2, \bm{q} \in \Theta_0\right)\\
	= P\left( |S_n|> K_2, \bm{q} \in \Theta_0\right) \le \frac{\tau}{8}.
\end{multline*}
The claim now follows by taking a union bound over the eight choices for $k,l$ and $r$. 
\bbox

\begin{theorem} \label{thm:ls.asy.dist.qonly}
Assume the errors $\epsilon_{ij}, 1 \le i \le n, 1 \le j \le m_i$ are i.i.d with mean zero, variance $\sigma^2$ and for some $\delta>0$  satisfy $E|\epsilon_{ij}|^{2+\delta} =\tau^{2+\delta}<\infty$. Assume that \eqref{eq:summi->infty} holds and that the limit $\Gamma$ as given in \eqref{def:Gamma_n.nu} exists.
Then for $\mathcal{U}_\ell(\bm{q})$ given by \eqref{def:J and U},
\begin{align*}
b_\ell\mathcal{U}_\ell(\bm{q}_0) \rightarrow_d \mathcal{N}\left( {\bm 0}, \sigma^2 \Gamma \right) \qmq{where} b_\ell= \sqrt{\sum_{i=1}^n m_i}.
\end{align*}
\end{theorem}

\noindent {\em Proof:} We verify \eqref{eq:MultLind} of Lemma \ref{lem:mult.clt}. The first condition holds by \eqref{eq:Gamma_n.1} of Lemma \ref{lem:U.V.props}, and that the limit $\Gamma$ in \eqref{def:Gamma_n.nu} exists. For the second condition of \eqref{eq:MultLind}, by \eqref{eq:U.and.U'.in.V}, write
\begin{align*}
b_\ell {\mathcal U}_\ell(\bm{q}_0)= \sum_{i=1}^n \sum_{j=1}^{m_i}{\bm X}_{ij} \qmq{where}
	{\bm X}_{ij} = 
-\frac{1}{\sqrt{\sum_{i=1}^n m_i}}   \partial_{\bm q}f_{ij}({\bm q}_0)  \epsilon_{ij}. 
\end{align*}
By the assumption $\mathbb{E}|\epsilon_{ij}|^{2+\delta}\le \tau^{2+\delta}$ and Lemma \ref{lem:partialsfij.bounded} there exists $C$ such that 
\begin{align*} 
 \sum_{1 \le i \le n,1 \le j \le m_i, 1 \le k \le 2}   E|X_{ij,k}|^{2+\delta} \le \frac{C \tau^{2+\delta}}{(\sum_{i=1}^n m_i)^{\delta/2}}, 
\end{align*}
which tends to zero by \eqref{eq:summi->infty}. 
\qed

We conclude this section with:\\ 
{\em Proof of Theorem \ref{thm:ls.cons.asy.dist.qonly}:} 
Theorems \ref{thm:Gamma}  and \ref{thm:ls.asy.dist.qonly} show that the hypotheses of Theorems \ref{theorem-consistent-root} and \ref{mle-normal} are satisfied, yielding the claims for consistency and asymptotic normality. 

It remains to prove the claims on the consistency of the variance estimator. 
By \eqref{eq:asynom+varestimator}, and recalling $m=\sum_{i=1}^n m_i$, we have 
\begin{multline*}
	\widehat{\sigma}_\ell^2 = \frac{1}{m}\sum_{i=1}^n \sum_{j=1}^{m_i} \left(\epsilon_{ij} + f_{ij}(\bm{q}_0)- f_{ij}(\bm{\widehat{q}}_\ell) \right)^2 \\= \frac{1}{m}\sum_{i=1}^n \sum_{j=1}^{m_i}\left( 
	\epsilon_{ij}^2 +2 \epsilon_{ij} (f_{ij}(\bm{q}_0)- f_{ij}(\bm{\widehat{q}}_\ell)+(f_{ij}(\bm{q}_0)- f_{ij}(\bm{\widehat{q}}_\ell)^2\right).
\end{multline*}
The first term tends to $\sigma^2$ in probability by the weak law of large numbers.

To handle the second term, letting
$$
R=\frac{1}{m}\sum_{i=1}^n \sum_{j=1}^{m_i} |\epsilon_{ij}| \qmq{we have}
E[R]=\frac{1}{m}\sum_{i=1}^n \sum_{j=1}^{m_i} E|\epsilon_{ij}|\le \frac{1}{m}\sum_{i=1}^n \sum_{j=1}^{m_i} \sqrt{E\epsilon_{ij}^2}=\sigma.
$$
With any $r \in (0,1)$ such that the ball $B_r$ of radius $r$ centered at $\bm{q}_0$ is contained in $\mathcal{Q}$, 
Lemmas \ref{lem:NajfeldHave} and \ref{lem:partialsfij.bounded} show that the first derivatives of $f_j(\bm{q})$ are uniformly bounded for $(\bm{q},t) \in B_r \times [0,T]$,  and hence, that there exists some $K>0$ such that over this set 
\begin{align} \label{eq:fbounds.for.variance.estimator}
|f_{ij}(\bm{q}_0)- f_{ij}(\bm{q})| \le K \|\bm{q}_0- \bm{q}\|.
\end{align}
Let $\delta \in (0,K)$ be arbitrary, $F=\{|\bm{q}_0 - \bm{\widehat{q}}_\ell | \le \delta/K \}$ and note that
$$
S=\Bvert \frac{1}{m}\sum_{i=1}^n \sum_{j=1}^{m_i} \epsilon_{ij} (f_{ij}(\bm{q}_0)- f_{ij}(\bm{\widehat{q}}_\ell)\Bvert \qmq{satisfies} S{\bm 1}_F \le \delta R{\bm 1}_F.
$$

By Markov's inequality, for any $\tau>0$, 
\begin{multline*}
P(S\ge \tau) \le  P(S{\bf 1}_F \ge \tau)+ P(F^c)\le P(\delta R{\bm 1}_F \ge \tau) + P(F^c) \le \frac{\delta E[ R{\bm 1}_F]}{\tau} + P(F^c)\\
\le \frac{\delta E[R]}{\tau} + P(F^c)\le  \frac{\delta \sigma}{\tau}+ P(F^c) \rightarrow  \frac{\delta \sigma}{{\tiny }\tau}, 
\end{multline*}
using the non-negativity of $R$ in the fourth inequality, and the consistency of $\bm{q}_\ell$ when taking the limit. As $\delta$ can be made arbitrarily small we conclude that $P(S\ge \tau) \rightarrow 0$, and as $\tau$ is arbitrary, that $S \rightarrow_p 0$. 

Similarly decomposing the third term on the good event $F$ and its complement, on $F$ the inequality  \eqref{eq:fbounds.for.variance.estimator} shows that this term is bounded as
\begin{align*}
	\frac{K^2}{m}\sum_{i=1}^n \sum_{j=1}^{m_i}
	\|\bm{q}_0-\bm{\widehat{q}}_\ell \|^2 =K^2 \|\bm{q}_0-\bm{\widehat{q}}_\ell \|^2, 
\end{align*}
which tends to zero in probability in view of the consistency of 
$\bm{\widehat{q}}_\ell $. \bbox

\begin{section}{Inference on the BrAC curve}\label{sec:InferenceBrAC}

		In this section we obtain confidence bounds on the BrAC curve generated by a drinking episode of a subject in the field, and estimated using $\n$ TAC observations and an estimate ${\bm q}_m$ computed from $m$ measurements in a previous calibration experiment. Our notation here differs from that used in previous sections, the parameter $n$ now being absorbed in the number $m$ of total observations for calibration. Our uniform confidence bounds for the 
	reconstructed BrAC curve are obtained by applying a variation on the standard multivariate delta method, as given in Theorem 7 in \citet{Ferguson17}, using the properties provided by Theorem \ref{thm:ls.cons.asy.dist.qonly} on ${\bm q}_m$, and the assumed properties of the TAC measurement error.

	We begin by specifying in detail how we obtain our estimate of the BrAC curve. Independently of ${\bm q}_m$, $\n$ TAC observations $y_{\n,1},\ldots,y_{\n,\n}$ are collected from a drinking episode at the increasing times $0 \le s_{\n,1}< \cdots<s_{\n,\n} \le S$, given by
	\begin{align} \label{eq:yj.for.BrAC.est}
		y_j=f_\mu(s_{\n,j}; {\bm q}_0)+\epsilon_{\n,j}  \qmq{where} f_\mu(s;{\bm q})= \int_0^s Ce^{A(s-u)}B \mu(u)du 
	\end{align}
	as in \eqref{eq:fij.q.only}, where $\mu$ is the unknown BrAC curve to be estimated,  the matrices 
	$A$ and $B$ depend on ${\bm q}$ as in \eqref{def:A,B}, and $C^\transpose$ is a given fixed vector.

	To start, we assume only that the errors $\epsilon_{\n,1},\ldots,\epsilon_{\n,\n}$ are uncorrelated and have mean zero. We will allow for the possibility that the device used in the field may have different characteristics than the one used for calibration, and for now only
	impose the condition that the field noise variances are uniformly bounded above by $\overline{\sigma_f^2}$, some positive constant.

	Assume $\nu_\n, \n \ge 1$, the empirical probability measure of the TAC observation times, has weak limit $\nu_0$. For a given resolution level $p \in \mathbb{N}$ we select a basis of $p$ integrable functions  
	$\{\phi_k, 1 \le k \le p \}$ on $[0,S]$. The finite basis approximation of $\mu$ of order $p$ at time $s \in [0,S]$, with coefficient vector $\bsy{\beta} \in \mathbb{R}^p$, is given by
	\begin{align} \label{def:mup}
		\widehat{\mu}(s;\bsy{\beta}) = {\bm \phi}(s)^\transpose \bsy{\beta} \qmq{where} \quad {\bm \phi}(s) = [\phi_1(s) ,\ldots,\phi_p(s)]^\transpose \in \mathbb{R}^p.
	\end{align}
	
	Substitution of $\mu$ by the approximation $\widehat{\mu}(s;\bsy{\beta})$ into the integral in \eqref{eq:yj.for.BrAC.est} yields the predicted TAC values given at $s \in [0,S]$ of
	\begin{align} \label{eq:fmu.basisp}
		f_{\widehat{\mu}}(s;{\bm q})= \left( \int_0^s Ce^{A(s-u)}B {\bm \phi}(u)^\transpose du \right) \bsy{\beta}:=\bsy{\psi}(s;{\bm q})^\transpose \bsy{\beta} \qmq{where} \psi(s;{\bm q}) \in \mathbb{R}^p.
	\end{align}
	
	
	Now dropping the double index notation for simplicity, letting ${\bm s}_\n=(s_1,\ldots,s_\n)^\transpose$, for a given ${\bm q}$ and a sequence of symmetric, non-negative definite matrices $M_\n \in R^{p\times p}$, selected to promote regularization, we choose
	$\bsy{\beta}=\bsy{\beta}({\bm s}_\n,{\bm q}) \in \mathbb{R}^p$ to minimize the objective function
	\begin{align} \label{eq:Jbeta.con.M}
		J(\bsy{\beta}) =   \frac{1}{\n}\|{\bm Y}-X({\bm s}_\n;{\bm q})\bsy{\beta}\|^2 +\bsy{\beta}^\transpose M_\n\bsy{\beta} =\|{\bm{V}}-W({\bm s}_\n;{\bm q})\bsy{\beta}\|^2,
	\end{align}
	where
	\begin{align*}
		{\bm Y}=\left[
		\begin{array}{c}
			y_1\\
			y_2\\
			\cdot\\
			\cdot\\
			y_\n
		\end{array}
		\right],
		\quad X({\bm s};{\bm q})=\left[
		\begin{array}{c}
			\bsy{\psi}(s_1,{\bm q})^\transpose \\
			\bsy{\psi}(s_2,{\bm q})^\transpose \\
			\cdots\\
			\cdots\\
			\bsy{\psi}(s_\n,{\bm q})^\transpose
		\end{array}
		\right] \in \mathbb{R}^{\n \times p},
	\end{align*}
	and
	\begin{align*}
		{\bm V}= \left[\frac{1}{\sqrt{\n}}{\bm Y},0\right]^\transpose \qmq{and} W({\bm s}_\n;{\bm q})=\left[\frac{1}{\sqrt{\n}}X({\bm s}_\n;{\bm q}),\sqrt{M_\n}\right]^\transpose.
	\end{align*} 
	By standard results in least squares estimation, when $W({\bm s}_\n;{\bm q})^\transpose W({\bm s}_\n;{\bm q})$ is full rank, the unique minimizer of $J(\bsy{\beta})$ is given by
	\begin{multline} \label{eq:beta.ls.orth.basis}
		\bsy{\beta}({\bm s}_\n,{\bm q}) = 
		(W({\bm s}_\n;{\bm q})^\transpose W({\bm s}_\n;{\bm q}))^{-1}W({\bm s}_\n;{\bm q})^\transpose {\bm V}\\=
		\Big(\left[\frac{1}{\sqrt{\n}}X({\bm s}_\n;{\bm q})^{\transpose},\sqrt{M_\n}\right]\left[
		\begin{array}{c}
			\frac{1}{\sqrt{\n}}X({\bm s}_\n;{\bm q})\\
			\sqrt{M_\n}
		\end{array}
		\right]\Big)^{-1}\left[\frac{1}{\sqrt{\n}}X({\bm s}_\n;{\bm q})^{\transpose},\sqrt{M_\n}\right]\left[
		\begin{array}{c}
			\frac{1}{\sqrt{\n}}\bm Y\\
			0
		\end{array}
		\right]\\=\left(\frac{1}{\n}X({\bm s}_\n;{\bm q})^\transpose X({\bm s}_\n;{\bm q})+M_\n\right)^{-1}\frac{1}{\n}X({\bm s}_\n;{\bm q})^\transpose {\bm Y}.
	\end{multline}
	
	We may also write these equations in a somewhat more convenient form. For $\n \ge 1$ let
	\begin{multline} \label{def:GnZn.etan}
		G_\n({\bm q}) = \int_0^S \bsy{\psi}(s;{\bm q}) \bsy{\psi}(s;{\bm q})^\transpose d\nu_\n(s),\\ {\bm Z}_\n({\bm q}) = \int_0^S \bsy{\psi}(s;{\bm q}) f_\mu(s;{\bm q}_0)d\nu_\n(s)
		\qmq{and} \bsy{\epsilon}_\n({\bm q}) = \frac{1}{\n}\sum_{j=1}^\n \bsy{\psi}(s_j;{\bm q}) \epsilon_j,
	\end{multline}
	and let these same formulas hold for $\n=0$ upon setting ${\bm \epsilon}_0=0$. 
	Then, using the alternative notation $\bsy{\beta}_\n({\bm q})$ for $\bsy{\beta}({\bm s}_\n,{\bm q})$, we recover \eqref{eq:beta.ls.orth.basis} from
	\begin{align} \label{eq:defbsybetan}
		\bsy{\beta}_\n({\bm q})= (G_\n({\bm q})+M_\n)^{-1}\left( {\bm Z}_\n({\bm q})+ \bsy{\epsilon}_\n({\bm q})\right) \qm{$\n \ge 0$,}
	\end{align}
	where, now assuming that the sequence of matrices $M_\n$ has limit $M_0$, we also define $\bsy{\beta}_0({\bm q})$ by \eqref{eq:defbsybetan} applying the stated convention that $\bsy{\epsilon}_0 = {\bm 0}$. We note $G_\n({\bm q})+M_\n$ will be invertible whenever $M_\n$ is positive definite. For notational simplicity in what follows, let
	\begin{align} \label{eq:defHn} 
		H_\n({\bm q})=G_\n({\bm q})+M_\n \quad \mbox{for $\n \ge 0$.}
	\end{align}
	
	When basing inference on the estimate ${\bm q}_m$ obtained from a calibration session, as in \eqref{def:mup}, the estimated BrAC curve is given by $\mu_\n(s;{\bm q}_m)$, where \begin{align}\label{eq:est.BrAC.basis}
		\mu_\n(s;{\bm q}) ={\bm \phi}(s)^\transpose \bsy{\beta}_\n({\bm q}) \qm{$\n \ge 0, s \in [0,S].$}
	\end{align}
	
	Next, define the Lipschitz (semi)norm of a real valued function $g$ with domain $\mathcal{D} \subset \mathbb{R}$ by
	\begin{align*}
		\|g\|_{\rm Lip} = \sup_{x \not = y, \{x,y\} \subset \mathcal{D}}\frac{|g(x)-g(y)|}{|x-y|}.
	\end{align*}
	In order to control the variation in the estimate $\bsy{\beta}_\n({\bm q})$ caused by that in ${\bm q}_m$, we introduce the following assumption. 
	
	\begin{assumption}\label{ass:nun.to.nu.C.alpha}
		For a given sequence of empirical probability measures $\nu_\n, \n \ge 1$ with weak limit $\nu_0$, all supported on $[0,S]$, there exists a constant $C$ that may depend on $S$, and a sequence $r_\n, \n \ge 1$ of real numbers tending to zero as $\n \rightarrow \infty$ such that
		\begin{align} \label{assumption.diff.integrals}
			\sup_{\|g\|_{\rm Lip} \le L}\Bvert \int_0^S g(u)d\nu_\n - \int_0^S g(u)d\nu_0 \Bvert \le LC r_\n \qmq{for all $n \ge 1$.}
		\end{align}
	\end{assumption}
	As $\nu_\n([0,S])=\nu_0([0,S])$ the difference over which the supremum is taken in  \eqref{assumption.diff.integrals} is unchanged by replacing $g(x)$ by $g(x)+c$ for any constant $c$, and hence we may assume that $g(0)=0$. In particular, for $x  \in [0,S]$ we then have that 
	\begin{align}\label{eq:sup.g.via.Lip}
		\|g\| = \sup_{x \in [0,S]}|g(x)|=\sup_{x \in [0,S]}|g(x)-g(0)|\le \sup_{x \in [0,S]}|x|\|g\|_{\rm Lip}=S\|g\|_{\rm Lip}.
	\end{align}

	When $\nu_\n$ is the empirical probability measure of the $\n$ equally spaced observations made at times $s_i=Si/\n, i=1,\ldots,\n$, then the limit measure $\nu_0$ is the uniform probability measure over $[0,S]$. In this case, 
	\begin{multline*}
		\Bvert \int_0^S g(u)d\nu_\n - \int_0^S g(u)d\nu_0 \Bvert = \Bvert \frac{1}{\n}\sum_{i=1}^\n g(s_i)- \int_0^S g(u)d \nu_0 \Bvert \\
		= \Bvert \sum_{i=1}^{\n-1} \left( \int_{s_i}^{s_{i+1}} \left( g(s_i)- g(u) \right) d\nu_0 \right) +\frac{g(S)}{\n}-\int_0^{s_1} g(u)du  \Bvert \\ \le \|g\|_{\rm Lip} \sum_{i=1}^{\n-1} \int_{s_i}^{s_{i+1}}|s_i-u| du + \frac{S\|g\|_{\rm Lip}}{\n} + \frac{S^2\|g\|_{\rm Lip}}{\n}\\
		= \frac{(\n-1)S^2\|g\|_{\rm Lip}}{2\n^2}+ \frac{(S+S^2)\|g\|_{\rm Lip}}{\n} 
		\le C\|g\|_{\rm Lip} r_\n,
	\end{multline*}
	and Assumption \ref{ass:nun.to.nu.C.alpha} holds with $C=3S^2/2+S,$ and $r_\n=1/\n$.
	
	Alternatively, when $\nu_\n$ is the empirical measure of times $X_1,\ldots,X_\n$, independent with common distribution $\nu_0$ supported on $[0,S]$, then Assumption \ref{ass:nun.to.nu.C.alpha} holds with $r_\n=1/\sqrt{\n}$ with high probability. In particular, Theorem 8.2.3 of  \citet{Versh18} shows that there exists an absolute constant $C$ such that 
	\begin{align} \label{eq:bdEZ}
		E[S_\n] \le CSLr_\n \qmq{where} S_\n = \sup_{\|g\|_{\rm Lip} \le L}	\Bvert \int_0^S g(u)d\nu_\n - \int_0^S g(u)d\nu_0 \Bvert.
	\end{align}
	By \citet{Tal96} and \citet{Bos03}, as applied in Theorem 3.3.9 of \citet{GiNi16}, wt $U=\sigma=SL$ by virtue of \eqref{eq:sup.g.via.Lip}, and hence $\nu_n= 2UE[S_\n]+\n \sigma^2$ as there, we have
	\begin{align*}
		P\left(S_\n \ge E[S_\n] + \sqrt{(4SL\, E[S_\n]  + 2\n S^2L^2)x}+SLx/3\right) \le e^{-x} \qmq{for all $x \ge 0$.}
	\end{align*}
	Using the bound \eqref{eq:bdEZ}, and recalling \eqref{eq:sup.g.via.Lip},  with $C$ being a constant not necessarily the same at each occurrence, we obtain
	\begin{multline*}
		E[S_\n] + \sqrt{(4SL\, E[S_\n]  + 2\n S^2L^2)x}+SL x/3 \\ \le
		SL \left( C \sqrt{\n} + \sqrt{(C \sqrt{\n}+ 2\n)x} + x/3\right) 
		\le SLC \left(\sqrt{\n} +\sqrt{\n x} + x \right), 
	\end{multline*}
	implying that
	\begin{align*}
		P\left(S_\n \ge SLC\left(1 + \sqrt{x} + x \right)r_\n  \right)\le P\left(S_\n \ge SLC\left(1 + \sqrt{x} + x/\sqrt{\n} \right)r_\n \right) \le e^{-x}.
	\end{align*}
	Hence, given $\alpha \in (0,1)$, inequality 
	\eqref{assumption.diff.integrals} in 
	Assumption \ref{ass:nun.to.nu.C.alpha} holds with probability at least $1-\alpha$ for some constant $C$ depending only on $\alpha$ and $r_\n=1/\sqrt{\n}$.

	We next pause to prove some technical results that will be invoked in Theorems \ref{thm:beta.consistent} and \ref{thm:n=1.asy.dist.BrAC}; the partials inside the integral in \eqref{eq:partialipsi} can be computed applying \eqref{d1eAuB} of Lemma \ref{lem:partialsfij.bounded}.

	\begin{lemma} \label{lem:beta.dervs}
		The partial derivatives of $\bsy{\psi}(s,{\bm q})$ in \eqref{eq:fmu.basisp}
		with respect to the $i^{th}$ component of ${\bm q}$ for $i=1,2$ 
		exist and are given by 
		\begin{align}\label{eq:partialipsi}
			\partial_i  \bsy{\psi}(s,{\bm q})^\transpose = \int_0^s C(\partial _i e^{A(s-u)}B) {\bm \phi}(u) ^\transpose du, 
		\end{align}
		are bounded and continuous as a function of $s \in [0,S]$ and continuous in ${\bm q}$ on $\mathcal{Q}$ as given in \eqref{def:Q0}, and for any $\mathcal{C}$ whose closure lies in  $\mathcal{Q}$ there exists a finite constant $L$ such that on $[0,S]$
		\begin{align} \label{eq:psi+derv.bdd}
			\sup_{{\bm q} \in \mathcal{C}}
			\|\bsy{\psi}(\cdot;{\bm q})\| <\infty	\qmq{and} \sup_{{\bm q} \in \mathcal{C}}
			\|\bsy{\psi}(\cdot;{\bm q})\|_{\rm Lip} \le L.
		\end{align}
		Further, for any  ${\bm q} \in \mathcal{Q}$ and $\n \ge 0$, $G_\n({\bm q})$ and ${\bm Z}_\n({\bm q})$  given in \eqref{def:GnZn.etan} are continuous and
				\begin{multline} \label{eq:partial_iZ0G0}
			\partial_i G_0({\bm q}) = \int_0^S \left( \partial_i \bsy{\psi}(s,{\bm q}) \bsy{\psi}(s,{\bm q})^\transpose + \bsy{\psi}(s,{\bm q}) \partial_i \bsy{\psi}(s,{\bm q})^\transpose \right) d\nu_0 \qm{and}\\
			\partial_i {\bm Z}_0({\bm q})=\int_0^S \partial_i \bsy{\psi}(s,{\bm q})  f_\mu(s,{\bm q_0}) d\nu_0
		\end{multline}
		exist and are continuous. For $\bsy{\beta}$ in \eqref{eq:defbsybetan}, the partials
		\begin{align}
			\label{eq:partial_ibeta_0}
			\partial_i \bsy{\beta}_0({\bm q}) = H_0(\bm q)^{-1} \partial_i {\bm Z}_0({\bm q})
			-H_0(\bm q)^{-1}(\partial_i G_0(\bm q))H_0(\bm q)^{-1}{\bm Z}_0({\bm q})
		\end{align}
		exist and are continuous at any ${\bm q} \in \mathcal{Q}$ for which $H_0(\bm q)^{-1}$ exists. 
	\end{lemma}

	\noindent {\em Proof:} The claims for  $\bsy{\psi}(s,{\bm q})$ and its partial derivatives follow directly from Lemma \ref{lem:partialsfij.bounded}, the integrability of $\phi_k(u), k=1,\ldots,p$ on $[0,S]$ and the fact that continuous functions on compact sets are bounded. The claims on $G_\n({\bm q})$ and $Z_\n({\bm q})$ and their partials follow from the  properties of $f_\mu(s,{\bm q}_0)$ over $s \in [0,S]$ as provided by Lemma
	\ref{lem:partialsfij.bounded}, the demonstrated 
	properties of $\bsy{\psi}(s,{\bm q})$ and the dominated convergence theorem. The well known formula for differentiating matrix inverses yields \eqref{eq:partial_ibeta_0}
	and the final claim, noting that the map taking a matrix to its inverse is continuous. \bbox
	
	\begin{lemma} \label{eq:quantities.have.rate.orn}
Let Assumption \ref{ass:nun.to.nu.C.alpha} hold.	Then for $G_\n({\bm q})$ and $Z_\n({\bm q})$ as in \eqref{def:GnZn.etan} for $\n \ge 0$, for any set $\mathcal{C}$ whose closure lies in $\mathcal{Q}$ and is compact,  there exists a constant $C$ such that	\begin{multline}\label{eq:diffinverseGs.Zs}
		\sup_{{\bm q} \in \mathcal{C}} \|G_\n({\bm q})-G_0({\bm q})\| \le Cr_\n,  \,\, \sup_{{\bm q} \in \mathcal{C}}\|Z_\n({\bm q})-Z_0({\bm q})\| \le C r_\n \\ \qmq{and} \sup_{{\bm q} \in \mathcal{C}} \|{\bm Z}_\n({\bm q})\| < \infty.
	\end{multline}
	
When the field error variables $\epsilon_1,\ldots,\epsilon_\n$ have mean zero and are uncorrelated and have variances uniformly dominated by $\overline{\sigma_f^2}$, then
\begin{align} \label{eq:var.eta.dominated}
	\sup_{{\bm q} \in \mathcal{Q}_0} \|{\rm Var}(\bsy{\epsilon}_\n({\bm q}))\| 
	\le \overline{\sigma_f}^2 \sup_{s \in [0,S],{\bm q} \in \mathcal{C}}\|\bsy{\psi}(s,{\bm q})\|^2/\n.
\end{align}

If  $H_0(\bm q)^{-1}$ exists for some ${\bm q}_0 \in \mathcal{Q}$ then there exists an open  set $\mathcal{Q}_0$  containing ${\bm q}_0$ whose closure $\overline{\mathcal{Q}_0}$ lies in $\mathcal{Q}$ and such that $H_0(\bm q)^{-1}$ exists for all ${\bm q} \in \overline{\mathcal{Q}_0}$.
If there exists a constant $C$ such that $\|M_\n-M_0\| \le Cr_\n$,
then there exists a constant $C$ such that for $\n=0$ and all $\n$ sufficiently large 
		\begin{align} \label{eq:supGs.Zs}
	\sup_{{\bm q} \in \mathcal{Q}_0} \|H_\n(\bm q)^{-1}-H_0(\bm q)^{-1}\| \le C r_\n \qmq{and}		\sup_{{\bm q} \in \mathcal{Q}_0} \|H_\n(\bm q)^{-1}\|<\infty.
		\end{align}

	\end{lemma}
	
	\noindent {\em Proof:} 	The first two claims of \eqref{eq:diffinverseGs.Zs} follow by Lemmas \ref{lem:beta.dervs} and and \ref{lem:partialsfij.bounded}, that show the integrands $g$ in \eqref{assumption.diff.integrals} of Assumption \ref{ass:nun.to.nu.C.alpha}
	for the two cases here are uniformly Lipschitz. The final claim then holds by the triangle inequality and those same lemmas, which verify the case $\n=0$.  The  claim \eqref{eq:var.eta.dominated} follows similarly directly from definition \eqref{def:GnZn.etan} and the stated assumptions on the error terms. 
	
Denoting the eigenvalues $\lambda_1(N_i) \le \cdots \le \lambda_p(N_i)$  of non-negative definite symmetric matrices $N_1,N_2 \in \mathbb{R}^{p \times p}$, by Weyl's theorem (see Theorem 4.3.1 of \citet{HoJo12}), 
	\begin{align} \label{eq:weyl}
		|\lambda_i(N_1)-\lambda_i(N_2))| \le \| N_1 - N_2 \| \qmq{for $i=1,\ldots,p$.}
	\end{align}
	The matrix $G_0({\bm q})$ is continuous in ${\bm q}$ by Lemma \ref{lem:beta.dervs},  hence $H_0(\bm q)$ is likewise continuous. When $H_0(\bm q)$ is invertible at ${\bm q}_0$, the continuity of $\lambda_1(\cdot)$ provided by \eqref{eq:weyl}
	yields the existence of $\epsilon>0$ such that $\lambda_1(H_0(\bm q)) > \epsilon$ for all ${\bm q}$ in some bounded open neighborhood $\mathcal{Q}_0$ centered at $\bm{q}_0$, and with radius $\epsilon$ taken small enough that the closure of $\mathcal{Q}_0$ lies in  $\mathcal{Q}$. By continuity this same inequality holds in the non-strict sense over the closure, thus showing the claim made following \eqref{eq:var.eta.dominated}.

By the first claim in \eqref{eq:diffinverseGs.Zs} and under our assumption bounding $\|M_\n-M_0\|$, we conclude that there exists a constant $C$ such that $\|H_\n({\bm q})-H_0({\bm q})\| \le C r_\n$.  As $r_\n \rightarrow 0$, we have $r_\n < \epsilon/2$ for all $\n$ sufficiently large. Hence, for such $\eta$ 
by \eqref{eq:weyl} we have $\inf_{{\bm q} \in \mathcal{Q}_0}\lambda_1(H_\n(\bm q)) > \epsilon/2$, and therefore, for all $\bm{q} \in \mathcal{Q}_0$, 
\begin{multline*}
\|H_\n^{-1}({\bm q})-H_0^{-1}({\bm q})\| = 
\|H_\n^{-1}({\bm q})(H_\n({\bm q})-H_0({\bm q}))H_0^{-1}({\bm q})\| \\ \le \|H_\n^{-1}({\bm q})\| \,\|H_\n({\bm q})-H_0({\bm q})\|\, \|H_0^{-1}({\bm q})\| \le \frac{2C}{\epsilon^2}r_\n,
\end{multline*}
this demonstrating the first claim in \eqref{eq:supGs.Zs}. The second follows from the first, and the triangle inequality. 	\bbox
	
	Now moving to the properties of  $\bsy{\beta}_\n({\bm q}_m)$, note the decomposition
	\begin{align} \label{eq:beta.pm}
		\bsy{\beta}_\n({\bm q}_m)-\bsy{\beta}_0({\bm q}_0)=
		(\bsy{\beta}_\n({\bm q}_m)-\bsy{\beta}_0({\bm q}_m)) +
		( \bsy{\beta}_0({\bm q}_m)-\bsy{\beta}_0({\bm q}_0)),
	\end{align}
	and for the first term, applying \eqref{eq:defbsybetan} we may write
	\begin{multline} \label{eq:triangle.for.diff.qm}
		\bsy{\beta}_\n({\bm q}_m)-\bsy{\beta}_0({\bm q}_m) \\ =
		\left( H_\n(\bm q_m)^{-1}
		- H_0(\bm q_m)^{-1}\right)
		{\bm Z}_\n({\bm q}_m)+H_0(\bm q_m)^{-1}\left({\bm Z}_\n({\bm q}_m) -  {\bm Z}_0({\bm q}_m)\right)\\
		+ H_\n(\bm q_m)^{-1} \bsy{\epsilon}_\n({\bm q}_m).	\end{multline}

	We now prove a consistency result for $\bsy{\beta}_\n({\bm q}_m)$, and apply it to show that the BrAC curve estimate converges uniformly in probability to $\mu_0(s;{\bm q}_0)$, given in \eqref{eq:est.BrAC.basis}.
	\begin{theorem}\label{thm:beta.consistent}
		Suppose that ${\bm q}_m$ is consistent for ${\bm q}_0$ as $m \rightarrow \infty$, that $H_0(\bm q_0)$ is invertible, and that the error variables $\epsilon_1,\ldots,\epsilon_\n$ have mean zero, and are uncorrelated with variances dominated by $\overline{\sigma_f}^2$. In addition, let Assumption \ref{ass:nun.to.nu.C.alpha} hold. Then when $m$ and $\n$ tend to infinity
		\begin{align*}
			\bsy{\beta}_\n({\bm q}_m) \rightarrow_p \bsy{\beta}_0({\bm q}_0), 
		\end{align*}
		and when $\sup_{k \ge 1}\|\phi_k\|<\infty$, the supremum norm $\|\mu_\n(\cdot;{\bm q}_m)  - \mu_0(\cdot;{\bm q}_0)\| \rightarrow_p 0$.
	\end{theorem}
	
	\noindent {\em Proof:} By the triangle inequality, to show $\bsy{\beta}_\n({\bm q}_m)$ is consistent for  $\bsy{\beta}_0({\bm q}_0)$, it suffices to verify that the two terms on the right side of tend to zero in probability. The first term 
	converges to zero in probability by the consistency of ${\bm q}_m$ for ${\bm q}_0$,  which in particular guarantees that ${\bm q}_m$ will lie in the set $\mathcal{Q}_0$ given by Lemma \ref{eq:quantities.have.rate.orn} with probability tending to one, and then also invoking
	\eqref{eq:triangle.for.diff.qm}, Lemma \ref{eq:quantities.have.rate.orn} and Assumption \ref{ass:nun.to.nu.C.alpha}.
	The second term tends to zero by virtue of the consistency of ${\bm q}_m$ and the continuity of the function $\bsy{\beta}_0(\cdot)$. 
	
	The last claim follows from the first, using that ${\bm \phi}(s)$ is bounded on $[0,S]$, and from \eqref{eq:est.BrAC.basis} yielding 
	\begin{align*}
		\sup_{s \in [0,S]}|\mu_\n(s;{\bm q}_m) - \mu_0(s;{\bm q}_0)| \le \|{\bm \phi}\| \|\bsy{\beta}_\n({\bm q})-\bsy{\beta}_0({\bm q}_0)\|.
	\end{align*}
	\bbox
	
	We prove the following distributional limit theorem the first term in \eqref{eq:beta.pm}.
	\begin{lemma}\label{lem:eta_n.d.limit} Assume  $H_0(\bm q)^{-1}$ exists for ${\bm q}_0 \in \mathcal{Q}$. In addition, let the errors $\epsilon_i,i \ge 1$ be independent, mean zero with common variance $\sigma_f^2 \in (0,\infty)$ and independent of $\bm{q}_m$, and suppose for some $\delta>0$ and $\tau>0$ that $E|\epsilon_i|^{2+\delta} \le \tau^{2+\delta}$ for all $i \ge 1$. 
		If $\sqrt{m}({\bm q}_m - {\bm q}_0)=O_p(1)$, Assumption \ref{ass:nun.to.nu.C.alpha} holds, and that $\n,m\rightarrow \infty$ so that $mr_\n^2/\n \rightarrow 0$, then, with $Y\sim \mathcal{N}(0,\sigma_f^2 H_0(\bm q_0)^{-1}G_0({\bm q}_0)H_0(\bm q_0)^{-1})$,

		\begin{multline} \label{eq:for.asy.ind}
			 H_\n(\bm q_m)^{-1}  \bsy{\epsilon}_\n(\bm{q}_m)= H_0(\bm q_0)^{-1} \bsy{\epsilon}_\n(\bm{q}_0)+o_p(1/\sqrt{m}) \quad \mbox{and}\\  \sqrt{\n} H_0(\bm q_0)^{-1} \bsy{\epsilon}_\n(\bm{q}_0) \rightarrow_d Y \qmq{where}  Y\sim \mathcal{N}(0,\sigma_f^2 H_0(\bm q_0)^{-1}G_0({\bm q}_0)H_0(\bm q_0)^{-1}).
		\end{multline}
		If $\n/m$ is bounded away from infinity and $\sqrt{\n} r_\n \rightarrow 0$, then 
		\begin{align} \label{eq:eta_n.d.limit}
			\sqrt{\n} (\bsy{\beta}_\n({\bm q}_m)-\bsy{\beta}_0({\bm q}_m))\rightarrow_d Y.
		\end{align}
	\end{lemma}

	We note that the condition that $\sqrt{m}({\bm q}_m - {\bm q}_0)=O_p(1)$ is implied by the conditions of Theorem \ref{thm:ls.cons.asy.dist.qonly}, as they provide the stronger conclusion that $\sqrt{m}({\bm q}_m - {\bm q}_0)$ converges in distribution. Note as well that $mr_\n^2/\eta \rightarrow 0$ whenever $m$ and $\n$ are of comparable size, and $r_\n \rightarrow 0$.

	\noindent {\em Proof:} Since $\sqrt{m}({\bm q}_m - {\bm q}_0)=O_p(1)$ we have ${\bm q}_m  \rightarrow_p {\bm q}_0$, and we may assume without loss of generality that $\bm{q}_m$ is contained in the set $\mathcal{Q}_0$ given in Lemma \ref{eq:quantities.have.rate.orn}. Writing
	\begin{multline} \label{eq:3rdterm.eta.decomp}
		H_\n(\bm q_m)^{-1} \bsy{\epsilon}_\n({\bm q}_m)\\  = 
			(H_\n(\bm q_m)^{-1} - H_0(\bm q_m)^{-1})\bsy{\epsilon}_\n({\bm q}_m)
			+ H_0(\bm q_m)^{-1} (\bsy{\epsilon}_\n({\bm q}_m)-\bsy{\epsilon}_\n({\bm q}_0))\\
		+ (H_0(\bm q_m)^{-1}-H_0(\bm q_0)^{-1})\bsy{\epsilon}_\n({\bm q}_0)
			+ H_0(\bm q_0)^{-1}\bsy{\epsilon}_\n({\bm q}_0),
	\end{multline}
	we show \eqref{eq:for.asy.ind} by demonstrating that the first three terms tend to zero in probability after scaling by $\sqrt{m}$. The squared norm of the covariance matrix of first term scaled by $m$ and conditional on $\bm{q}$, by applying the given assumptions on the errors, can be bounded by
	\begin{align*}
	m\|(H_\n(\bm q_m)^{-1} - H_0(\bm q_m)^{-1}){\rm Var}(\bsy{\epsilon_\n}(\bm{q}_m)|\bm{q}_m)(H_\n(\bm q_m)^{-1} - H_0(\bm q_m)^{-1})\| \le Cmr_\n^2/\n
	\end{align*}
	using the first inequality in \eqref{eq:supGs.Zs} and \eqref{eq:var.eta.dominated}; the upper bound clearly bounds the squared norm of the scaled unconditional covariance matrix of the first term.

	For the second term, properly scaled, define
	$$
	A_{\n,m}=\sqrt{m}(\bsy{\epsilon}_\n(\bm{q}_m)-\bsy{\epsilon}_\n(\bm{q}_0)) =  \frac{\sqrt{m}}{\n}\sum_{j=1}^\n (\psi(s_j,\bm{q}_m)-\psi(s_j,\bm{q}_0))\epsilon_j.
	$$
	Let $\tau \in (0,1)$ be given and $\mathcal{Q}_0$ be as in Lemma \ref{eq:quantities.have.rate.orn}. Since $\sqrt{m}({\bm q}_m - {\bm q}_0)=O_p(1)$, there exists $M$ such that 
	$$
	\liminf_m P(\Omega_{M,m}) \ge 1-\tau \qmq{where} \Omega_{M,m} = \{ \sqrt{m}\|{\bm q}_m-{\bm q}_0\| \le M
	\},
	$$
	and we take $m$ sufficiently large so that the closed ball of radius $M/\sqrt{m}$ centered at $\bm{q}_0$ lies in the set  $\mathcal{Q}_0$ given in Lemma \ref{lem:beta.dervs}.
	
	By the independence between ${\bm q}_m$ and $\epsilon_j,j=1,\ldots,\n$,
	\begin{align} \label{eq:cond.exp.A.zero}
		E(A_{\n,m}{\bm 1}_{\Omega_{M,m}}|{\bm q}_m)=0.
	\end{align}
	Hence, 
	via the conditional variance formula, and that $\Omega_{M,m}$ is measurable with respect to ${\bm q}_m$, we obtain
	\begin{multline*}
		{\rm Var}\left( A_{\n,m}{\bm 1}_{\Omega_{M,m}}\right)
		= E\left( {\rm Var}(A_{\n,m}{\bm 1}_{\Omega_{M,m}}|{\bm q}_m)\right) \\= \frac{m \sigma_f^2}{\n^2} E \left[
		\sum_{j=1}^\n \left( \psi(s_j,\bm{q}_m)-\psi(s_j,\bm{q}_0)\right)^2{\bm 1}_{\Omega_{M,m}} \right] 
		\le \frac{ \sigma_f^2L^2}{\n}E[m\|{\bm q}_m-{\bm q}_0\|^2{\bm 1}_{\Omega_{M,m}}] \\ \le \frac{ \sigma_f^2L^2}{\n}M^2 \rightarrow 0 \qmq{as $\n \rightarrow \infty$,}
	\end{multline*}
	where we used Lemma \ref{lem:beta.dervs} in the first inequality. Hence, as $E[A_{\n,m}{\bm 1}_{\Omega_{M,m}}]=0$ by \eqref{eq:cond.exp.A.zero}, and that $\tau>0$ is arbitrary, we have $A_{\n,m} \rightarrow 0$ as $\n \rightarrow \infty$. By 
		\eqref{eq:supGs.Zs} of Lemma \ref{lem:beta.dervs} and that $\bm{q}_m \rightarrow_p \bm{q}_0$, the second term is $o_p(1)$.

	For the third term, first note that when $\bm{q}_m \in \mathcal{Q}_0$, which occurs with probability tending to 1 as $m \rightarrow \infty$, there exists a constant $C$ such that
	\begin{multline*}
		\|H_0^{-1}({\bm q}_m)-H_0^{-1}({\bm q}_0)\| = 
		\|H_0^{-1}({\bm q}_m)(H_0({\bm q}_m)-H_0({\bm q}_0))H_0^{-1}({\bm q}_0)\| \\ \le \|H_0^{-1}({\bm q}_m)\| \,\|H_0({\bm q}_m)-H_0({\bm q}_0)\|\, \|H_0^{-1}({\bm q}_0)\| \\= \|H_0^{-1}({\bm q}_m)\| \,\|G_0({\bm q}_m)-G_0({\bm q}_0)\|\, \|H_0^{-1}({\bm q}_0)\| \le C\|G_0({\bm q}_m)-G_0({\bm q}_0)\|,
	\end{multline*}
	by \eqref{eq:supGs.Zs} and \eqref{eq:psi+derv.bdd} of Lemma \ref{lem:beta.dervs}. Now by \eqref{eq:partial_iZ0G0} of Lemma \ref{lem:beta.dervs} there exists $C$ such that
	$$
	\sqrt{m}\|G_0({\bm q}_m)-G_0({\bm q}_0)\| \le C\sqrt{m}\|{\bm q}_m-{\bm q}_0\| = O(1). 
	$$
	Hence, by \eqref{eq:var.eta.dominated}, this term will tend to zero as $\eta \rightarrow \infty$. The proof of \eqref{eq:for.asy.ind} is complete.

To verify the distributional convergence claimed in
\eqref{eq:for.asy.ind} it suffices that 
	$$
	\sqrt{\n}\bsy{\epsilon}_n({\bm q}_0) = \sum_{i=1}^\n X_{\n,i}\rightarrow_d \mathcal{N}(0,\sigma_f^2 G_0(\bm{q}_0))
	\qmq{where}
	X_{\n,i}=\frac{1}{\sqrt{\n}}\psi(s_i;{\bm q_0})\epsilon_i.
	$$
	We apply Lemma \ref{lem:mult.clt}, noting that the first condition in \eqref{eq:MultLind} holds, as $G_\n({\bm q}_0)$ converges to $G_0({\bm q}_0)$ by Lemma \ref{eq:quantities.have.rate.orn}.
	
	It remains to verify the second condition in \eqref{eq:MultLind} to complete the proof of \eqref{eq:for.asy.ind}.
	The vector ${\bm \psi}(s,{\bm q})$ is uniformly bounded over $s \in [0,S]$ and ${\bm q} \in \mathcal{Q}_0$ by \eqref{eq:psi+derv.bdd} of Lemma \ref{lem:beta.dervs}. Hence, for some constant $C$, as $\n \rightarrow \infty$,
	$$
	\sum_{i=1}^\n \E \|X_{\n,i}\|^{2+\delta} \le \frac{1}{\n^{1+\delta/2}}\sum_{i=1}^\n
	\|\psi(s_i;{\bm q_0})\|^{2+\delta}E|\epsilon_i|^{2+\delta} \le C^{2+\delta}\n^{-\delta/2}\tau^{2+\delta} \rightarrow 0, 
	$$
	completing the proof of \eqref{eq:for.asy.ind}.

		For the final claim, after scaling by $\sqrt{\n}$, we see by the triangle inequality that the first two terms in \eqref{eq:triangle.for.diff.qm} tend to zero by the consistency of $\bm{q}_m$ for $\bm{q}_0$, Lemma \ref{eq:quantities.have.rate.orn} and that $\sqrt{\n}r_n \rightarrow 0$.  The final claim \eqref{eq:eta_n.d.limit} now follows from \eqref{eq:for.asy.ind}, applying the assumption that $\n/m$ is bounded. 
	\bbox

Lastly, we determine the asymptotic distribution of the estimated BrAC curve, properly scaled. Using that results, we show in Remark \ref{rem:3.conf.intervals} how confidence intervals for certain functionals of interest, and a 
uniform confidence bound, for the reconstructed curve can be constructed asymptotically. An expression for the partials required for the computation of the matrix $K$ defined in \eqref{eq:beta.n.q.m.convergence} is provided by \eqref{eq:partial_ibeta_0} and \eqref{eq:partial_iZ0G0}.
	
	\begin{theorem}\label{thm:n=1.asy.dist.BrAC} Suppose that 
		\begin{align*}
			\sqrt{m} \left({\bm q}_m-{\bm q}_0 \right) \rightarrow_d {\mathcal N}({\bm 0},\sigma^2 \Gamma^{-1}) \qmq{as $m \rightarrow \infty$}
		\end{align*}
		for some invertible matrix $\Gamma$,  that $H_0({\bm q}_0)$ is invertible,  Assumption \ref{ass:nun.to.nu.C.alpha} holds, that 
		$\epsilon_i, i=1,\ldots,\n$ are mean zero random variables with common variance $\sigma_f^2$ and uniformly bounded $2+\eta$ moments for some $\eta>0$, independent of each other and of $\bm{q}_m$, and that $\sup_{k \ge 1}\|\phi_k\| < \infty$.  Assume also that $\sqrt{\n}r_\n \rightarrow 0$.
		
		If $m/\n \rightarrow \rho \in [0,\infty)$  then
		\begin{multline} \label{eq:beta.n.q.m.convergence}
			\sqrt{m}(\bsy{\beta}_\n({\bm q}_m)-\bsy{\beta}_0({\bm q}_0))
			\rightarrow_d \mathcal{N}(0,\sigma^2 K^\transpose \Gamma^{-1} K + \rho \sigma_f^2  H_0^{-1}(\bm{q}_0) G_0(\bm{q}_0) H_0^{-1}(\bm{q}_0))\\ \mbox{where} \quad K = \partial_{\bm q} \bsy{\beta}_0({\bm q}_0)^\transpose,
		\end{multline}
		and 
		\begin{multline}\label{def:Wm}
			W_m(s)=\sqrt{m} \left(\mu_\n(s;{\bm q}_m)-\mu_0(s;{\bm q}_0) \right) \rightarrow_d W_{\sigma,\rho}(s)\\
			\mbox{where} \quad W_{\sigma,\rho}(s)=\bsy{\phi}(s)^\transpose \left( \sigma K^\transpose \Gamma^{-1/2} {\bm Z}_1 + \sqrt{\rho}\sigma_f H_0^{-1}(\bm{q}_0)G_0^{1/2}(\bm{q}_0){\bm Z}_2 \right)
		\end{multline}
		as processes on the space $C[0,S]$ of continuous functions on $[0,S]$, endowed with the supremum norm, where 
		$\Gamma^{-1/2}$ and $G_0^{1/2}(\bm{q}_0)$ are 
		the unique positive definite square roots of $\Gamma^{-1}$ and $G_0(\bm{q}_0)$ respectively, and 
		${\bm Z}_1 \sim \mathcal{N}_2({\bm 0}, {\bm I})$ and ${\bm Z}_2 \sim \mathcal{N}_p({\bm 0}, {\bm I})$ are independent.  
		
				If $m/\n \rightarrow \rho = \infty$  then  \eqref{eq:beta.n.q.m.convergence} and  \eqref{def:Wm} hold with the scaling $\sqrt{m}$ replaced by $\sqrt{\n}$ and the parameters of the limiting distributions in those displays set to $(\sigma,\rho)=(0,1)$.

	\end{theorem}

\begin{remark}
	As $K$ and $\Gamma$ in \eqref{def:Wm} depend on the unknown ${\bm q}_0$, in practice these quantities can be estimated by their values along a sequence of consistent estimates ${\bm q}_m, m \ge 1$. As $K$ and $\Gamma$ are continuous at ${\bm q}_0$, these resulting estimates will likewise be consistent. Similar remarks apply as to the estimation of $\sigma_f^2$ and ${\bm G}_0(\bm{q}_0)$.
\end{remark}

	\begin{remark}
		The boundary case $\rho=0$ corresponds to the situation where the number of observations taken in the field is so large that the variability of the resulting BrAC estimate depends only on the uncertainty in the parameter estimate $\bm{q}_0$, hence asymptotically equivalent to the situation where the field observations are taken without noise. 
		
		At the other extreme, the case $\rho=\infty$ reflects the situation where the number of observations taken in the calibration experiment in the lab is so large that for the purposes of BrAC estimation, the parameter $\bm{q}_0$ is, in a practical sense, known. 
	\end{remark}

\begin{remark} \label{rem:3.conf.intervals}
When the distributional convergence in \eqref{def:Wm} holds then for any continuous $F:C[0,S] \rightarrow \mathbb{R}$ we have
\begin{align} \label{eq:FWm.to.FWlim}
F(W_m(s)) \rightarrow_d F(W_{\sigma,\rho}(s)) \qmq{as $m \rightarrow \infty$.}
\end{align}
Consider the following examples:
\begin{enumerate}
	
\item For $\mu \in C[0,S]$ and $t \in [0,S]$ let 
$$
F_t(\mu) =  \int_0^t \mu(s)ds,
$$
that is, the area under $\mu$ at time $t$. Of special interest is the case $t=S$, the total area under the curve. 
In this case \eqref{def:Wm} and the linearity of the integral yields
$$
\sqrt{m} \left(\int_0^t \mu_\eta(s:\bm{q}_m)ds-\int_0^t\mu_0(s;\bm{q}_0)ds \right) \rightarrow_d \int_0^t W_{\sigma,\rho(s)ds}
$$

\item Let 
$$
F(\mu)=\frac{1}{S}\int_0^S |\mu(s)|ds, \qmq{that is} F(\mu)= \|\mu\|_1,
$$
the scaled $L_1$ norm of $\mu$ on $[0,S]$. From \eqref{eq:FWm.to.FWlim} we obtain
\begin{align} \label{eq:L1.conf}
 \frac{\sqrt{m}}{S}\int_0^S \vert  \mu_\eta(s;\bm{q}_m)-\mu_0(s;\bm{q}_0)\vert ds
\rightarrow_d \frac{1}{S}\int_0^S |W_{\sigma,\rho}(s)|ds=\|W_{\sigma,\rho}\|_1.
\end{align}
We may also consider the scaled, or unscaled such norm on any non-empty sub-interval $[a,b]\subset [0,S]$.

\item Taking
$$
F(\mu)=\sup_{s \in [0,S]} |\mu(s)|, \qmq{that is} F(\mu)= \|\mu\|_\infty
$$
the $L_\infty$ norm, yields
\begin{align} \label{eq:unif.conf}
\sqrt{m}\Bvert \Bvert  \mu_\eta(\cdot:\bm{q}_m)-\int_0^t\mu_0(\cdot;\bm{q}_0) \Bvert \Bvert_\infty
\rightarrow_d \|W_{\sigma,\rho}\|_\infty.
\end{align}

\end{enumerate}
			
The performance of the asymptotic confidence intervals these examples yield for the total area under the curve, and $L^1$ and uniform confidence bounds on the reconstructed curve, respectively, are considered in Section \ref{sec:sim.and.data}. 

For $\alpha \in (0,1)$  and $L$ and $U$ having the distributions of 
$\|W_{\sigma,\rho}\|_1$ and  $\|W_{\sigma,\rho}\|_\infty$ respectively, 
let $l_{1-\alpha}$ be the infimum over all $t$ such that $P( L \le t) \ge 1-\alpha$, and define $u_{1-\alpha}$ similarly using $U$ in place of $L$. Then the $L^1$ and uniform $1-\alpha$ confidence sets centered at $\mu_0$ are given by ${\rm C}^q_{l_{1-\alpha}}$ for $q=1$ and $q=\infty$ respectively, where
$$
{\rm C}^q_\beta=\{\mu: \|\mu-\mu_0\|_q \le \beta \}.
$$
Since
$$
\|W_{\sigma,\rho}\|_1 \le \|W_{\sigma,\rho}\|_\infty \qmq{we have} L \le_{\rm st} U, 
$$
where $\le_{\rm st}$ denotes stochastic dominance between distributions.  We conclude that $l_{1-\alpha} \le  u_{1-\alpha}$, and hence we may derive
$$
C_{l_{1-\alpha}}^1\subset C_{u_{1-\alpha}}^1 \qmq{and that} C_{u_{1-\alpha}}^\infty \subset C_{u_{1-\alpha}}^1
$$
but no implications as to relations between $C_{l_{1-\alpha}}^1$ and
$C_{u_{1-\alpha}}^\infty$.

\end{remark}
	
	\noindent {\em Proof:}  By the delta method as in Theorem 7, Chapter 7 in \citet{Ferguson17}, using that $\partial_{\bm q}\bsy{\beta}({\bm q})$ is continuous in a neighborhood of ${\bm q}_0$ by Lemma \ref{lem:beta.dervs}, we obtain 
	\begin{align} \label{eq:beta_0.diffqs.normal.limit}
		\sqrt{m}(\bsy{\beta}_0({\bm q}_m)-\bsy{\beta}_0({\bm q}_0))\rightarrow_d \sigma U \sim \mathcal{N}(0,\sigma^2 K^\transpose \Gamma^{-1} K).
	\end{align}

	Now suppose that $m/\n \rightarrow \rho \in [0,\infty)$.  By Lemma \ref{lem:eta_n.d.limit}, adopting the notation in \eqref{eq:eta_n.d.limit},
	\begin{align}\label{eq:useSlutsky}
		\sqrt{m}(\bsy{\beta}_\n({\bm q}_m)-\bsy{\beta}_0({\bm q}_m)) = \sqrt{\frac{m}{\n}} \left( \sqrt{\n}(\bsy{\beta}_\n({\bm q}_m)-\bsy{\beta}_0({\bm q}_m)\right) \rightarrow_d \sqrt{\rho}Y,
	\end{align}
and by \eqref{eq:for.asy.ind} we see that $Y$ is the distributional limit of a quantity not depending on $\bm{q}_m$, plus a term that tends to zero in probability, thus showing that $U$ and $Y$ are asymptotically independent. Hence, adding the expressions in \eqref{eq:beta_0.diffqs.normal.limit} and \eqref{eq:useSlutsky}, we find that 
	\begin{align}\label{eq:alpha.U+sqrtrho.Y}
		\sqrt{m}(\bsy{\beta}_\n({\bm q}_m)-\bsy{\beta}_0({\bm q}_0)) \rightarrow_d \sigma U+\sqrt{\rho}Y,
	\end{align}
	completing the proof of \eqref{eq:beta.n.q.m.convergence}.

	Letting $\bsy{\alpha}(\n,m)=\sqrt{m}(\bsy{\beta}_\n({\bm q}_m)-\bsy{\beta}_0({\bm q}_0))$, by the definition of $W_m$ in \eqref{def:Wm}, $\mu_\n$ in \eqref{eq:est.BrAC.basis}, and the convergence in \eqref{eq:alpha.U+sqrtrho.Y}, for $d \ge 1$ the finite dimensional distributions of $W_m$ at the $d$ arbitrarily chosen times points $0 \le s_1 < \ldots < s_d \le S$ converge to those of $W_0$, as
	\begin{multline} \label{eq:ffd}
		[W_m(s_1), \ldots,W_m(s_d)]^\transpose = [\bsy{\phi}(s_1), \ldots,\bsy{\phi}(s_d)]^\transpose \bsy{\alpha}(\n,m)\\ \rightarrow_d [\bsy{\phi}(s_1), \ldots,\bsy{\phi}(s_d)]^\transpose \left(
		\sigma U+\sqrt{\rho}Y
		\right)
		=_d [W_{\sigma,\rho}(s_1), \ldots,W_{\sigma, \rho}(s_d)]^\transpose .
	\end{multline}
	
	Define the modulus of continuity of a continuous function $\phi(s)$ on $[0,S]$ by
	\begin{align*}
		\Omega_\phi(\delta) = \sup_{|s-t|<\delta, 0 \le s,t \le S}|\phi(s)-\phi(t)| \qm{for $0 < \delta  \le S$.}
	\end{align*}
	By Theorems 8.1 and 8.2 of \citet{Billingsley13}, the proof will be complete upon showing the  following two properties that, together, imply $\{W_m, m \ge 1\}$ is tight: for every $\tau \in (0,1]$ there exists $a$ such that 
	\begin{align} \label{eq:tight1}
		P(|W_m(0)| > a) \le \tau \qm{for all $m \ge 1$,}
	\end{align}
	and for every $\tau \in (0,1]$ and positive $\epsilon$, there exists $\delta>0$ and an integer $m_0$ such that
	\begin{align} \label{eq:tight2}
		P( \Omega_{W_m}(\delta) \ge \epsilon) \le \tau \qm{for all $m \ge m_0$.}
	\end{align}
	Condition \eqref{eq:tight1} follows from \eqref{eq:ffd} with $d=1$ and $s_1=0$; as $W_m(0)$ converges in distribution, the sequence $W_m(0)$ is tight. 
	
	Let $\tau \in (0,1]$ and $\epsilon>0$ be given. As 
	$\bsy{\alpha}(\n,m)$ converges in distribution there exists $C$ such that $P(\|\bsy{\alpha}(\n,m)\| \le C) \ge 1-\tau$ for all $m \ge 1$. Let
	\begin{align*}
		\delta= \inf \{\zeta: \Omega_{\phi_k}(\zeta)< \epsilon/Cp, 1 \le k \le p\}; 
	\end{align*}
	this quantity will be positive for all $\epsilon>0$ as each basis function in the set $\{\phi_k, k=1,2,\ldots,p\}$ is continuous on $[0,S]$, and therefore this set of functions is uniformly continuous. Thus, with probability at least $1-\tau$, for $|s-t| < \delta, 0 \le s,t \le S$ and all $m \ge 1$, 
	\begin{multline*}
		|W_m(s)-W_m(t)| = |({\bm \phi}(s)-{\bm \phi}(t))^\transpose \bsy{\alpha}(\n,m)| 
		\le \|{\bm \phi}(s)-{\bm \phi}(t)\|\, \|\bsy{\alpha}(\n,m)\| \\ \le C \|{\bm \phi}(s)-{\bm \phi}(t)\| \le C \sum_{k=1}^p |\phi_k(s)-\phi_k(t)| \le C \sum_{k=1}^p \Omega_{\phi_k}(\delta) \le \epsilon, 
	\end{multline*}
	from which \eqref{eq:tight2} follows with $m_0=1$.
	
	Lastly, consider the case $m/\n \rightarrow \infty$.  Scaling  \eqref{eq:beta.pm}, by $\sqrt{\n}$,  for the second term, using Slutsky's theorem, we see that 
	$$
	\sqrt{\n}\left( \bsy{\beta}_0({\bm q}_m)-\bsy{\beta}_0({\bm q}_0) \right)=
	\sqrt{\frac{\n}{m}}\sqrt{m}
	\left(
	\bsy{\beta}_0({\bm q}_m)-\bsy{\beta}_0({\bm q}_0)
	\right) \rightarrow_p 0, 
	$$
	as the term in parenthesis convergences in distribution. Hence, the only term contributing in the decomposition \eqref{eq:beta.pm} is \eqref{eq:eta_n.d.limit},  and the argument for the previous case carries through with essentially no modification. 
	\bbox

	\begin{remark}
		The regularization matrix $M_n$ in the objective function \eqref{eq:Jbeta.con.M} is used to avoid numerical instability;  details on the relevant choice of $M_n$ used here can be found in Section \ref{sec:regularization}. However, regularization can induce bias. To illustrate, assume for some $p$ that the true BrAC curve lies in the span of the basis functions $\phi_1(s),\ldots,\phi_p(s)$,  so that 
		\begin{align} \label{eq:def.beta.star}
			\mu(s)= {\bm \phi}(s)^\transpose \bsy{\beta}_\star \qmq{for some $\bsy{\beta}_\star \in \mathbb{R}^p$.}
		\end{align}  In the limiting case $n=0$ of \eqref{eq:defbsybetan} for ${\bm q}={\bm q}_0$, in light of \eqref{def:mup},  \eqref{eq:fmu.basisp} and \eqref{eq:def.beta.star}, we obtain 
		\begin{align*} 
			\bsy{\beta}_0({\bm q}_0) = (G_0({\bm q}_0)+M_0)^{-1}{\bm Z}_0({\bm q}_0) = (G_0({\bm q}_0)+M_0)^{-1}G_0({\bm q}_0) \bsy{\beta}_\star. 
		\end{align*}
		In particular, the limiting coefficient vector $\bsy{\beta}_0({\bm q}_0)$ may be biased for the true $\bsy{\beta}_\star$ unless $M_0=0$. 
	\end{remark}

\section{Basis and Regularization Details} \label{sec:regularization}
For a BrAC curve estimate $\mu$ in the span of a basis 
$\{\phi_i,i=1,\ldots,p\}$
of absolutely continuous functions on $[0,S]$, there exists a unique vector ${\bm \beta}=[\beta_1,\ldots,\beta_p]^\transpose \in \mathbb{R}^p$ such that \begin{align*}
	\mu(t)= \sum_{i=1}^p \beta_i \phi_i(t)={\bm \phi(t)^\transpose}{\bm \beta} \qmq{and hence} \mu'(t)= \sum_{i=1}^p \beta_i \phi_i'(t)={\bm \phi'(t)^\transpose}{\bm \beta}
\end{align*}
where ${\bm \phi}(t) = [\phi_1(t) ,\ldots,\phi_p(t)]^\transpose \in \mathbb{R}^p$.
In this case, we can express the $L^2$ norm of $\mu$ and its derivative as quadratic forms involving matrices $R$ and $Q$ respectively, given by
$$
\int_0^S \mu^2(t) dt=\int_0^S ({\bm \phi(t)^\transpose}{\bm \beta})^\transpose {\bm \phi(t)^\transpose}{\bm \beta}dt={\bm \beta}^\transpose\left[\int_0^S {\bm \phi(t)}{\bm \phi(t)}^\transpose dt\right] {\bm \beta}=: {\bm \beta}^\transpose R {\bm \beta},
$$
and likewise
$$
\int_0^S [\mu'(t)]^2dt=\int_0^S ({\bm \phi'(t)^\transpose}{\bm \beta})^\transpose {\bm \phi'(t)^\transpose}{\bm \beta}dt={\bm \beta}^\transpose\left[\int_0^S {\bm \phi'(t)}{\bm \phi'(t)}^\transpose dt\right] {\bm \beta}=:{\bm \beta}^\transpose Q {\bm \beta},
$$
and then regularize via $M$, resulting in a linear combination of these penalty terms, defined by
\begin{align}\label{eq:M.mu.lambda.regularized}
M= \lambda \int_0^S \mu^2(t)dt+ \mu \int_0^S [\mu'(t)]^2 dt = \lambda R + \mu Q. 
\end{align}

The $B$-spline basis is one convenient choice. These functions are constructed by first choosing an integer $N \ge 0$, the number of interior knots, non-decreasing real numbers 
$$
t_0 \le t_1 \le \cdots \le t_N \le t_{N+1},
$$
and the degree $n \ge 1$, 
of the spline, from which one defines the augmented knot set 
$$
t_{-n}=\cdots = t_0 \le t_1 \le \cdots \le t_{N+1}=\cdots=t_{N+n+1}.
$$
Now, for each augmented knot $t_i,i=-n,\ldots,N+n+1$, let
$$
B_{i,0}(x) = \left\{
\begin{array}{cc}
	1 & t_i \le x < t_{i+1}\\
	0 & \mbox{otherwise,}
\end{array}
\right.
$$
and for $j=0,\ldots,n-1$
recursively define $i^{th}$ $B$-spline basis function $B_{i,j}$ of order $j$ by
$$
B_{i,j+1}(x) = \alpha_{i,j+1}(x)B_{i,j}(x)+[1-\alpha_{i+1,j+1}(x)]B_{i+1,j}(x), \quad -n+j+1 \le i \le N+n+1
$$
where 
$$
\alpha_{i,j}(x) = \left\{
\begin{array}{cc}
	\frac{x-t_i}{t_{i+j}-t_i} & t_i \not = t_{i+j}\\
	0 & \mbox{otherwise,}
\end{array}
\right.
$$
see \citet{DeBo78}. For these functions we cannot in general write explicit forms for the integrals that produce $R$ and $Q$ that form the regularizing matrix $M$.

\ignore{
We specialize to the chapeau (hat) functions, defined as follows.
Letting
$$
\psi_-(t)=(1+t){\bm 1}_{[-1,0]}
\quad \mbox{and} \quad \psi_+(t)=(1-t){\bm 1}_{[0,1]},
$$
and letting $r=S/p$, set
$$
\phi_j(t)=\psi_-(t/r-j){\bm 1}_{j \ge 1}+\psi_+(t/r-j){\bm 1}_{j \le p-1} \qmq{for $0 \le j \le p$.}
$$
We claim the components of $R$ and $Q$ are given by
\begin{align*}
	R_{ij}=\langle \phi_i, \phi_j \rangle = \left\{
	\begin{array}{cl}
		r/3\left({\bm 1}_{j \ge 1} + {\bm 1}_{j \le p-1}\right).& i=j \\
		r/6 & |i-j|=1\\
		0 & |i-j| \ge 2,
	\end{array}
	\right.
\end{align*}
and
\begin{align*}
	Q_{ij}=\langle \phi_i', \phi_j' \rangle = \left\{
	\begin{array}{cl}
		r^{-1}({\bm 1}_{j \ge 1} + {\bm 1}_{j \le p-1}) & i=j \\
		-r^{-1} & |i-j|=1\\
		0 & |i-j| \ge 2.
	\end{array}
	\right.
\end{align*}
We note that the value of zero when $|i-j| \ge 2$ follows immediately from the fact that $\phi_i$ and $\phi_j$ have disjoint support in this case. 

It is easy to verify that 
\begin{multline} \label{eq:i=jphi}
	\int_0^1 \psi_+^2(t)dt = \int_{-1}^0 \psi_-^2(t)dt = 1/3 \\
	\mbox{implying} \quad \int_0^S \psi_+^2(t/r-j)dt = \int_0^S \psi_-^2(t/r-j)dt = r/3, 
\end{multline}
and that 
\begin{multline} \label{eq:i=j-1phi}
	\int_0^1 \psi_-(t-1)\psi_+(t)dt = \int_0^1 t(1-t)dt = 1/6\\
	\mbox{implying} \quad \int_0^S \psi_-(t/r-(j-1))\psi_+(t/r-j)dt = r/6.
\end{multline}

For $i = j$, by \eqref{eq:i=jphi} we obtain
\begin{align*}
	\langle \phi_i,\phi_i \rangle	=& \int_0^S \phi_i^2 (t)  dt\\
	= &\int_0^S \psi_-(t/r-j)^2{\bm 1}_{j \ge  1} dt + \int_0^S \psi_+(t/r-j)^2 {\bm 1}_{j \le p-1}dt\\
	= & \frac{r}{3}{\bm 1}_{j \ge 1} + \frac{r}{3}{\bm 1}_{j \le p-1}.
\end{align*}

For the remaining case $|i-j|=1$,  by relabelling we may assume $i = j - 1, \;\; j = 1, \ldots, p$. In this case, by \eqref{eq:i=j-1phi} we have
\begin{align*}
	\langle \phi_{j-1}, \phi_j \rangle = \int_0^S \phi_{j-1}(t) \phi_j(t)dt
	= \int_0^S \psi_-(t/r-(j-1)) \psi_+(t/r-j)dt = r/6.
\end{align*}
For the derivative, as $\psi_-'(t)={\bm 1}_{[0,1]}$ and $\psi_+'(t)=-{\bm 1}_{[0,1]}$, we obtain
\begin{align} \label{eq:phij.prime}
	\phi_j'(t) = r^{-1}({\bm 1}_{[(j-1)r,jr]}{\bm 1}_{j \ge 1} - {\bm 1}_{[jr,(j+1)r]}{\bm 1}_{j \le p-1}),
\end{align}
which yields
\begin{align*}
	\langle \phi_i',\phi_i' \rangle = \int_0^S [\phi_i'(t)]^2dt =
	r^{-2}
	\int_0^S
	({\bm 1}_{[(j-1)r,jr]}{\bm 1}_{j \ge 1} + {\bm 1}_{[jr,(j+1)r]}{\bm 1}_{j \le p-1})dt
	\\
	= r^{-1}\left( 
	{\bm 1}_{j \ge 1}+{\bm 1}_{j \le p-1}
	\right),
\end{align*}
and for $j=1,\ldots,p$, noting that
\begin{multline*}
	({\bm 1}_{[(j-2)r,(j-1)r]}{\bm 1}_{j \ge 2} - {\bm 1}_{[(j-1)r,jr]}{\bm 1}_{j \le p})  \times ({\bm 1}_{[(j-1)r,jr]}{\bm 1}_{j \ge 1} - {\bm 1}_{[jr,(j+1)r]}{\bm 1}_{j \le p-1})\\
	= -{\bm 1}_{[(j-1)r,jr]}{\bm 1}_{1 \le j \le p},
\end{multline*}
we obtain
\begin{align*}
	\langle \phi_{j-1}',\phi_j' \rangle = -r^{-2}\int_0^S {\bm 1}_{[(j-1)r,jr]}{\bm 1}_{1 \le j \le p} dt
	= 	-r^{-1}{\bm 1}_{1 \le j \le p}.
\end{align*}
}

\ignore{
\subsubsection{Linear Independence of Derivatives}

By the Cauchy-Schwarz inequality the determinant of $\Gamma$ is nonnegative, and it vanishes if and only if $f(t;\bm{q})$ and $\partial_1 f(t;\bm{q})$ are linearly dependent (in $t$). To determine conditions under which we can guarantee these functions are linearly independent, we start with the following result. 

\begin{lemma} \label{lem:dif.convolution.integral}
Let $\mu: [0,T] \rightarrow [0,\infty)$ be continuous at 0, and suppose that t valued function $X:[0,T] \rightarrow \mathbb{R}$ is differentiable on $[0,T]$, and that there exists a real valued function $Y(s)$ such that
\begin{align*}
|X'(t-s)| \le Y(s) \qmq{for all $0\le s \le t \le T$ and} \int_0^T Y(s)\mu(s) ds < \infty. 
\end{align*}
Then for all $t \in [0,T]$
\begin{align*}
g(t)=\int_0^t X(t-s)\mu(s)ds
\end{align*}
is differentiable in $[0,T]$ with
\begin{align*}
g'(t) = \int_0^{t} \partial X(t-s)
\mu(s)ds+X(0)\mu(t).
\end{align*}
\end{lemma}

\noindent {\em Proof:}
For $h \not = 0$, 
\begin{align*}
	\frac{g(t+h)-g(t)}{h} &= \frac{1}{h}\left( \int_0^{t+h}X(t+h-s)\mu(s)ds - \int_0^{t}X(t-s)\mu(s)ds\right) \nonumber \\
	&= \int_0^{t}\frac{X(t+h-s)-X(t-s)}{h}\mu(s)ds+\frac{1}{h}\int_t^{t+h}X(t+h-s)
	\mu(s)ds.
\end{align*}
The result follows by applying the dominated convergence theorem in the first integral, and using the continuity of the integrand at $s=t$ in the second. 
\bbox

Thus, recalling \eqref{eq:fij.q.only} and setting $X(s)=Ce^{sA}B$, using Lemma \ref{lem:dif.convolution.integral} we obtain
\begin{equation}\label{part.f.t}
\partial_t f(t) =  \int_0^{t}Ce^{A(t-s)}AB\mu(s)ds+CB\mu(t)
\end{equation}
and similarly,
\begin{equation*}
\partial_t^2 f(t) =  \int_0^{t}Ce^{A(t-s)}A^2B\mu(s)ds+CAB\mu(t)+CB\mu'(t).
\end{equation*}
By Taylor expansion around zero, we have
\begin{align}
f(t)&=f(0)+t\partial_tf(0)+t^2\partial_t^2f(0)/2+O(t^3) \nonumber \\
&= tCB\mu(0)+t^2(CAB\mu(0)+CB\mu'(0))/2+O(t^3).\label{ft.taylor}
\end{align}

According to \cite[Theorem~3.5]{Hall03}, 
\begin{multline}\label{d1eAt-s.ser}
\partial_1 e^{A(t-s)} = e^{A(t-s)}\left\{(t-s)D-\frac{[(t-s)A,(t-s)D]}{2!}\right.\\
\left.+\frac{[(t-s)A,[(t-s)A,(t-s)D]]}{3!}-\ldots\right\},
\end{multline}
where $[M_1,M_2]=M_1M_2-M_2M_1$ is the commutator of square matrices $M_1, M_2$. Using bilinearity of $[\cdot,\cdot]$, the series in curly brackets in \eqref{d1eAt-s.ser} can be written
\begin{equation*}
\sum_{i=1}^\infty (-1)^{i+1}(t-s)^i D_i\qmq{where}D_1=D\qmq{and} D_{i+1}=[A,D_i],\quad i\ge 1.
\end{equation*}
Then 
\begin{align*}
\partial_1 f(t) &=\int_0^t  C \partial_1(e^{A(t-s)}B)\mu(s)ds\\
& = \int_0^t  C (\partial_1 e^{A(t-s)})B\mu(s)ds\\
&= \int_0^t  Ce^{A(t-s)} \left\{ \sum_{i=1}^\infty (-1)^{i+1}(t-s)^i D_i B\right\} \mu(s)ds.
\end{align*}
Then, letting
\begin{align*}
G(s)=A\left(\sum_{i=1}^\infty (-1)^{i+1}s^i D_i B\right) + \sum_{i=1}^\infty i(-1)^{i+1}s^{i-1} D_i B,
\end{align*}
applying Lemma \ref{lem:dif.convolution.integral} yields
\begin{align*}
\partial_t\partial_1 f(t) = 
 \int_0^t  Ce^{A(t-s)} G(t-s) \mu(s)ds \label{dt1f.integral},
\end{align*}
and
\begin{equation*}
\partial_t^2\partial_1 f(t) =  C\int_0^t \partial (e^{A(t-s)}G(t-s))\mu(s)ds 
\end{equation*}
Taylor expansion around zero now yields
\begin{align}
\partial_1 f(t)&=\partial_1 f(0)+t\partial_t \partial_1 f(0)+t^2\partial_t^2 \partial_1 f(0)/2+O(t^3) \nonumber \\
&= O(t^3). \label{d1f.taylor}
\end{align}
Comparing \eqref{ft.taylor} and \eqref{d1f.taylor} we see that these functions are linearly independent.
}

\section{Transdermal blood alcohol monitoring: Simulations and data analysis} \label{sec:sim.and.data}

In both the simulation and real data study presented below we investigate the case where data are collected from single drinking episodes. 
The computations were carried out in MATLAB and the optimization producing the estimate of the parameter ${\bm q}$ was solved using the Optimization Toolbox routine FMINCON.\\

\subsection{Simulation studies}
In Section \ref{subsub:q} we validate our theoretical results on the consistency and asymptotic normality of the parameter estimate given in Theorem \ref{thm:ls.cons.asy.dist.qonly}, and also illustrate the practical impact of the number of observations on its behavior. In \ref{subsub:brac} we evaluate the behavior of the BrAC curve estimates, and the associated confidence bounds. 

\subsubsection{Estimation of the $q$ vector}\label{subsub:q}
To reflect a simple real-world situation, BrAC was simulated using  a small but realistic drinking diary that consists of a single drink 6 minutes after the beginning of the drinking session. BrAC was computed using the Michaelis-Menten approach \citep[see][]{Dai16} that models the metabolic effects of the ethanol specific enzymes ADH and ALDH typically found in the liver, and also known to be present in trace amounts in the skin. 

For simplicity, we set ${\bm q_0}=(1,1)$ to be the true value of the parameter ${\bm q}$ and $T=1$ hour to be the duration of the drinking session. We choose the vectors and matrices in \eqref{eq:k.is.conv} and \eqref {eq:AB.by.q} to be
\begin{align*}
D={\bm I}_2, \quad E={\bm O}_2, \quad  C=(1,0) \qmq{and} F=(1,0)^\transpose.
\end{align*}
Equally spaced TAC measurement were calculated after adding independent error terms to the expression in (\ref{eq:fij.q.only}), thus resulting in \eqref{eq:y=f+e.model}, with each error term distributed as $\mathcal{N}(0,\sigma^2)$ with $\sigma=0.01$. Calculating the theoretical limiting covariance matrix in Theorem  \ref{thm:ls.cons.asy.dist.qonly} we obtain
\begin{align*}
\Sigma = \left(
\begin{array}{cr}
16.4404 & -7.2947 \\
-7.2947 & 3.4586 
\end{array}
\right).
\end{align*}

Table \ref{tbl:cov_matrices} displays the results of three simulations, each of 100 ${\bm q}$ estimates, for three experiments having 20, 60 and 100 TAC samples. A comparison of the results obtained to the true values of the parameters agrees with our theoretical consistency results.

\begin{center}
\begin{tabular}{|c|c|c|} 
 \hline
 Number of TAC observations& Mean Parameter Estimate & Scaled Sample Covariance Matrix\\ 
\hline
 20 & $\begin{pmatrix}  0.9447  \pm 0.1549    \\  1.0597  \pm 0.0684\end{pmatrix}$ & $\begin{pmatrix}  12.6231  & -5.2525 \\
   -5.2525  &  2.4586 \end{pmatrix}$ \\ 
\hline
 60 & $\begin{pmatrix} 1.0375  \pm 0.0997    \\ 1.0042 \pm 0.0435\end{pmatrix}$&$\begin{pmatrix} 15.6790  & -6.6024 \\
   -6.6024  &  2.9912 \end{pmatrix}$ \\ 
\hline
100&$\begin{pmatrix} 0.9762 \pm 0.0805    \\ 1.0228 \pm 0.0381 \end{pmatrix}$&$\begin{pmatrix} 17.0397 &  -7.8260 \\
   -7.8260 &   3.8215 \end{pmatrix}$ \\
\hline
\end{tabular}
\label{tbl:cov_matrices}
\end{center}
\begin{center}
Table \ref{tbl:cov_matrices} Sample mean and covariance matrices from 100 simulation replicates of $\widehat{\bm q}$.
\end{center}

To experimentally validate the limiting bivariate normal distribution obtained in Theorem \ref{thm:ls.cons.asy.dist.qonly} for $\widehat{\bm q}$, 
Figures \ref{fgr:20obs}, \ref{fgr:60obs}, and \ref{fgr:100obs} plot the vector $\mathbb{R}^2$ value of 100 estimators calculated from the synthetic data for 20, 60 and 100 observations, respectively, along with levels curves of the corresponding limiting bi-variate normal distribution. The amount of probability mass contained in the level curves pictured is indicated by the color coding on the right hand side of the figure, which when all taken together show increasingly good fits to the theoretical result as the number of TAC samples increases. 

\begin{figure}[H]
    \centering
    \includegraphics[width=16cm,height=8cm]{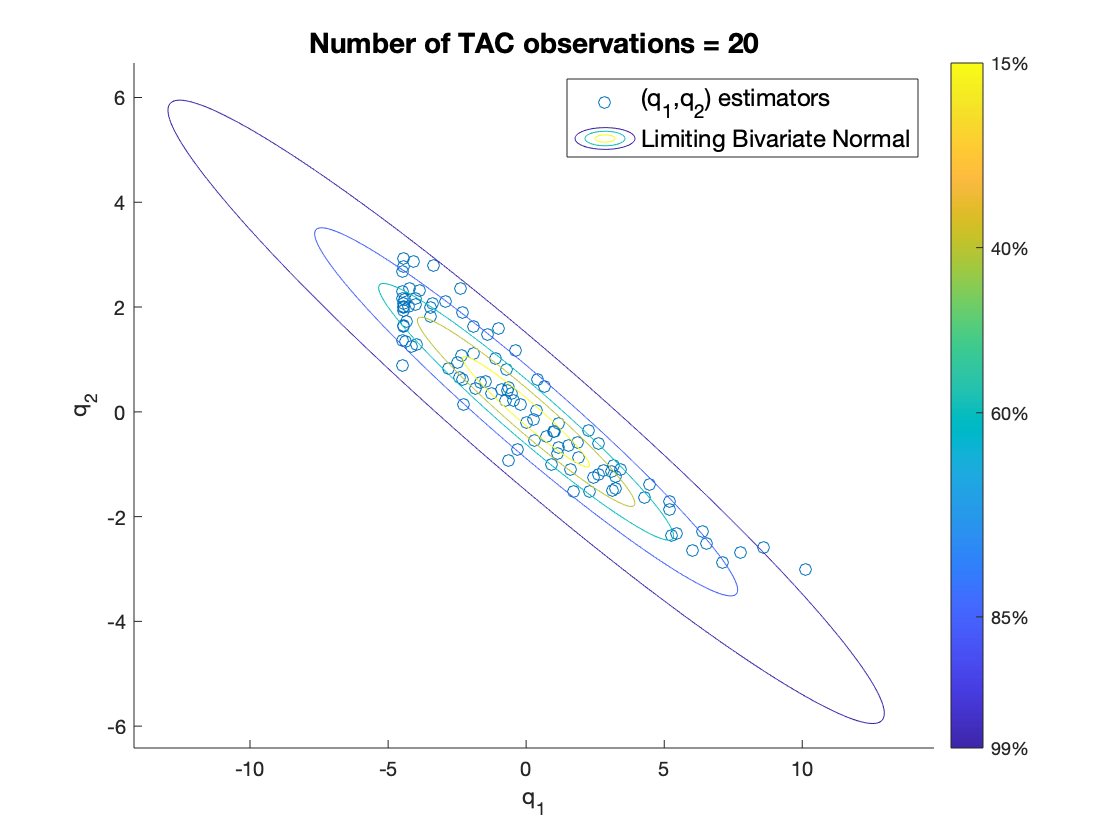}
    \caption{Values of the $\widehat{\bm q}$ estimators obtained when using 20 TAC observations over $T=1$ hour.}
\label{fgr:20obs}
\end{figure}

\begin{figure}[H]
    \centering
    \includegraphics[width=16cm,height=8cm]{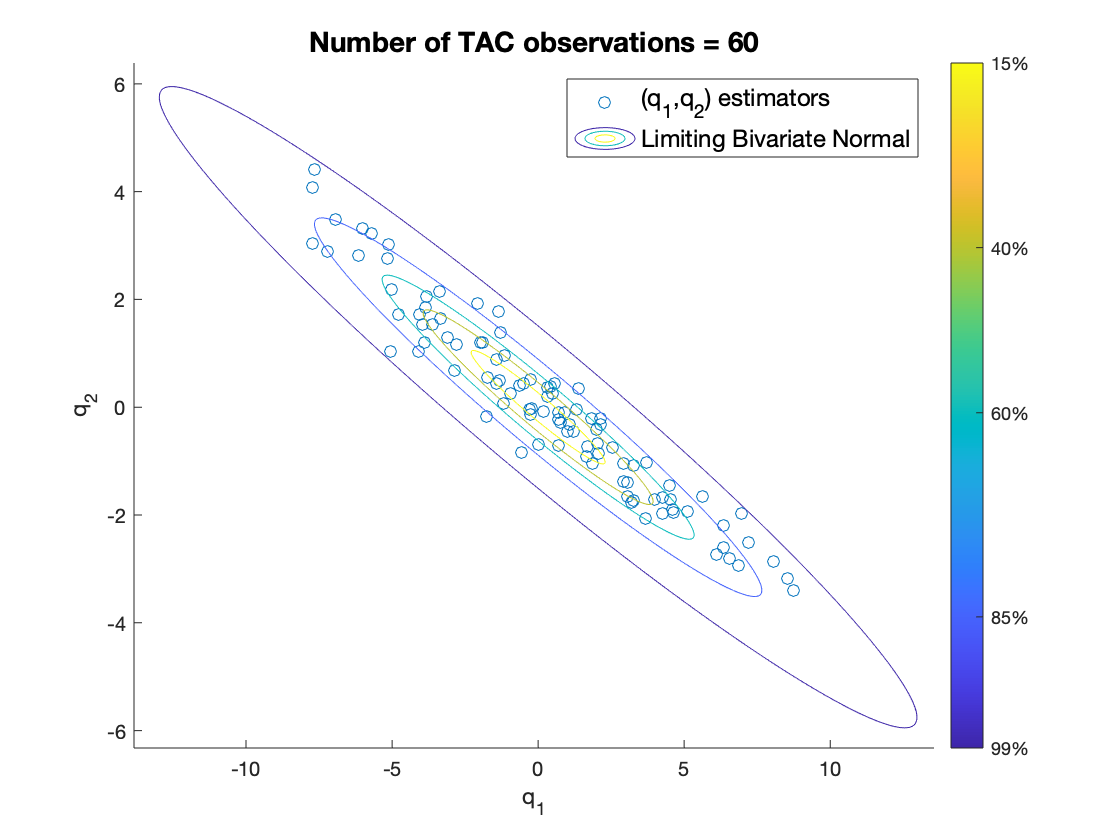}
    \caption{Values of the $\widehat{\bm q}$ estimators obtained when using 60 TAC observations over $T=1$ hour.}
\label{fgr:60obs}
\end{figure}

\begin{figure}[H]
    \centering
    \includegraphics[width=16cm,height=8cm]{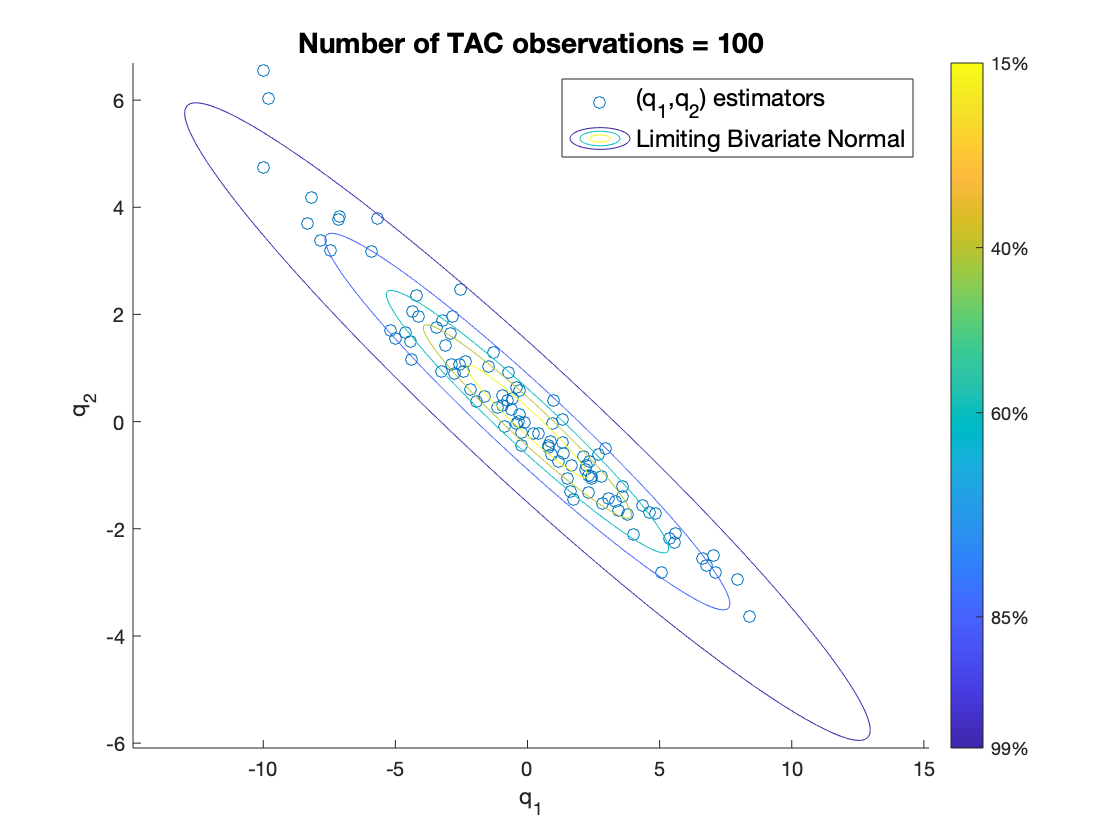}
    \caption{Values of the $\widehat{\bm q}$ estimators obtained when using 100 TAC observations over $T=1$ hour.}
\label{fgr:100obs}
\end{figure}


The running time of these experiments may be long due to the computation of the matrix exponential of $A$ in \eqref{eq:k.is.conv}, which in general is not symmetric. For that reason, we note how speed can be improved using the following diagonalization procedure. 
From \cite{Rosen14}, see also \cite{Sirlanci17}, with $\omega_i, i=1, \ldots, k$ the basis for the finite dimensional approximation in \eqref{xt.diff.soln}, define matrices $K_1,K_2$ and $V$ in $\mathbb{R}^{k \times k}$ by
$$
K_{1,ij}=\int_0^1 \omega_i'(u) \omega_j'(u) du, \quad K_{2,ij}=\omega_i(0) \omega_j(0) \qmq{and} V_{ij}=\int_0^1 \omega_i(u) \omega_j(u) du; 
$$
we note that the matrices $V$ and $K_1,K_2$ are symmetric.
Then we obtain
\begin{align}\label{A.M.K}
D=-V^{-1}K_1, E=-V^{-1}K_2 \qmq{and hence}
A=q_1D+E=-V^{-1}(q_1K_1+K_2).
\end{align}
Multiplying the final expression for $V$ in (\ref{A.M.K}) on the left and right by $V^{1/2}$ and $V^{-1/2}$ respectively yields
\begin{align}\label{A.mult}
V^{1/2}AV^{-1/2}=-V^{-1/2}(q_1K_1+K_2)V^{-1/2}.
\end{align}
As the right hand side of \eqref{A.mult} is symmetric we may apply the spectral theorem to write
\begin{align}\label{A.diag}
V^{1/2}AV^{-1/2}=S_{q_1}L_{q_1}S_{q_1}^{-1} \qmq{and so, for $t \in \mathbb{R}$} V^{1/2}tAV^{-1/2}=S_{q_1}tL_{q_1}S_{q_1}^{-1}
\end{align}
where $S_{q_1}$ is invertible and $L_{q_1}$ is diagonal with real entries. Using \eqref{A.diag} in the final equality, for any non-negative integer power $p$ we have
\begin{align*}
V^{1/2}(tA)^pV^{-1/2}={(V^{1/2}(tA)V^{-1/2})}^p
={S_{q_1}}(tL_{q_1})^pS_{q_1}^{-1},
\end{align*}
which, by substitution into the power series for the exponential function, yields
\begin{align*}
V^{1/2}e^{tA}V^{-1/2}=S_{q_1}e^{tL_{q_1}}S_{q_1}^{-1} 
\qmq{and hence} e^{tA}=V^{-1/2}S_{q_1}e^{tL_{q_1}}S_{q_1}^{-1}V^{1/2},
\end{align*}
thus allowing the computation of the matrix exponential of $tA$ to require the exponentiation of the diagonal matrix $L_{q_1}$ only. 

\subsubsection{BrAC Curve Estimation and Confidence Bands}
 \label{subsub:brac}
 We explore the reconstruction of the BrAC curve from TAC measurements, with a view to the practicality of applying Theorem \ref{thm:n=1.asy.dist.BrAC} and Remark \ref{rem:3.conf.intervals}, with particular attention to the uniform and $L^1$ norm confidence bands, and confidence intervals for the total area under the true BrAC curve. In brief, the lesson here is that due to the inherent numerical instability of the inverse problem being solved, the theory works well for TAC errors of a magnitude well under those that can be obtained by the measurement tools currently available, having an estimated standard deviation of $2.5 \times 10^{-3}$.

Following the approach of \citet[][Section~2]{Dai16}, we used the discretization level of $k = 4$ in \eqref{eq:fij.q.only} and computed the matrices $C,D,E$ and $F$ as there. The curves in this section are standardized to be on the time interval $[0,1]$, and as described at the start of Section \ref{sec:InferenceBrAC}, with a $B$-spline basis used in \eqref{def:mup}, as detailed in Section \ref{sec:regularization}, with degree $n$, $N$ interior knots at $i/(N+1), i=1,\ldots,N$ and boundary knots at $\{0,1\}$, thus producing the curves $B_{i,n}, 1=1,\ldots,N+n+1$. Removing the first and last to accommodate the zero boundary conditions results in the curves indexed by $i=2,\ldots,N+n$. We set $n=3$ and $N=8$, yielding $N+n+1-2=10$ curves, of which we take $B_{3,3}$ as the true BrAC curve.

To set a `best case' baseline instance, we first consider the case where the TAC errors are set to zero, which results in the single, non-random `Estimated-BrAC' curve shown in Figure \ref{fgr:0}, that overlaps with the true BrAC curve without the need for any regularization.
\begin{figure}[H]
	\centering
	\includegraphics[width=16cm,height=8cm]{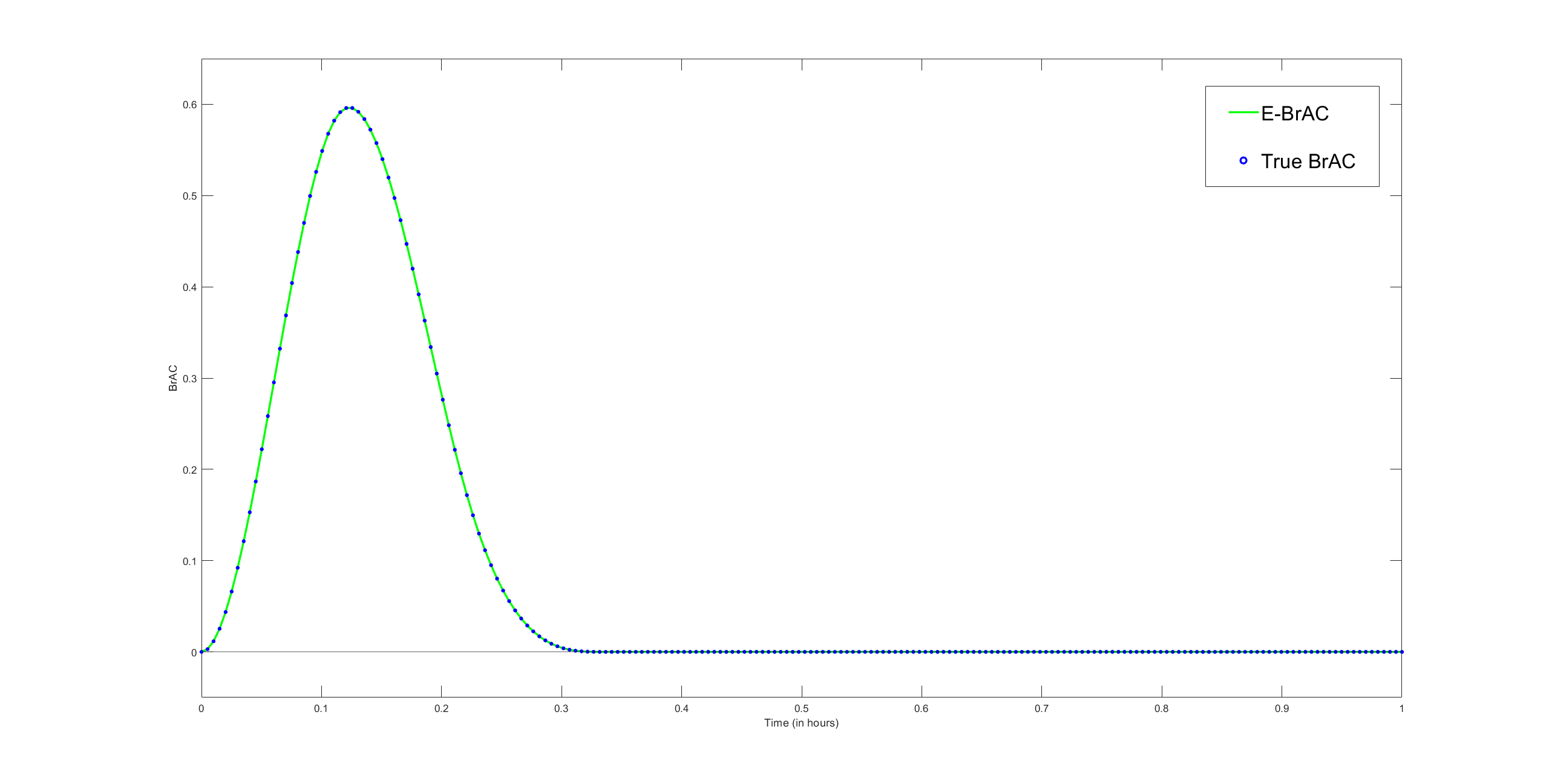}
	\caption{TAC errors set to zero, no regularization}
	\label{fgr:0}
\end{figure}
This figure, and an appeal to continuity, suggests that the confidence regions provided by Theorem \ref{thm:n=1.asy.dist.BrAC} should hold in practice over some substantial time interval in $[0,1]$
in settings where the TAC errors have  sufficiently small standard deviations.

Six experiments were run, all with non-trivial independent Gaussian TAC noise, having varying magnitudes of the standard deviations of the lab and field TAC errors and of the regularization parameters $\mu$ and $\lambda$ in \eqref{eq:M.mu.lambda.regularized}. In each experiment 10 BrAC curve estimates were generated, and the results are compared to the 70\% 
uniform and $L^1$ confidence bands, and for one case, the 70\% confidence interval for the total area under the curve, as provided by Theorem  \ref{thm:n=1.asy.dist.BrAC} and 
Remark \ref{rem:3.conf.intervals}.

In the first such experiment, whose outcome is shown in Figure \ref{fgr:2c}, both TAC error standard deviations are set to be $2.5 \times 10^{-5}$. The small value of the error allows for the problem to be sufficiently numerically stable over a large enough portion of the interval that regularization is not needed, and both regularization parameters were set to zero. The uniform confidence band is represented by a pink band in the figure, which is more easily seen in the enlargement over a smaller time interval visualized by the interior box. Of the 10 BrAC curves generated, 7 are completely contained within the 70\% uniform confidence band, and there are also 7 within the 70\% $L^1$ confidence band,
over the truncated time interval $[0,0.63]$; note the instability in the reconstructed curve towards the end of the interval. From the detail provided by the interior box of the figure, we see that two curves exit the uniform confidence band at the given time threshold. This experiment, where the confidence bands perform at the level desired over the truncated interval, serves as a  reference case for experiments that increasingly approach the more realistic levels of TAC errors having two full orders of magnitude larger. Hence, in these later cases we will consider the same time interval as the one here in order to better make comparisons. 
 
\begin{figure}[H]
	\centering
	\includegraphics[width=16cm,height=8cm]{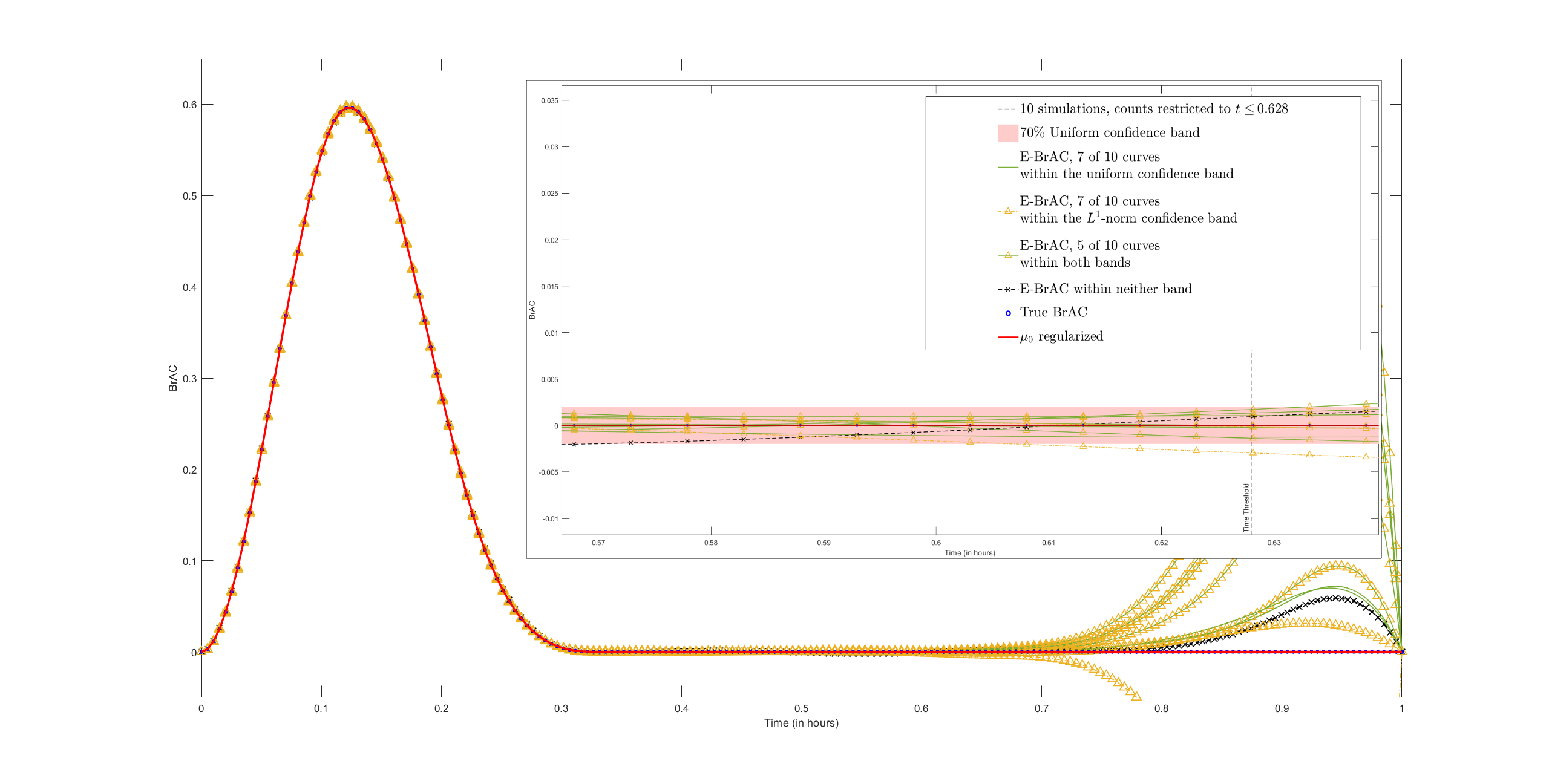}
	\caption{TAC errors $2.5 \times 10^{-5}$, no regularization.}
	\label{fgr:2c}
\end{figure}

In the next experiment, illustrated in Figure \ref{fgr:6b}, we increase the lab error by a factor of 10 and maintain the same field TAC error. Over the same time interval considered in Figure \ref{fgr:2c} we see that the uniform 70\% confidence band contains only 5 of the 10 generated BrAC curves, while the $L^1$ confidence band continues to hold 7, bolstering the intuition that the $L^1$ band should be more robust. Nevertheless, the interior box in the figure illustrates that an increased standard deviation of the lab error causes the curve to have more variability around its peak.

\begin{figure}[H]
	\centering
	\includegraphics[width=16cm,height=8cm]{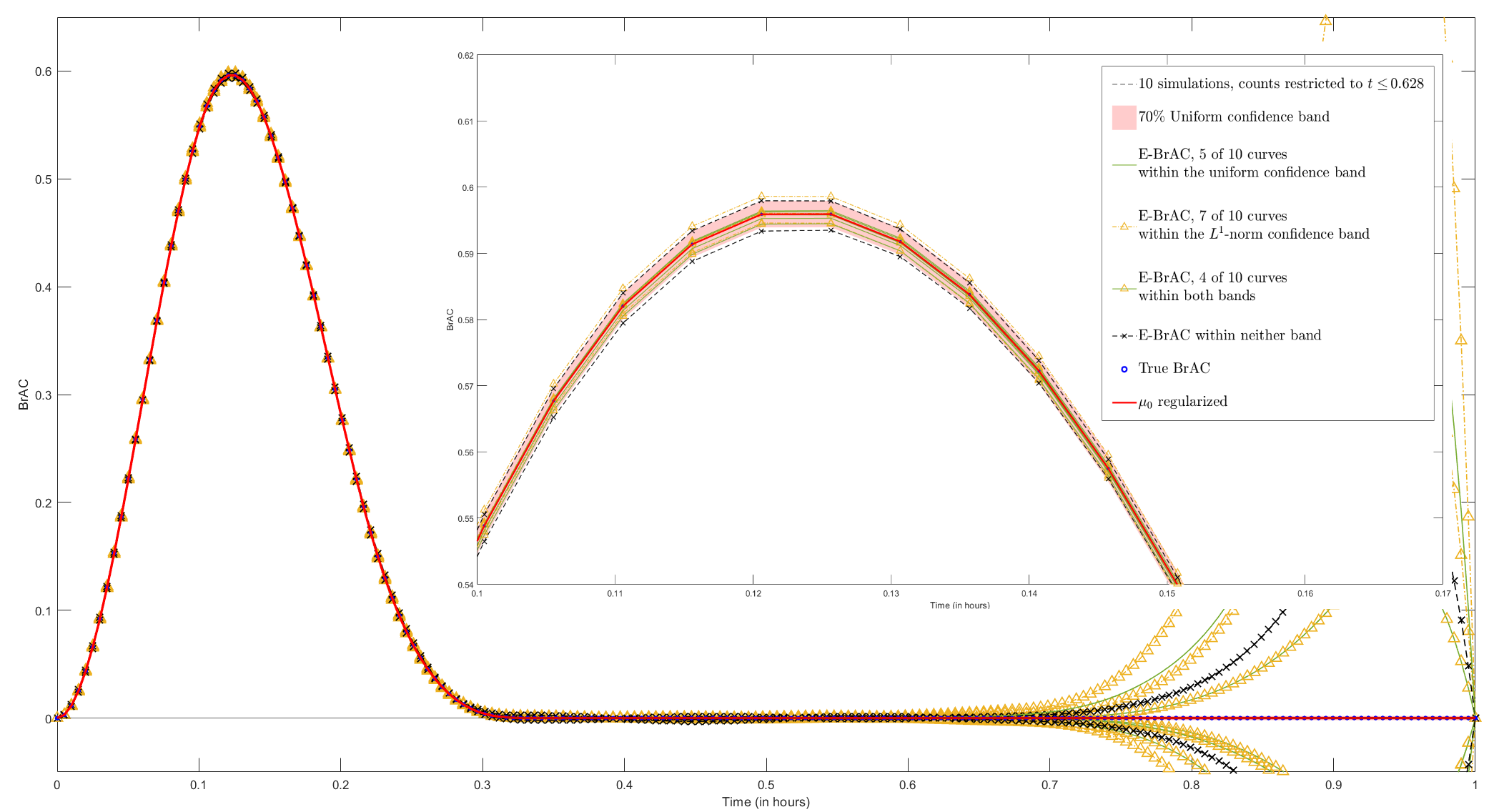}
	\caption{Lab and field TAC error $2.5 \times 10^{-4}$ and $2.5 \times 10^{-5}$, no regularization.}
	\label{fgr:6b}
\end{figure}

Increasing the noise level further, Figure \ref{fgr:7x} shows the result of setting both TAC error standard deviations at 10 times the levels of the previous experiment. Adding a small amount of regularization, there are 7 of 10, and 8 of 10 curves, respectively, within the uniform band and the $L^1$ band. However, its clear from the figure that the curves have much more variability due to the higher noise levels, and fluctuate strongly within wider confidence bands.

\begin{figure}[H]
	\centering
	\includegraphics[width=16cm,height=8cm]{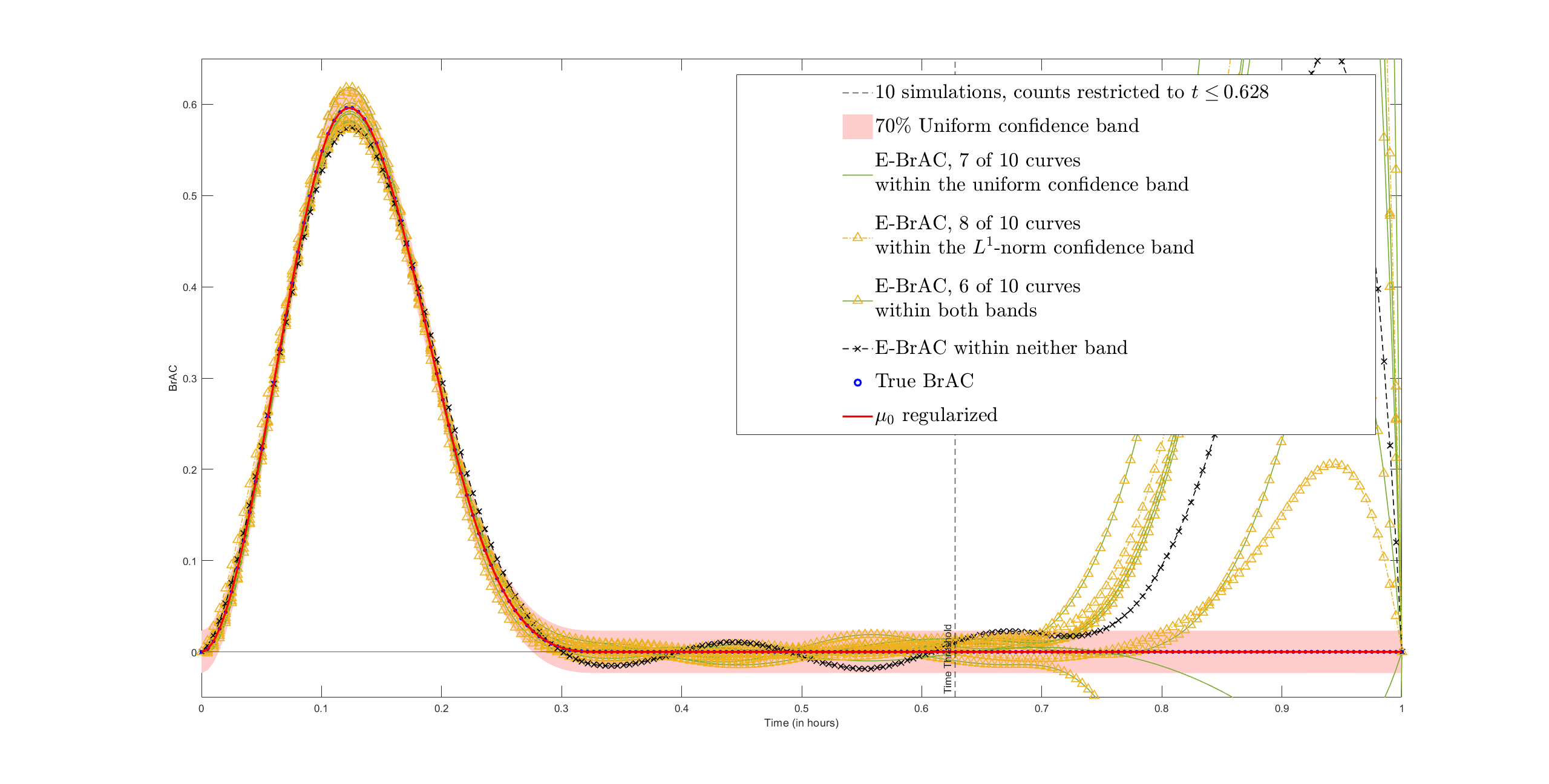}
	\caption{Lab and field TAC errors $2.5 \times 10^{-3}$ and $2.5 \times 10^{-4}$, both regularization parameters set to $2.5 \times 10^{-10}$.}
	\label{fgr:7x}
\end{figure}

For this experiment, having the largest noise levels so far, Figure \ref{fgr:AreaCI} additionally notes the performance of the 70\% confidence interval for the total area under the curve; in particular, for 8 of the 10 curves generated the confidence interval covered the true area value. 
\begin{figure}[H]
	\centering
	\includegraphics[width=16cm,height=8cm]{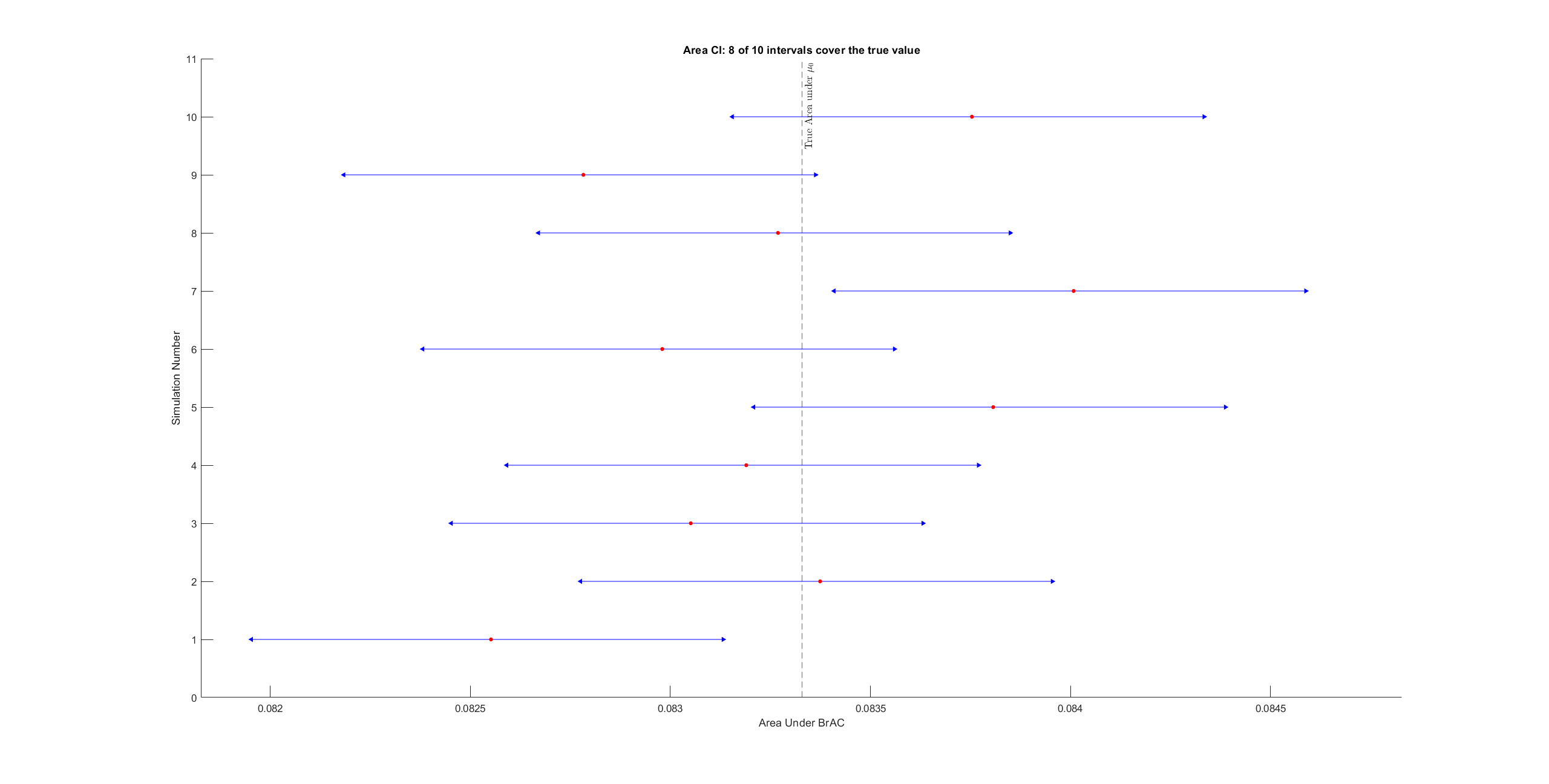}
	\caption{Confidence Interval: Area under the curve.}
	\label{fgr:AreaCI}
\end{figure}

Lastly, we present the results of an experiment that illustrates the effect of regularization, demonstrating why it becomes necessary in certain instances. The TAC error values are the same in the following two plots, the only difference being that no regularization was applied when generating Figure \ref{fgr:8a}, and a very small amount of regularization was used for Figure \ref{fgr:8b}. Instability in the first instance causes the width of the uniform 70\% confidence band to be overly wide, while the  band obtained when regularizing in the second plot captures 7 of the 10 generated curves upon restricting to the interval $[0,0.63]$.
\begin{figure}[H]
	\centering
	\includegraphics[width=16cm,height=8cm]{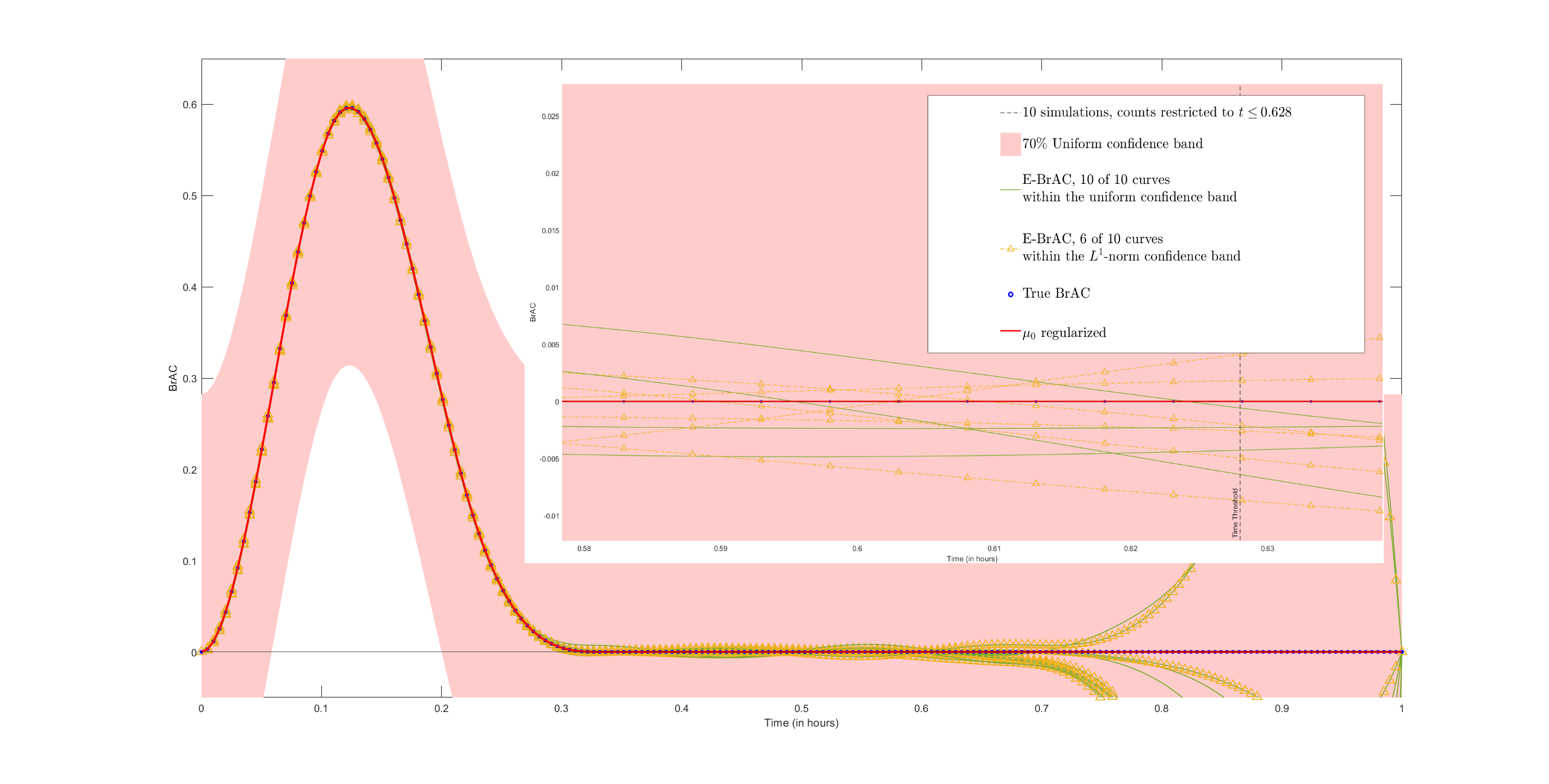}
	\caption{Lab and field TAC error  $2.5 \times 10^{-4}$ and $7.5 \times 10^{-5}$, no regularization.}
	\label{fgr:8a}
\end{figure}

\begin{figure}[H]
	\centering
	\includegraphics[width=16cm,height=8cm]{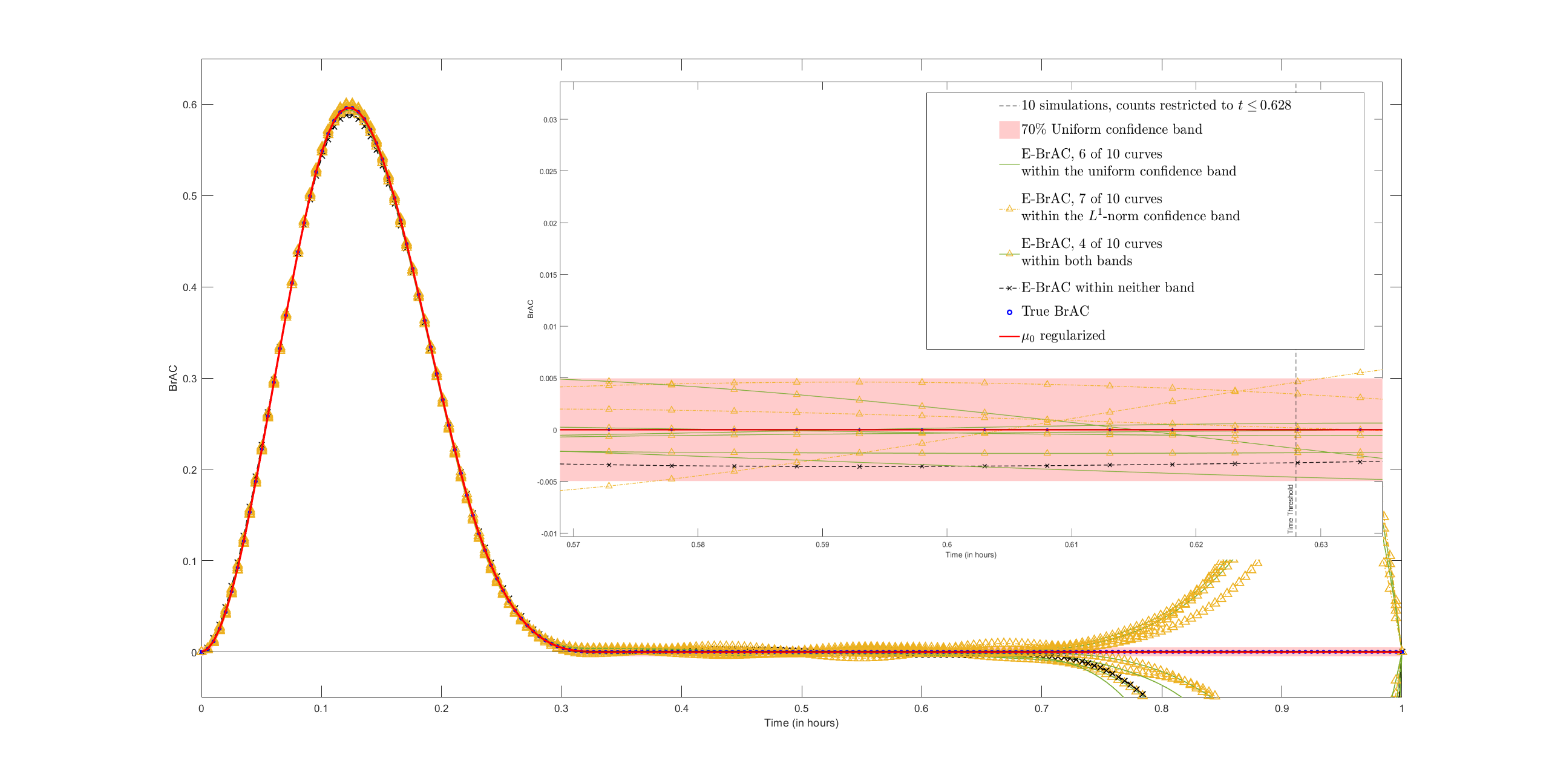}
	\caption{Lab and field TAC error $2.5 \times 10^{-4}$ and $7.5 \times 10^{-5}$, both regularization parameters set to $2.5 \times 10^{-10}$.}
	\label{fgr:8b}
\end{figure}

Increasing errors further results in an instability that requires an amount of  regularization that may induce significant bias. Again, we recall that the sensors on which these experiments are based were designed only to be abstinence monitors for individuals under house arrest, that is, made only to test the presence of alcohol in an individual's system. We anticipate that improvements of measurement technology currently in progress will allow the methods developed here to soon be of practical use. In particular, confidence bands for the data plot in Figure \ref{fgr:real} are not considered. 

\subsection{Real Data Analysis}\label{sec:real.data}
The data analyzed were collected during a single drinking session that was conducted in Dr. Susan Luczak’s laboratory at the University of Southern California as part of a larger study involving 40 participants. This human subjects research was approved by the USC Institutional Review Board and written informed consent was obtained prior to starting the drinking episode. In this session, a participant drank alcohol evenly over a 15-minute period at a dose designed to reach a peak of approximately .05mg\%. TAC was first measured 30 minutes prior to the first BrAC measurement of .000 (just as drinking commenced) and then BrAC readings continued until BrAC returned to 0.000 and TAC readings continued until TAC returned to 0.000. Once drinking began, TAC and BrAC observations were taken approximately every 10 minutes except while drinking (due to mouth alcohol interfering with the breathalyzer's ability to accurately capture BrAC). After BrAC returned to .000, BrAC was no longer measured and TAC observations were taken every 30 minutes. The first non-zero TAC observation was 67 minutes after the first non-zero BrAC observation. TAC and BrAC observations were taken over 6.3 hours.

A total of 70 TAC measurements were taken with two TAC devices, one worn on each arm of the subject, and 28 BrAC measurements were taken with the breath analyzer.  Because of the variability in how the TAC devices trigger readings and the time it takes for the breath analyzer to be ready to take a reading, the TAC and BrAC times may not have temporally coincided exactly, but were within minutes of one another while BrAC was greater than 0.000.

In general our theory accommodates the situation where multiple devices are used, as it assumes only that the observations are unbiased with independent Gaussian errors having the same variance. 
\begin{figure}[H]
	\centering
	\includegraphics[width=16cm,height=8cm]{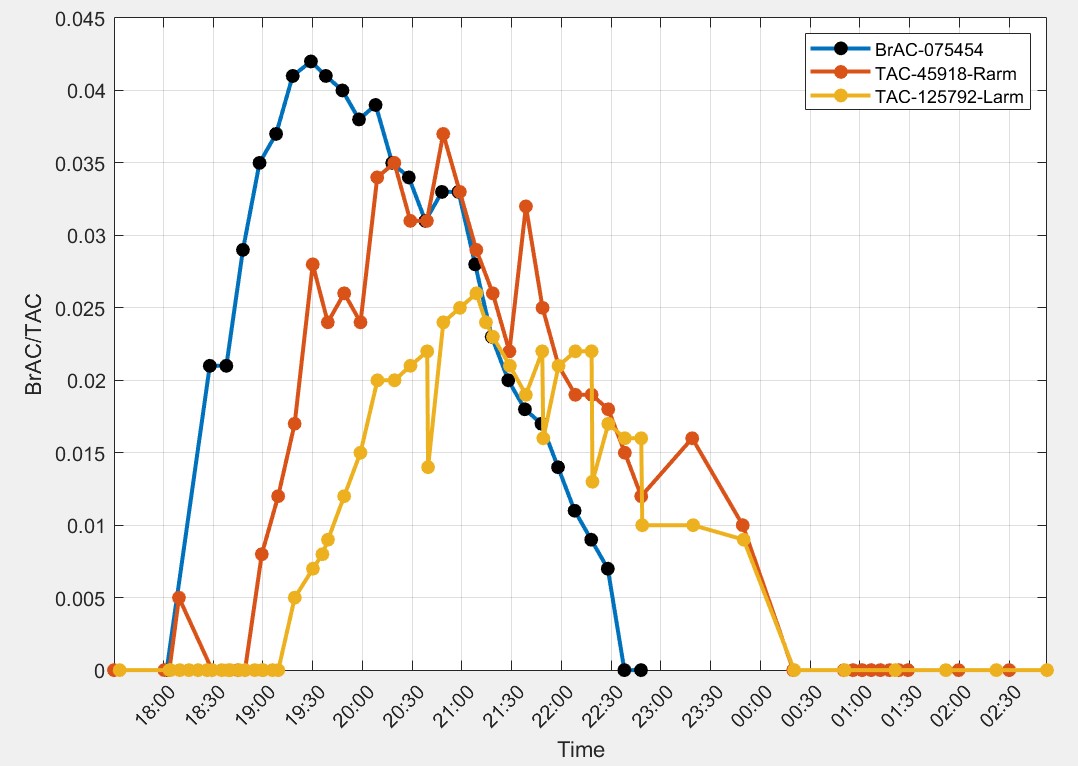}
	\caption{BrAC from recordings of two TAC device observations of dataset BT311 Session1 .}
	\label{fgr:TwoDevices}
\end{figure}

Nevertheless, as indicated in Figure \ref{fgr:TwoDevices}, the two devices used in this experiment exhibited some variation in the magnitude of their responses, which likely inflated the estimate of the standard deviation, thus also enlarging the confidence bands and contributed to the numerical instability of the deconvolution. In particular, we anticipate improved performance of our results when any multiple devices used in a single experiment are more equally calibrated. 


Minimizing \eqref{def:J and U} for this data resulted in the estimate $\widehat{\bm q}=(0.5611, 0.7655)$. The cyan curve in the figure was computed as for the experiments in Section \ref{subsub:brac}, that is, using  \eqref{eq:defbsybetan}, and thus with no constraints when least squares optimizing. The purple curve was created by further constraining all basis coefficients to be non-negative, thus producing a non-negative BrAC curve, which is visibly close to the true BrAC.

%


\begin{figure}[H]
	\centering
	\includegraphics[width=16cm,height=8cm]{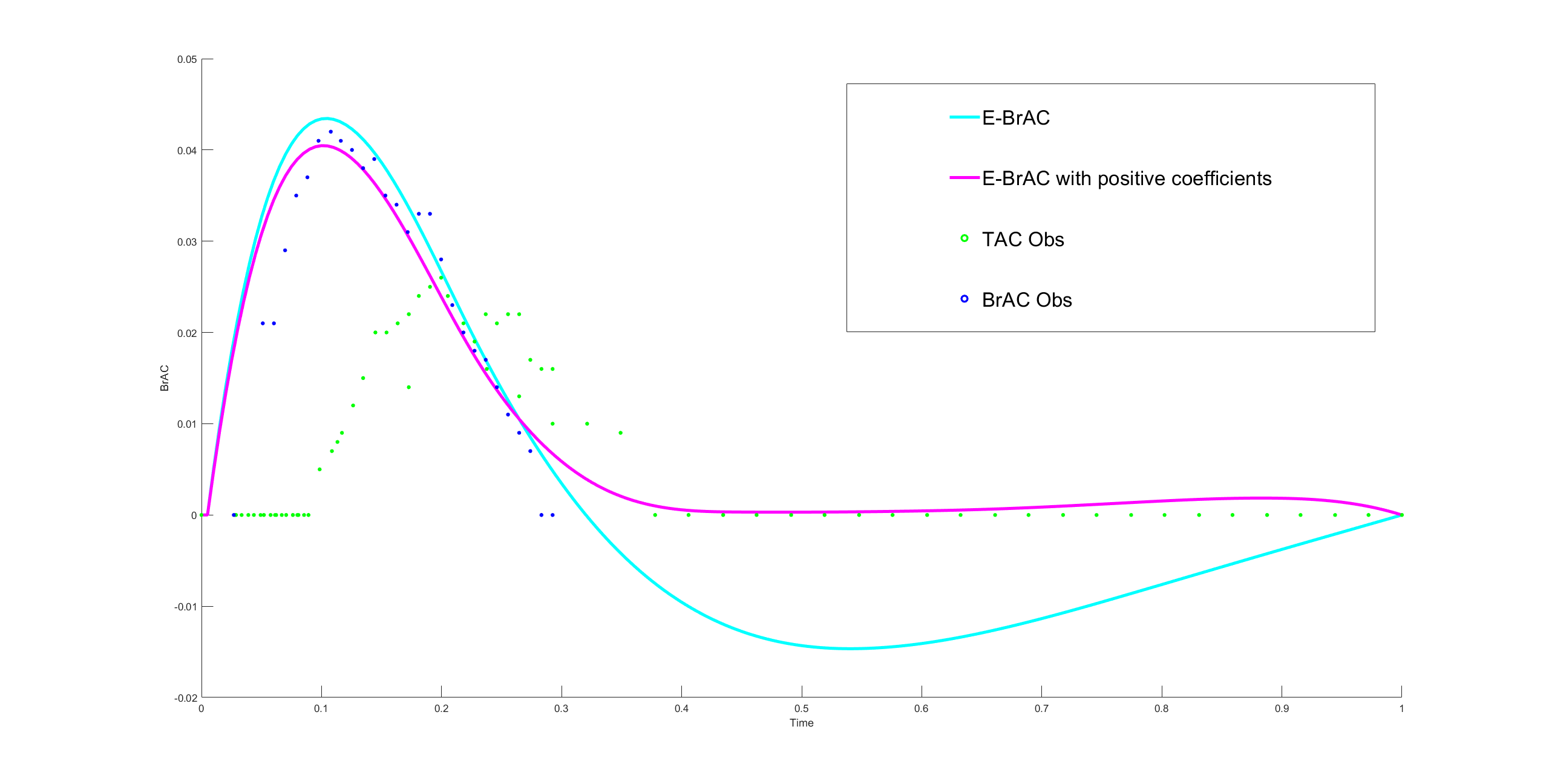}
	\caption{BrAC from TAC observations of dataset BT311 Session1 06132019 and Estimated BrAC that results from using the minimizer $\widehat{\bm q}=(0.5611, 0.7655)$. Estimated standard deviation of TAC errors is $2.5 \times 10^{-3}$, both regularization parameters set to $0.006$.}
	\label{fgr:real}
\end{figure}
Minimizing \eqref{def:J and U} for this data resulted in the estimate $\widehat{\bm q}=(0.5611, 0.7655)$. The cyan curve in the figure was computed as for the experiments in Section \ref{subsub:brac}, that is, using  \eqref{eq:defbsybetan}, and thus with no constraints when least squares optimizing. The purple curve was created by further constraining all basis coefficients to be non-negative, thus producing a non-negative BrAC curve, which is visibly close to the true BrAC.

\ignore{
\subsection{Fourier Methods}
Before proceeding to the proof of Theorem  \ref{thm:n=1.asy.dist.BrAC} let us introduce some notation. Recall that from (\ref{eq:fij.q.only}) we have
\begin{align}\label{def:kernel}
f_{\mu}(t;\bm{q})=\int_0^t Ce^{At}B\mu(s)ds=\int_0^t K(t-s;\bm{q})\mu(s)ds \qmq{where} K(t;\bm{q})=Ce^{At}B
\end{align}
Letting  $\hat{K}(t;\bm{q})$, $\hat{f}(t;\bm{q})$ and $\hat{\mu}(t)$ be the Laplace transforms of $K(t;\bm{q})$,$ f_{\mu}(t;\bm{q})$ and $\mu(t)$ respectively we obtain the following results .

\begin{lemma}\label{lem:Laplace.tr.K}
Let  $K(s;\bm{q})$ be given in (\ref{def:kernel}) where $A$ and $B$ are given in (\ref{def:A,B})and $t\in[0,T]$ then 
\begin{align*}
\hat{K}(t;\bm{q})=-C \sum_{n=0}^{\infty}\sum_{k=0}^n \frac{A^nT^{n-k}e^{-sT}}{(n-k)!s^{k+1}}B
\end{align*}
\end{lemma}
}

\section*{Acknowledgements} This work was supported by National Institutes of Health grant R01-AA026368.  We thank the members of the Luczak laboratory, including Emily Saldich, for their assistance with data collection and management.


\def\cprime{$'$}

\end{document}